%% file: ms.tex
\documentclass[twocolumn]{aastex631}

\usepackage{amsmath,amssymb,bm}
\usepackage{braket}

\shorttitle{CMB Line-of-Sight Distortion Field Analysis}
\shortauthors{{\keckarray} and {\bicep} collaboration}

\graphicspath{{./}{figures/}}

\begin{document}

\input{defs}

\title{BICEP/$Keck$ XVII:  Line of Sight Distortion Analysis: Estimates of Gravitational Lensing, Anisotropic Cosmic Birefringence, Patchy Reionization, and Systematic Errors}

\input{authors}

\begin{abstract}
We present estimates of line-of-sight distortion fields derived from the 95\,GHz and 150\,GHz data taken by BICEP2, BICEP3, and \textit{Keck Array} up to the 2018 observing season, leading to cosmological constraints and a study of instrumental and astrophysical systematics.
Cosmological constraints are derived from three of the distortion fields concerning gravitational lensing from large-scale structure, polarization rotation from magnetic fields or an axion-like field, and the screening effect of patchy reionization.
We measure an amplitude of the lensing power spectrum $A_L^{\phi \phi}=0.95\pm 0.20$.
We constrain polarization rotation, expressed as the coupling constant of a Chern-Simons electromagnetic term $g_{a\gamma} \leq 2.6\times10^{-2}/H_I$, where $H_I$ is the inflationary Hubble parameter, and an amplitude of primordial magnetic fields smoothed over $1$Mpc $B_{1\text{Mpc}} \leq 6.6 \;\text{nG}$ at 95\,GHz.
We constrain the root mean square of optical-depth fluctuations in a simple "crinkly surface" model of patchy reionization, finding $A^\tau<0.19$ ($2\sigma$) for the coherence scale of $L_c=100$.
We show that all of the distortion fields of the 95\,GHz and 150\,GHz polarization maps are consistent with simulations including lensed-$\Lambda$CDM, dust, and noise, with no evidence for instrumental systematics.
In some cases, the {\it EB} and {\it TB} quadratic estimators presented here are more sensitive than our previous map-based null tests at identifying and rejecting spurious {\bmode}s that might arise from instrumental effects.
Finally, we verify that the standard deprojection filtering in the {\bicep/\keck} data processing is effective at removing temperature to polarization leakage.
\end{abstract}

\keywords{Observational cosmology (1146) --- Cosmic microwave background radiation (322) --- Weak gravitational lensing (1797) --- Primordial magnetic fields (1294) --- Reionization (1383)}

\section{Introduction} \label{sec:intro}
Even with many orders of magnitude improvement in the precision of measurements, primordial CMB fluctuations remain statistically isotropic, such that their statistics are well described by angular power spectra. On the other hand, multiple secondary effects after recombination distort the primary CMB fluctuations inducing new correlations among observed CMB fluctuations. Examples include gravitational lensing by large-scale structure \citep{Zaldarriaga1998}, patchy re-ionization that modulates the amplitude of the CMB fields \citep{Hu2000,HuPatchyScreening}, and cosmic birefringence that rotates the CMB polarization angle \citep{AxionCosmologyrev, Yadav2012}. There are also various instrumental systematics that can generate spurious {\bmode}s by distorting the incoming $T,Q,U$ fields, most notably the temperature to polarization ($T$ to $P$) leakage caused by beam and gain mismatches \citep{biceptwoIII, biceptwoXI}, and $E$ to $B$ leakage from errors in polarization angle calibration. A comprehensive investigation of the statistical properties of the temperature and polarization maps can be used as a powerful tool to distinguish the sources of the observed {\bmode}s, deciding whether they are cosmological or instrumental. \\

The secondary and instrumental effects listed above are similar in that they can be described as distortion effects that mix the Stokes $T$, $Q$, and $U$ fields along or around each line-of-sight direction $\hn$. 
\cite{Yadav2010} characterize distortions of  the primordial CMB fluctuations by introducing 11 distortion fields which depend on the line-of-sight direction. The {\bmode}s generated by these map distortions would have correlations with {\it E} or {\it T} that do not exist in the primordial signal in standard \lcdm\  cosmology. Thus {\it EB} and {\it TB} correlations can  be used to reconstruct the distortion fields and study the physical processes and instrumental systematic effects that are associated with specific types of distortions. \\

In this paper, we reconstruct the 11 distortion fields by applying the minimum variance {\it EB} and {\it TB} quadratic estimators derived in \cite{Yadav2010} and \cite{Hu2002} to our observed {\bmode} signal and use their power spectra $\hat{C}_L^{DD}$ to constrain cosmological models and systematics.
We will be referencing the previous publications from the {\bicep}/{\keck} (BK) experiments, (\cite{biceptwoI}, hereafter BK-I; \cite{biceptwoII}, hereafter BK-II; \cite{biceptwoIII}, hereafter BK-III; \cite{biceptwoVII}, hereafter BK-VII; \cite{biceptwoVIII}, hereafter BK-VIII; \cite{biceptwoIX}, hereafter BK-IX; \cite{biceptwoX}, hereafter BK-X; \cite{biceptwoXIII}, hereafter BK-XIII).\\

This paper is organized as follows: In Section~\ref{Section: Introduction to the distortion fields}, we give an overview of the different distortion fields and some background on the cosmological effects that correspond to some of the distortions. In Section~\ref{section: Data and Simulations}, we describe our data and simulations used for the distortion field analysis. In Section~\ref{section: Analysis of the Distortion Fields}, we outline the analysis method, including how to go from $Q$ and $U$ maps to an unbiased distortion field power spectrum and how to combine the distortion power spectra from two data sets. In Section~\ref{sec: Cosmological signal corresponding to Distortion Fields}, we use the power spectra of three of the reconstructed distortion fields to set constraints on gravitational lensing, patchy reionization, and cosmic birefringence.
In Section~\ref{section: mbs: Distortion Fields as systematics check}, we discuss the instrumental effects that could produce distortion effects in our data and test for residual systematic effects in the {\bicep}/{\keck} data with the distortion field spectra.

\section{Introduction to the distortion fields} \label{Section: Introduction to the distortion fields}

In \cite{Hu2003}, systematic effects in CMB polarization maps are described as modifications to the Stokes $Q$ and $U$ maps by distortions along the line-of-sight $\hn$. Following \cite{Yadav2010}, we model these distortions with 11 distortion fields as
\begin{widetext}
\begin{equation}
\begin{split}
\delta [Q\pm iU](\hn) &= [\tau\pm i2\alpha](\hn)[\Qt \pm i\Ut](\hn) + [f_1\pm i f_2](\hn)[\Qt \mp i\Ut](\hn) + [\gamma_1\pm i\gamma_2](\hn)\Tt(\hn) \\     
& + \sigma \bm{p}(\hn)\cdot\nabla[\Qt \pm i\Ut](\hn) + \sigma[d_1\pm id_2](\hn)[\partial_1\pm i\partial_2]\Tt(\hn) + \sigma^2q(\hn)[\partial_1\pm i\partial_2]^2\Tt(\hn) \,,
\label{Eq.mbs: Characterization of 11 distortion fields}
\end{split}
\end{equation}
\end{widetext}
where $\Tt$, $\Qt$ and $\Ut$ stand for the un-distorted primordial CMB intensity and polarization fields.
There are 11 terms; $\bm{p}(\hn)$ is a two-dimensional vector. Quantities in the top line correspond to distortions along a unique line of sight while the second line shows field mixing in the neighborhood of a single line $\hn$. The quantity $\sigma$ denotes a chosen length scale for these terms and makes the distortion fields unitless.
The operators $\partial_1$ and $\partial_2$ represent the covariant derivatives along the RA and Dec. directions, and $\nabla[\Qt \pm i\Ut]$ is the gradient with components $\partial_i [\Qt \pm i\Ut]$. \\

\cite{Yadav2010} further show that these distortion fields can be estimated 
directly using quadratic combinations of the data. The filter weights $f^{EB}_{\ell_1,\ell_2}$ and $f^{TB}_{\ell_1,\ell_2}$ used in the construction of each of the eleven distortions from power spectra are shown in Table \ref{table: TB/EB quadratic estimator weights}. Note that this table differs from a similar table in \cite{Yadav2010} in that we use a different notation for pixel-space and harmonic-space quantities, denoting the latter with alphabetical instead of numerical subscripts. Also, the weights to construct perturbations to {\it E} and {\it B} from these distortions have been omitted, since we do not use them. See Appendix A and Section \ref{section: Analysis of the Distortion Fields} for more details. \\
 
Each of the distortion fields can be matched with a specific source, offering a rich phenomenology. The $\tau(\hn)$ field describes a modulation of the amplitude of the 
polarization maps, $\alpha(\hn)$ describes rotation of the polarization angle, $f_1(\hn)$ and $f_2(\hn)$ describe the coupling between the two polarization field spin states, the components of $\bm{p}(\hn)$, $p_1(\hn)$ and $p_2(\hn)$ describe the change in photon direction, $\gamma_1(\hn)$ and $\gamma_2(\hn)$ describe monopole $T$ to $P$ leakage, $d_1(\hn)$ and $d_2(\hn)$ describe dipole $T$ to $P$ leakage, and $q(\hn)$ describes quadrupole $T$ to $P$ leakage. All 11 distortion fields can correspond to specific potential instrumental systematic effect. We will discuss them in depth in that context in Section~\ref{section: mbs: Distortion Fields as systematics check}. \\

Among the distortion fields in Eq.~\ref{Eq.mbs: Characterization of 11 distortion fields}, there are three that correspond to known or conjectured cosmological signals.
These are $\bm{p}(\hn)$: change of direction of the CMB photons, $\tau(\hn)$: amplitude modulation, and $\alpha(\hn)$: rotation of the plane of linear polarization. \\

CMB photons traveling from the last scattering surface are deflected by the intervening matter along the line of sight \citep{Zaldarriaga1998}. The change of photon direction, $\bm{p}(\hn)$,
is referred to as the weak gravitational lensing of the CMB. \\

The lensing potential is commonly decomposed into {\it gradient} and {\it curl} lensing potentials, $\Phi$ and $\Omega$ \citep{Hirata2003,Cooray2005,Namikawa2012}, such that the lensed $Q$ and $U$ maps can be described as
\begin{equation}
    [Q \pm iU](\hn) = [\Tilde{Q} \pm i\Tilde{U}](\hn +\nabla \Phi + \nabla\times\Omega) \,,
\end{equation}
where the gradient $\nabla \Phi$ has components $\partial_i \Phi$ and the curl $\nabla \times \Omega$ has components $\epsilon_{ij}\partial_j\Omega$, where $\epsilon_{ij}$ is the antisymmetric symbol.
To leading order, we obtain the map distortions
\begin{equation}
    \delta[Q \pm iU](\hn) = \nabla \Phi \cdot \nabla [\Tilde{Q} \pm i\Tilde{U}] + \nabla \times \Omega \cdot \nabla [\Tilde{Q} \pm i\Tilde{U}](\hn) \,,
\end{equation}
which allows us to identify $\nabla \times \Omega$ and $\nabla \Phi$ with the curl and gradient mode of $\bm{p}(\hn)$, respectively.
The gradient component of CMB lensing, $\Phi$, is generated by the linear order density perturbations, while the curl component, $\Omega$, is only generated by second order effects in scalar density perturbations or lensing by for example gravitational waves or cosmic strings \citep{Dodelson2003,Cooray2005,Yamauchi2012}. We expect these cosmological signals to be negligible \citep{Hirata2003,Pratten2016,Fabbian2018}. \\

The (gradient) CMB lensing potential power spectrum has been measured to high precision by many experiments using temperature, polarization, or both \citep{ACTPolLensing2017,Plancklensing2018,SPTPolLensing2019,polarbearlensing2020,Carron2022}. \\

The distortion field $\tau(\hn)$ (amplitude modulation) can be generated by inhomogeneities in the reionization process, also referred to as \textit{patchy reionization}.
In addition to the kinematic Sunyaev-Zeldovich (kSZ) signal generated by the peculiar motion of ionized gas \citep{SZ1970,SZ1980},
patchy reionization causes an uneven screening effect of photons \citep{HuPatchyScreening}.
The screening effect is described as
\begin{align}
    (Q\pm iU)(\hn) &= e^{-\tau_0(\hn)}(\Tilde{Q}\pm i\Tilde{U})(\hn) \label{Eq. patchy tau} \\
    \Rightarrow  \delta (Q\pm iU)(\hn) &\approx -\tau_0(\hn)(\Tilde{Q}\pm i\Tilde{U})(\hn) \,,
\end{align}
where $\tau_0(\hn)$ is the optical depth to recombination that varies for different line-of-sight directions $\hn$. Taylor expanding Eq.~\ref{Eq. patchy tau}, the screening effect from patchy reionization generates the distortion field $\tau(\hn)$. \\

The details of the patchy reionization process are still largely unknown. Recent searches for the redshifted 21-cm signal from neutral hydrogen by EDGES put a lower bound on the duration of reionization as $\Delta z \gtrsim 0.4$ \citep{EDGES2017}. The kSZ power obtained from the South Pole Telescope prefers $\Delta z \lesssim 4.1$ \citep{SPTkSZ2021,Gorce2022}. The constraints from {\planck} CMB temperature and polarization power spectra suggest that reionization occured at $z_{\text re}\approx 8$ with a duration of $\Delta z \lesssim 2.8$ \citep{PlanckEOR2016}. Previous work that studies patchy reionization through $\tau(\hn)$ reconstructions with CMB temperature and polarization include \cite{Gluscevic2013} and \cite{Namikawa2018}. We constrain the same {\it crinkly-surface} model of patchy reionization where the power spectrum of the optical-depth is given by
\begin{equation}
    C_L^{\tau \tau} = \left( \frac{A^\tau}{10^4}\right) \frac{4\pi}{L_c^2} e^{-L^2/L_c^2} \,,
\label{Eq. mbs: patchy reionization fiducial model spectra}
\end{equation}
with the amplitude $A^\tau$ and the coherence length $L_c$ \citep{Gluscevic2013}.
\\

The distortion field $\alpha(\hn)$ can be generated by a cosmic birefringence field that rotates the primordial $\Tilde{Q}$ and $\Tilde{U}$ according to
\begin{align}
    [Q\pm iU](\hn) &= e^{\pm 2i\alpha(\hn)}[\Tilde{Q}\pm i \Tilde{U}](\hn), \\
    \Rightarrow   \delta (Q\pm iU)(\hn) &\approx \pm i2\alpha(\hn) [\tilde{Q}\pm i\tilde{U}](\hn) \,.
\end{align}
Two potential physical processes that can cause a rotation field $\alpha(\hn)$ are the coupling of CMB photons with pseudo-scalar fields through the Chern-Simons term, also described as parity-violating physics, and Faraday rotation of the CMB photons due to interactions with background magnetic fields. A massless axion-like pseudo-scalar field $a$ that couples to the standard electromagnetic term has the Lagrangian density \citep{Carroll1990}:
\begin{equation}
    \mathcal{L} \supset \frac{g_{a\gamma}}{4}aF_{\mu \nu}\Tilde{F}^{\mu \nu} \,,
\end{equation}
where $g_{a\gamma}$ is the coupling constant between the axion-like particles and photons, and $F_{\mu \nu}$ is the electromagnetic field tensor. The amount of rotation is given by: 
\begin{equation}
    \alpha = \frac{g_{a\gamma}}{2}\Delta a \,.
\end{equation}

When the pseudo-scalar field fluctuates in space and time, the change of the field integrated over the photon trajectory, $\Delta a$, varies across the sky and generates an anisotropic cosmic rotation field $\alpha(\hn)$. For a massless scalar field, the large-scale limit ($L \lesssim 100$) of the expected cosmic rotation power spectra is described by \citep{Caldwell2011}
\begin{equation}
    \sqrt{\frac{L(L+1)C_L^{\alpha \alpha}}{2\pi}}= \frac{H_{I}g_{a\gamma}}{4\pi} \,,
\label{Eq. mbs: ACB to axion like particles interaction g_ag}
\end{equation}
where $H_I$ is the inflationary Hubble parameter. \\

A second physical process that could generate a cosmic rotation field $\alpha(\hn)$ is Faraday rotation of the CMB photons by primordial magnetic fields (PMFs). In the large-scale limit ($L \lesssim 100$), the cosmic rotation power spectra generated by a nearly scale-invariant PMF is (\cite{De2013,Yadav2012}):
\begin{equation}
    \sqrt{\frac{L(L+1)C_L^{\alpha \alpha}}{2\pi}} = 
    1.9\times 10^{-4} \left( \frac{\nu}{150\text{GHz}} \right)^{-2} 
    \left( \frac{B_{\text{1Mpc}}}{1 \; \text{nG}} \right) \,,
\label{Eq. mbs: ACB constraints to PMF nG}
\end{equation}
where $\nu$ is the observed CMB frequency, and $B_{\text{1Mpc}}$ is the strength of the PMFs smoothed over 1~Mpc. The $\nu^{-2}$ frequency scaling of the Faraday rotation angle implies that a lower frequency offers better leverage for PMF measurement. \\

Observations from multiple CMB experiments have been employed to derive constraints on anisotropies of the cosmic birefringence using $\alpha(\hn)$ reconstructions, which includes WMAP \citep{Gluscevic2012WMAPACB}, POLARBEAR \citep{POLARBEARACB2015}, {\bicep/\keck} (BK-IX), {\planck} \citep{PlanckACB2017,Gruppuso2020,Bortolami2022}, the Atacama Cosmology Telescope (ACT) \citep{Namikawa2020}, and the South Pole Telescope (SPT) \citep{Bianchini2020}.\\

\section{Data and Simulations} \label{section: Data and Simulations}

In this paper, we use {\bicep}/{\keck} maps that use data up to and including the 2018 observing season, referred to as the BK18 maps. In particular, we will focus on the two deepest maps: the 150\,GHz map from 
\biceptwo\ and \keckarray\ data
which achieves 2.8\,$\mu$K-arcmin over an effective area of around 400\ square degrees, and the 95\,GHz map from BICEP3 which achieves 2.8\,$\mu$K-arcmin over an effective area of around 600\ square degrees. These two data sets with the lowest noise levels are the most interesting for studying both the cosmological and instrumental effects related to the distortion fields. \\

We construct an apodization mask that down-weights the noisier regions of the $T$, $Q$, and $U$ maps. For the polarized $Q$ and $U$ map, we use a smoothed inverse variance apodization mask similar to that used in BK-XIII. For the $T$ map, we add constant power of 10\,\uksq\ to the smoothed noise variance and invert it to construct the apodization mask. The mask is similar to a Wiener filter with flat weights in the central region dominated by sample variance and an inverse variance weight at the edges of the map. Additionally, for the analysis of $T$ to $P$ distortions, we mask the 20 point sources with the largest polarized fluxes from a preliminary SPT-3G catalog by applying a 0.5$\deg$ wide Gaussian divot at the location of each point source in the apodization mask. The effects of point sources are discussed in Section~\ref{subsection: Effects of point source contamination} and Appendix~\ref{appendix: Detection of point sources}. \\

We reuse the standard sets of simulations described in BK-XIII and previous papers: lensed {\lcdm} signal-only simulations constrained to the {\planck} $T$ map (denoted by lensed-\lcdm), sign-flip noise realizations, and Gaussian dust foreground simulations, each having 499 realizations. The details of the CMB signal and noise simulations are described in Section V of BK-I and the dust simulations are described in Section IV.A of \cite{bkp} and Appendix E of \cite{biceptwoVI}. For estimating the noise bias of the distortion spectra constructed with {\it TB} estimators (described in Section~\ref{section: mbs: estimating the distortion field power spectra}), an additional set of lensed CMB signal-only simulations with unconstrained temperature are generated. \\

In addition to the standard simulation sets, we also generate simulations of random Gaussian realizations of the distortion fields, $D(\hn)$, that are characterized by certain power spectra. For pipeline verification and calibration of the normalization factors (Eq.~\ref{Eq. mbs: compute normalization with sims}), we use simulations described by a scale-invariant distortion spectrum,
\begin{equation}
    \frac{L(L+1)}{2\pi}C_L^{DD} = A^D_\textrm{fid.} \,,
\label{Eq. mbs: red input spectra}
\end{equation}
with fiducial amplitudes, $A^D_\textrm{fid.}$, and their specific values for each distortion field type given in Table \ref{table: fiducial amplitudes}.
\begin{table}[ht!]
\centering
\begin{tabular}{c | c c c c c c c} 
 \hline \hline 
 \rule{0pt}{4ex}
$D$ & $\tau$ & $\alpha$ & $\gamma$ & $f$ & $d$ & $q$ & $p$ \\ [0.5ex] 
    \hline
    \rule{0pt}{3ex}
 $A^D_\textrm{fid.}$ & $10^{-3}$ & $10^{-4}$ & $10^{-6}$ & $10^{-4}$ & $10^{-10}$ & $10^{-14}$ & $10^{2}$ \\[0.5ex]
    \hline \hline
\end{tabular}
\caption{The fiducial amplitude for the scale-invariant power spectrum used as input for the Gaussian simulations for the calibration of the normalization.}
\label{table: fiducial amplitudes}
\end{table}

For the amplitude modulation field $\tau(\hn)$, we generate Gaussian simulations of $\tau(\hn)$ according to the power spectrum in Eq.~\ref{Eq. mbs: patchy reionization fiducial model spectra}. For comparing the sensitivity between quadratic estimators and {\it BB} power spectra for detecting distortion systematics, we generate Gaussian realizations of distortion fields with a scale-invariant spectrum within a narrow range of multipoles ($\Delta L=50$). \\

 The distortion field simulations and the unconstrained temperature simulations are generated with the \textit{observation matrix} $\mathcal{R}$ described in BK-VII.
 This matrix captures the entire map-making process including the observing strategy, the timestream filtering, and the deprojection of leading order beam systematics.
 Simulations are rapidly generated with matrix multiplications: 
\begin{equation}
\begin{pmatrix}
Q^{\text{obs}}\\
U^{\text{obs}}
\end{pmatrix}
= \mathcal{R}
\begin{pmatrix}
Q^{\text{in}}\\
U^{\text{in}}
\end{pmatrix}
+
\begin{pmatrix}
Q^{\text{noise}}\\
U^{\text{noise}}
\end{pmatrix}
\,,
\label{Eq. observation matrix}
\end{equation}
where $Q^{\text{in}}$ and $U^{\text{in}}$ are input signal maps, $Q^{\text{noise}}$ \& $U^{\text{noise}}$ are sign-flip noise realizations, and $Q^{\text{obs}}$ and $U^{\text{obs}}$ are "as observed" output maps. 
Because of $TE$ correlation in \lcdm\ such simulations are not fully accurate when the input $T$ sky is not the same as that assumed in the construction of the deprojection operation which is built into the observing matrix. \\

\section{Analysis of the Distortion Fields} 
\label{section: Analysis of the Distortion Fields}
\subsection{Quadratic Estimator Construction}

Since the BK-observed patch is relatively small (1-2\% of the total sky), we work in the flat-sky limit using Fourier transforms. A complex field, $D_1(\hn) \pm i D_2(\hn)$, of spin $s$ can be represented by its Fourier transform
\begin{equation}
  \notag
    [D_a \pm iD_b]_\Bl = (\pm 1)^s \int d\hn [D_1(\hn)\pm iD_2(\hn)] e^{\mp is\phi_\Bl} e^{-i\Bl\cdot \hn} \,,
\label{Eq. mbs: Harmonics component of distortion fields.}
\end{equation}
where $\phi_\Bl = \cos^{-1}(\hn \cdot \hat{\Bl})$. In particular, we note that $\tau$, $\omega$ and $q$ are spin\nobreakdash-0 fields, $p_1 \pm i p_2$ and $d_1\pm i d_2$ are spin\nobreakdash-1 fields, $\gamma_1\pm i\gamma_2$ are spin\nobreakdash-2 fields and $f_1\pm i f_2$ are spin\nobreakdash-4 fields. We transform between even-parity modes $D_a$ and odd-parity modes $D_b$, and modes aligned with the RA/Dec coordinate system of the underlying maps, $D_1$ and $D_2$, with a rotation
\begin{align}
    [D_a]_\Bl &=+[D_1]_\Bl \cos(s\phi_\Bl) + [D_2]_\Bl \cos(s\phi_\Bl)\\
    [D_b]_\Bl &=-[D_1]_\Bl \cos(s\phi_\Bl) + [D_2]_\Bl \cos(s\phi_\Bl)
\end{align}
for fields with even-valued spin or
\begin{align}
    [D_a]_\Bl &=-i [D_1]_\Bl \cos(s\phi_\Bl) + i [D_2]_\Bl \cos(s\phi_\Bl)\\
    [D_b]_\Bl &=+i [D_1]_\Bl \cos(s\phi_\Bl) + i [D_2]_\Bl \cos(s\phi_\Bl)
\end{align}
for fields with odd-valued spin. \\

In Appendix \ref{section: Appendix: Minimal QE}, we show that one can construct unbiased minimum variance {\it TB} and {\it EB} quadratic estimators for each distortion field given by
\begin{align}
    \bar{D}^{XB}_\Bl&={A}^{D,XB}_L \int \frac{d^2\bl_1}{(2\pi)^2}X_{\bl_1}B_{\bl_2} \frac{f^{D,XB}_{\bl_1, \bl_2}}{C_{\bl_1}^{XX} C_{\bl_2}^{BB}}\,, \label{Eq. mbs: general quadratic estimator}\\
    \hat{D}^{XB}_\Bl& = \bar{D}^{XB}_\Bl - \braket{\bar{D}^{XB}_\Bl} \,,
    \label{Eq. mbs: general quadratic estimator mean-field}
\end{align}
where $\Bl=\bl_1+\bl_2$, $X$ may be $T$ or $E$ and $C_{l_1}^{XX}$, $C_{l_2}^{BB}$ are the total observed power spectra including contributions from the noise and lensing. These estimators directly reconstruct the Fourier transform of the map distortions introduced in Eq.~\ref{Eq.mbs: Characterization of 11 distortion fields}, which are denoted by alphabetical subscripts $a$, $b$. The specific filter functions $f^{D,XB}_{\bl_1, \bl_2}$ for each distortion field $D$ and estimator $XB$ are listed in Table~\ref{table: TB/EB quadratic estimator weights}. Eq.~\ref{Eq. mbs: general quadratic estimator mean-field} shows the correction for the mean-field bias, which is estimated from simulations \citep{Namikawa2014}. \\
\begin{table}[ht!]
\centering
\begin{tabular}{c c} 
 \hline \hline 
 \rule{0pt}{4ex}
$D$ & $f^{XB}_{\bl_1,\bl_2}$   \\ [0.5ex] 
    \hline
    \rule{0pt}{3ex}
    $\tau$      & $\CXE \sin 2(\phi_{\bl_1}-\phi_{\bl_2})$ \\ [0.5ex] 
    $\alpha$    & $2 \CXE \cos 2(\phi_{\bl_1}-\phi_{\bl_2})$   \\[0.5ex] 
    $\gamma_a$  & $\CTX \sin 2(\phi_{\Bl}-\phi_{\bl_2})$   \\[0.5ex] 
    $\gamma_b$  & $\CTX \cos 2(\phi_{\Bl}-\phi_{\bl_2})$  \\ [0.5ex] 
    $f_a$       & $\CXE \sin 2(2\phi_{\Bl}-\phi_{\bl_1}-\phi_{\bl_2})$   \\[0.5ex] 
    $f_b$       & $\CXE \cos 2(2\phi_{\Bl}-\phi_{\bl_1}-\phi_{\bl_2})$  \\[0.5ex] 
    $d_a$       & $\CTX (\bl_1 \sigma) \cos 2(\phi_{\Bl}+\phi_{\bl_1}-2\phi_{\bl_2})$  \\[0.5ex] 
    $d_b$       & $-\CTX (\bl_1 \sigma) \sin 2(\phi_{\Bl}+\phi_{\bl_1}-2\phi_{\bl_2})$ \\[0.5ex]
    $q$         & $-\CTX (\bl_1\sigma)^2 \sin 2(\phi_{\bl_1}-\phi_{\bl_2})$ \\ [1ex] 
    $p_a = \Omega$    & $-\CXE \sigma (\bl_1 \times \hat{\Bl}) \sin 2(\phi_{\bl_1}-\phi_{\bl_2})$  \\[0.5ex] 
    $p_b = \Phi$ & $-\CXE \sigma (\bl_1 \cdot \hat{\Bl}) \sin 2(\phi_{\bl_1}-\phi_{\bl_2})$  \\[0.5ex] 
    \hline \hline
\end{tabular}
\caption{Filter functions $f^{D,XB}_{\bl_1, \bl_2}$ for the different distortion field estimators as introduced in Eq.~\ref{Eq. mbs: general quadratic estimator}. Here $X$ can either be $T$ or $B$ in order to obtain the filter functions for the $TB$ or $EB$ estimator, respectively. $\CTX$ and $\CXE$ are lensed CMB power spectra corresponding to our fiducial model. Note that for the distortion fields $p_a$ and $p_b$ we use the notation prevailing in CMB lensing, $\Omega$ and $\Phi$. }
\label{table: TB/EB quadratic estimator weights}
\end{table}

The analytical normalization factor is given by
\begin{equation}
    {A}^{D,XB}_L=\left[ \int \frac{d^2 \bl_1}{(2\pi)^2} \frac{(f^{D,XB}_{\bl_1,\bl_2})^2}{C_{l_1}^{XX} C_{l_2}^{BB}}   \right]^{-1} \,.
\end{equation}
In practice, we obtain the normalization factor empirically by running Monte Carlo simulations: 
\begin{equation}
    A^{D}_L = \frac{\braket{|D^{\text{in}}_\Bl|^2}}
    {\braket{D^{\text{in}}_\Bl(D^{\text{sim}}_\Bl)^*}} \,,
\label{Eq. mbs: compute normalization with sims}
\end{equation}
where $D^{\text{in}}_\Bl$ are the input distortion field Fourier modes, and $D^{\text{sim}}_\Bl$ are the un-normalized, reconstructed distortion modes. To obtain the input distortion Fourier modes $D^{\text{in}}_\Bl$, the same apodization mask for the $T$, $Q$, and $U$ maps are applied prior to the Fourier transform.
We use the scale-invariant distortion input simulations (Eq.~\ref{Eq. mbs: red input spectra}) to calibrate the normalization factor for all the distortion fields except for lensing ($\bm{p}(\hn)$ in Eq.~\ref{Eq.mbs: Characterization of 11 distortion fields}), where the standard lensed-\lcdm\ simulations are used. \\

We can construct {\it TB} estimators sensitive to the distortion fields only concerning polarization ($\tau$, $\alpha$, $f_1$, $f_2$, $\bm{p}$) due to the non-zero $TE$ correlation in the CMB. Therefore all 11 distortion fields can be probed by both the {\it EB} and {\it TB} estimators with the weights listed in Table~\ref{table: TB/EB quadratic estimator weights}. 
However, the polarization-only distortion fields are better measured with the {\it EB} estimators, while the {\it TB} estimators have higher sensitivity to the distortion fields involving $T$ to $P$ leakage ($\gamma_{1/2}$, $d_{1/2}$, $q$). \\

\subsection{Input E and B-modes for reconstruction}
\label{section: Input E and B modes for reconstruction}

As described in BK-VII, the mixing of $E$ and {\bmode}s due to map filtering and apodization are taken care of by the matrix-based purification method with purification matrices $\Pi_B$ and $\Pi_E$. The purified $E$ and $B$-mode-only maps are:
\begin{align}
\begin{pmatrix}
\hat{Q}^{E}\\
\hat{U}^{E}
\end{pmatrix}
      & =
\Pi^E
\begin{pmatrix}
Q^{\text{obs}}\\
U^{\text{obs}}
\end{pmatrix} \,,\\
\begin{pmatrix}
\hat{Q}^{B}\\
\hat{U}^{B}
\end{pmatrix}
       &=
\Pi^B
\begin{pmatrix}
Q^{\text{obs}}\\
U^{\text{obs}}
\end{pmatrix} 
\,,
\end{align}
where $Q^{\text{obs}}$ and $U^{\text{obs}}$ can be either a simulation or the real map. The Fourier transform of the purified $\hat{Q}^{E/B}$ and $\hat{U}^{E/B}$ are used to construct the purified $\hat{E}$ and $\hat{B}$ modes which are the input to the quadratic estimators. \\

Eq.~\ref{Eq. mbs: general quadratic estimator} minimizes the variance in the ideal case, ignoring beam smoothing, filtering, and apodization. In practice, transfer functions due to these effects must be compensated for in addition to $E$ and $B$-mode purification. The observed BK maps, the $\hat{E}$ and $\hat{B}$ Fourier modes are corrected by:
\begin{align}
    \Bar{X}_\bl &= t_{|\bl|}^X \hat{X}_\bl \,,   \\
    t_l^X &= \sqrt{\frac{C_l^{XX,\text{in}}}{C_l^{XX,\text{out}}}}
\,,
\end{align}
where $C_l^{XX,\text{in}}$ and $C_l^{XX,\text{out}}$ are the mean input and output spectra of the lensed-\lcdm\ signal-only simulations, and $t_l^X$ is the transfer function.\\

In practice, we find that using Fourier modes up to multipole of 600 yields the best signal-to-noise, whereas the $l=600-700$ modes are noisy and can worsen the signal-to-noise of the reconstruction due to a misestimation of $t^X$. Therefore, we use $l_{\text{max}}=600$ as our baseline reconstruction parameter. To avoid potential contamination by dust, we mask out the lowest multipoles and use $l_{\text{min}}^B=100$ for 95\,GHz and $l_{\text{min}}^B=150$ for 150\,GHz. The $l_{\text{min}}^B$ cutoff is chosen such that the observed dust {\bmode} power spectrum in the BK patch of sky is lower than the lensing {\bmode} power spectrum at $l > l_{\text{min}}^B$ using BK18 and {\planck} {\lcdm} best-fit {\bmode} power spectra. 

\subsection{Estimating the Distortion Field Power Spectra}
\label{section: mbs: estimating the distortion field power spectra}

The power spectrum of a distortion field can be estimated by squaring the estimator $\hat{D}(\Bl)$ from Eq.~\ref{Eq. mbs: general quadratic estimator}:
\begin{equation}
    \left< \left|\hat{D}_\Bl \right|^2 \right> = \hat{C}_\Bl^{DD} + \hat{N}_\Bl^{DD} \,,
\label{Eq. mbs: DL^2 = CL + NL}
\end{equation}
where $\hat{C}_\Bl^{DD}$ is the observed distortion field power spectrum and $\hat{N}_\Bl^{DD}$ is the noise bias. When there is no distortion field present, the main contribution for $\hat{N}_\Bl^{DD}$ is the disconnected $N^0$ bias, which can be estimated by the \textit{realization-dependent} method described in \cite{Namikawa2013} and BK-VIII: 
\begin{equation}
    \begin{split}
    \hat{N}^0_\Bl=&\braket{|\hat{D}^{E_1,\hat{B}}_\Bl + \hat{D}^{\hat{E},B_1}_\Bl|^2}_1 - \\
   & \frac{1}{2}\braket{|\hat{D}^{E_1,B_2}_\Bl + \hat{D}^{E_2,B_1}_\Bl|^2}_{1,2}
        \end{split}
        \,,
\label{Eq. mbs: rlz dependent N0 bias simple.}
\end{equation}
where $\hat{E}$ and $\hat{B}$ are the real $E$ and $B$ modes, or a given simulation realization. The 499 simulation realizations are divided into two sets of roughly equal size, and the subscripts 1 and 2 stand for the first and second sets of simulations. The first term is averaged over the first set of simulations, while the second term is averaged over the first and second set of simulations. \\

For the {\it TB} estimators that we use for systematics checks in Sec.~\ref{section: mbs: Distortion Fields as systematics check}, the realization-dependent bias is estimated in a similar manner to Eq.~\ref{Eq. mbs: rlz dependent N0 bias simple.} but with $T$ instead of $E$. Since the standard lensed-\lcdm\ simulations are generated with the temperature sky fixed to the Planck $T$ map (see BK-I), an additional set of simulations with unconstrained $T$ are used as the simulation sets 1 and 2 to be averaged over. The realization-dependent bias is evaluated for the observed $|\hat{D}_\Bl^{\hat{T}\hat{B}}|^2$ and for each of the reconstructed distortion bandpowers of the 499 standard constrained-$T$ simulations. See Sec.~\ref{subsection: Distortion field null tests} for more details.  \\

When there is a distortion field signal, apart from the disconnected $N^0$ bias, there is an additional bias term that is proportional to the amplitude of the distortion field signal, referred to as the $N^1$ bias \citep{KesdenLensingN1bias}. The $N^1$ bias can be estimated with two sets of simulations sharing the same distortion field realization \citep{Story2015}:
\begin{equation}
    \hat{N}^1_\Bl=\braket{|\hat{D}^{E_1,B_2}_\Bl + \hat{D}^{E_2,B_1}_\Bl|^2}_{1,2} - \braket{\hat{N}^0_\Bl} \,,
\end{equation}
where the subscripts 1,2 stand for the two sets of simulations with different CMB/noise realizations that share the same set of distortion field inputs, and $\braket{\hat{N}^0_\Bl}$ is the ensemble average of Eq.~\ref{Eq. mbs: rlz dependent N0 bias simple.}. \\

Higher-order bias terms are either mitigated by our choice of weights \citep{Hanson2011} or are found to be small for our sensitivity levels \citep{Beck2018,Boehm2018}. Likewise, we do not expect a significant bias from galactic foregrounds in our polarization-based estimators \citep{Beck2020} or from masking extragalactic sources in our temperature maps \citep{Lembo2021}. \\

Since the CMB signal contains gravitational lensing, the correlation between the lensing distortions and the various quadratic estimators can create a lensing bias $N^{\text{Lens}}$. This is estimated with the mean reconstructed distortion field spectrum $\braket{C_L^{DD}}$ of the lensed but otherwise un-distorted \lcdm\ simulations. \\

\begin{figure}[th]
\centering
\includegraphics[width=1.0\linewidth]{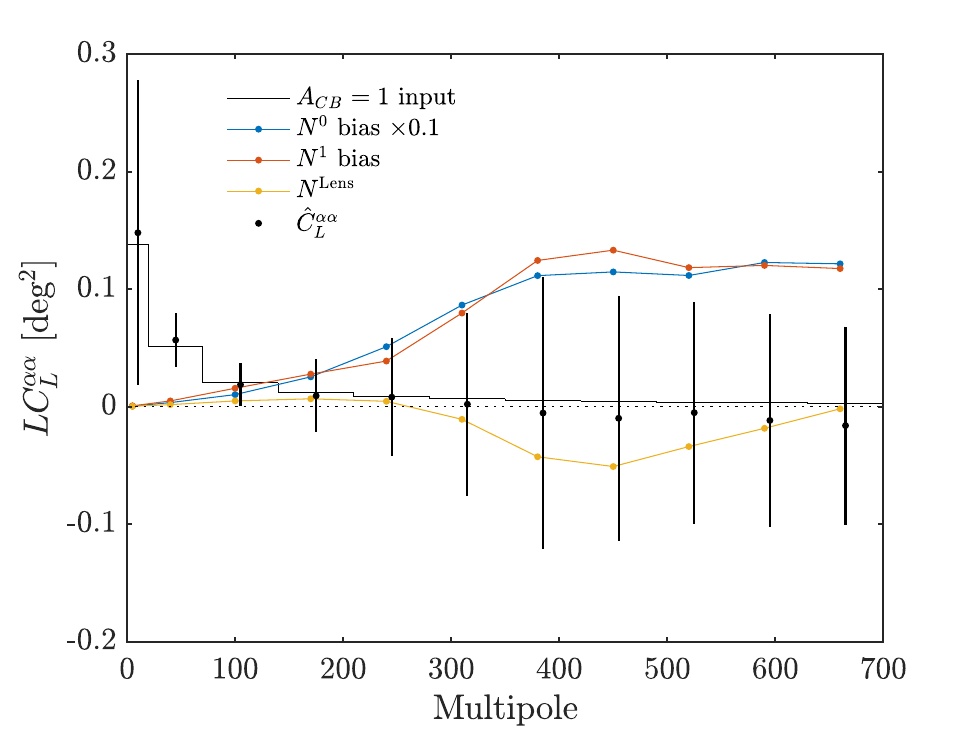}
\caption{Example of the distortion field power spectra pipeline verification for the polarization rotation field $\alpha(\hn)$. The horizontal lines show the binned theory input for an $A_{CB}=1$ spectrum. $\hat{C}_L^{\alpha \alpha}$, the mean simulation bandpowers, matches the input spectrum after accounting for $N^0$, $N^1$, and the lensing bias $N^{\text{Lens}}$. The error bars show the standard deviation of the simulation realizations. 
}
\label{Fig: omega distortion spectra with N0/N1 bias}
\end{figure}

We verified that after the normalization from Eq.~\ref{Eq. mbs: compute normalization with sims} and accounting for the $N^0$, $N^1$, and $N^{\text{Lens}}$, the input distortion spectra are recovered. Fig.~\ref{Fig: omega distortion spectra with N0/N1 bias} shows as an example the polarization rotation $\hat{C}^{\alpha \alpha}_L$ spectra for a scale-invariant distortion input (Eq.~\ref{Eq. mbs: red input spectra}) and its $N^0$, $N^1$ and $N^{\text{Lens}}$ biases. Since the $N^1$ bias is proportional to the distortion field spectra, it is included as part of the signal when we constrain the amplitudes of cosmological models with the distortion field power spectra. \\

We measure the distortion field power spectrum in multipole bins with widths of $\Delta L=70$, and the binned power spectrum values are referred to as {\it bandpowers}. The cosmological results in Section~\ref{sec: Cosmological signal corresponding to Distortion Fields} are derived from $L\in [1, 350)$ since the constraining power only comes from the low multipole modes, whereas in Section~\ref{section: mbs: Distortion Fields as systematics check} we use $L\in[1,700)$ to perform systematics checks. In certain applications where the lowest multipole modes are important, i.e.\ constraining cosmic birefringence models and performing distortion systematics tests, an additional $L\in [1,20)$ bin is separated out from the $L\in [1,70)$ bin. \\

\subsection{Joint analysis of two sets of maps}
\label{subsection: mbs: combine 95/150 trispectra}

When using the distortion fields as systematics checks, the two frequency maps are examined independently since the {\bicepthree} (95\,GHz) and {\biceptwo/\keck} (150\,GHz) maps may have different instrumental systematics. However, for studying cosmological signals, it is desirable to combine the results from the two frequencies into a single more powerful measurement. 
For the inference of cosmological information we will only consider the most sensitive {\it EB} estimators in the combination of the two frequency maps. \\

Our approach is to form distortion field estimators with all possible combinations of the $E$ and $B$ modes: $\hat{D}^{E_1, B_1}$, $\hat{D}^{E_1, B_2}$, $\hat{D}^{E_2, B_1}$, and $\hat{D}^{E_2, B_2}$, where 1,2 stands for 95\,GHz and 150\,GHz respectively. 
In our analysis for the tensor-to-scalar ratio $r$ (BK-I, BK-X, BK-XIII), we examine all possible auto and cross spectra from the multiple frequencies and experiments without forming a combined map. 
In this distortion field analysis, we follow a similar approach where we construct all combinations of $\hat{D}^{E_i, B_j}$, combine their auto and cross spectra, and derive a joint cosmological constraint. 
While the cross spectra approach might not necessarily yield the highest signal-to-noise compared to an analysis on the combined map, the different combinations of spectra can provide consistency checks between data sets. \\

In Eq.~\ref{Eq. mbs: DL^2 = CL + NL}, the squares of the distortion field estimators are used to derive the auto power spectra. Similarly, we take the four estimators $\hat{D}^{E_1, B_1}$, $\hat{D}^{E_1, B_2}$, $\hat{D}^{E_2, B_1}$, and $\hat{D}^{E_2, B_2}$ and compute all the possible auto- and cross-spectra. With 4 individual distortion field estimates, we get a total of 10 spectra: 4 auto-spectra and 6 cross-spectra. Similarly to Eq.~\ref{Eq. mbs: rlz dependent N0 bias simple.}, we compute the realization-dependent bias for the general scenario including the case of cross spectra. The more general form of the realization-dependent $N^0$ bias with the two $E$ maps $W/Y$ and the two $B$ maps $X/Z$ is (Eq. A17 of \cite{Namikawa2014}):
\begin{equation}
\begin{split}
        \hat{N}^0_\Bl  (\hat{W},\hat{X},\hat{Y},\hat{Z}) 
    & = \left< - |\hat{D}^{W_{1}, X_{2}}\hat{D}^{*Y_{2}, Z_{1}}|^2    \right.  \\
    & + |\hat{D}^{W_{1}, \hat{X}}\hat{D}^{*Y_1, \hat{Z}}|^2 
      + |\hat{D}^{W_{1}, \hat{X}}\hat{D}^{*\hat{Y}, Z_{1}}|^2 \\
    & + |\hat{D}^{\hat{W}, X_1}\hat{D}^{*Y_1, \hat{Z}}|^2 
      + |\hat{D}^{\hat{W}, X_1}\hat{D}^{*\hat{Y}, Z_1}|^2  \\
    &  - \left. |\hat{D}^{W_1, X_2}\hat{D}^{*Y_1, Z_2}|^2   \right>_{1,2} \,,
\end{split}
\label{Eq. mbs: general form of N0 bias.}
\end{equation}
where the subscripts 1,2 represent the two sets of simulations with different CMB/noise realizations, and $D^*$ is the complex conjugate of $D$. $\hat{W},\hat{Y}$ can be either the observed 95\,GHz or 150\,GHz $E$ modes, and $\hat{X},\hat{Z}$ can be either the observed 95\,GHz or 150\,GHz $B$ modes. It can be verified that Eq.~\ref{Eq. mbs: general form of N0 bias.} reduces to Eq.~\ref{Eq. mbs: rlz dependent N0 bias simple.} when all four maps $W,X,Y,Z$ come from the same frequency map. \\

The 10 possible auto and cross-spectra are combined linearly with appropriate weights so that the variance of the combined bandpowers is minimized, 
\begin{equation}
    C_b = \sum_i w_{b,i} C_{b,i} \,,
\label{Eq. 10 spectra combined into minimum variance bandpowers}
\end{equation}
where $i$ stands for the 10 spectra indices, $b$ stands for the bins of the bandpowers, and the weights $w_{b, i}$ are a function of both the bins and the spectra indices. The weights are:
\begin{equation}
    w_{b,i} = \frac{\text{mean}_k(\bar{C}_{b,k}) \sum_j \textbf{Cov}^{-1}_{b,ij} \bar{C}_{b,j}}
    {\sum_{jk} \bar{C}_{b,j} \textbf{Cov}^{-1}_{b,jk} \bar{C}_{b,k} } \,,
\end{equation}
where $\bar{C}_{b,i}$ is the mean power from the 499 simulations of spectrum $i$ and bin $b$, and $\textbf{Cov}^{-1}_{b,ij}$ is the covariance matrix of the bandpowers of bin $b$ from the 10 spectra. The minimum variance bandpowers $C_b$ from Eq.~\ref{Eq. 10 spectra combined into minimum variance bandpowers} combine the statistical power from the two frequency maps and are used to constraint the corresponding cosmological processes. \\ 
\vspace{1cm}

\section{Results: Cosmology from Distortion Fields}
\label{sec: Cosmological signal corresponding to Distortion Fields}

In the following subsections, we present the observed distortion field spectra and the derived cosmological constraints from $\Phi(\hn)$, $\tau(\hn)$, and $\alpha(\hn)$ corresponding to gravitational lensing, patchy reionization and cosmic birefringence.

\subsection{Gravitational Lensing}
In Table~\ref{table: TB/EB quadratic estimator weights}, the weights for reconstructing the gradient, $\Phi$, and the curl component, $\Omega$, are listed. We use the gradient part to constrain the amplitude of the lensing signal parametrized as $A_{L}^{\phi \phi}$, while using the curl part as a systematics check in Section~\ref{section: mbs: Distortion Fields as systematics check}. \\

It is often more convenient to work with the lensing-mass (convergence) field $\kappa$ since the lensing potential has a red spectrum while the lensing-mass field has a nearly flat spectrum \citep{PlanckLensing}. The lensing convergence $\kappa$ is related to the lensing potential $\Phi$ as:
\begin{equation}
    \kappa = -\frac{1}{2}\nabla^2 \Phi \,,
\end{equation}
For the Fourier transform, we have:
\begin{equation}
    \kappa_\Bl = \frac{L(L+1)}{2}\Phi_\Bl \,,
\end{equation}
where $L=|\Bl|$. Similarly, we also define the analogous quantity for the lensing rotation $\omega$ as:
\begin{equation}
    \omega_\Bl = \frac{L(L+1)}{2}\Omega_\Bl\,.
\end{equation} \\

In Eq.~\ref{Eq. mbs: compute normalization with sims} where we empirically calibrate the normalization factor, we correlate the input distortion field with the reconstruction. Before performing the Fourier transform and cross correlation, the inverse variance apodization masks are applied to the input distortion fields. Because the $\kappa$ spectrum is relatively flat compared to the red $\Phi$ spectrum, it is better to apply the apodization mask to the $\kappa$ map instead of to the $\Phi$ map to avoid mode mixing (BK-VIII):

\begin{equation}
    \Phi_\Bl = \frac{2}{L^2}\int d^2\hn e^{-i\hn\cdot\Bl}\kappa(\hn) \,.
\end{equation}

\begin{figure*}[thbp]
\centering
\includegraphics[width=1.0\linewidth]{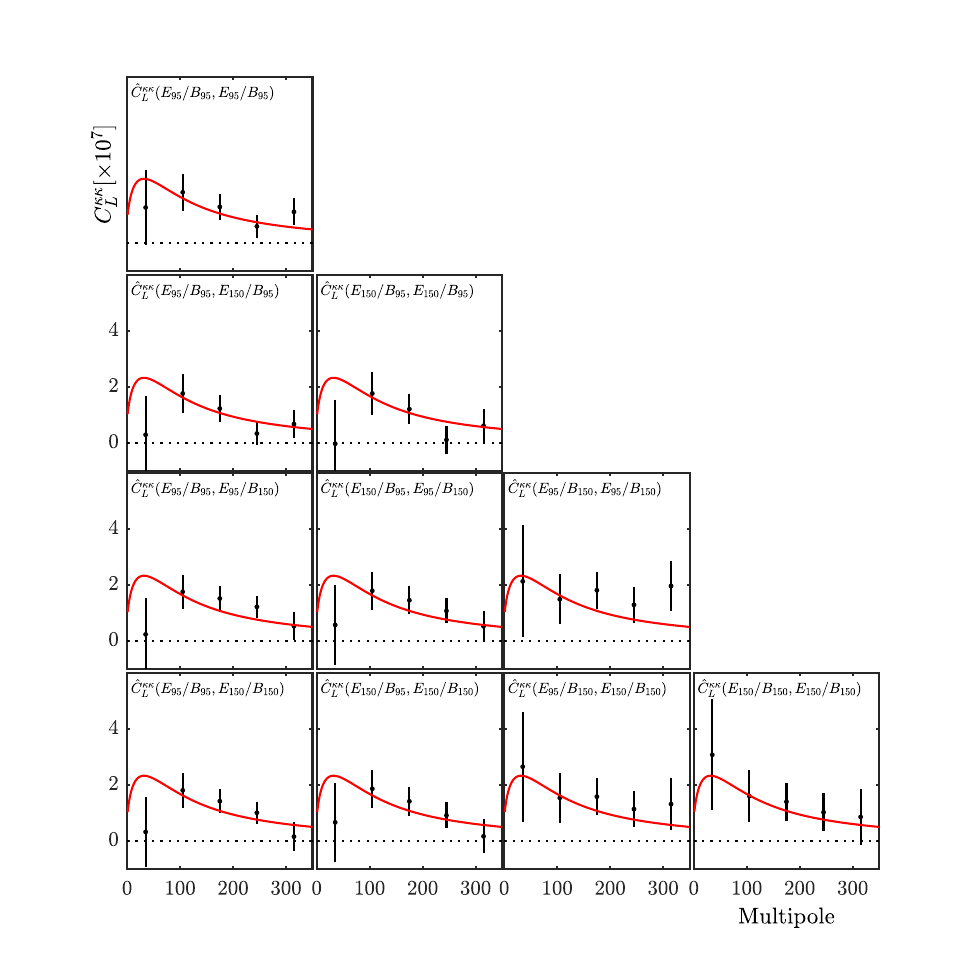}
\caption{The 10 different ways to combine two sets of $E$ and $B$ maps to form the lensing convergence spectrum estimator $\hat{C}_L^{\kappa \kappa}$. The red line is the theoretical lensing convergence spectrum corresponding to our fiducial model assuming Planck 2013 cosmological parameters \citep{Planck2013Cosmology}. The top left subplot is the auto spectrum from 95\,GHz, while the bottom right is the auto spectrum from 150\,GHz. The other subplots contain information from both 95\,GHz and 150\,GHz.
We examine the 10 individual spectra, all 10 spectra combined, and some other data combinations, and we find that they are all consistent with the lensed-{\lcdm} predictions.}
\label{Fig: mbs: 10 Lensing spectra combinations}
\end{figure*}

The 10 possible auto and cross-spectra for the lensing reconstruction are shown in Fig.~\ref{Fig: mbs: 10 Lensing spectra combinations}, where the lensing convergence spectrum $C_L^{\kappa \kappa} \approx 4/L^4 C_L^{\Phi \Phi}$ is plotted. The 4 diagonal subplots are the auto spectra from the 4 possible $\Phi^{E_i, B_j}$, while the other 6 are from cross correlating the different $\Phi^{E_i, B_j}$. The top left subplot is derived from only 95\,GHz, and the bottom right subplot is derived from only 150\,GHz. The other 8 subplots combine some information from both the 95\,GHz and 150\,GHz. \\

\begin{figure}[thb]
\centering
\includegraphics[width=1.0\linewidth]{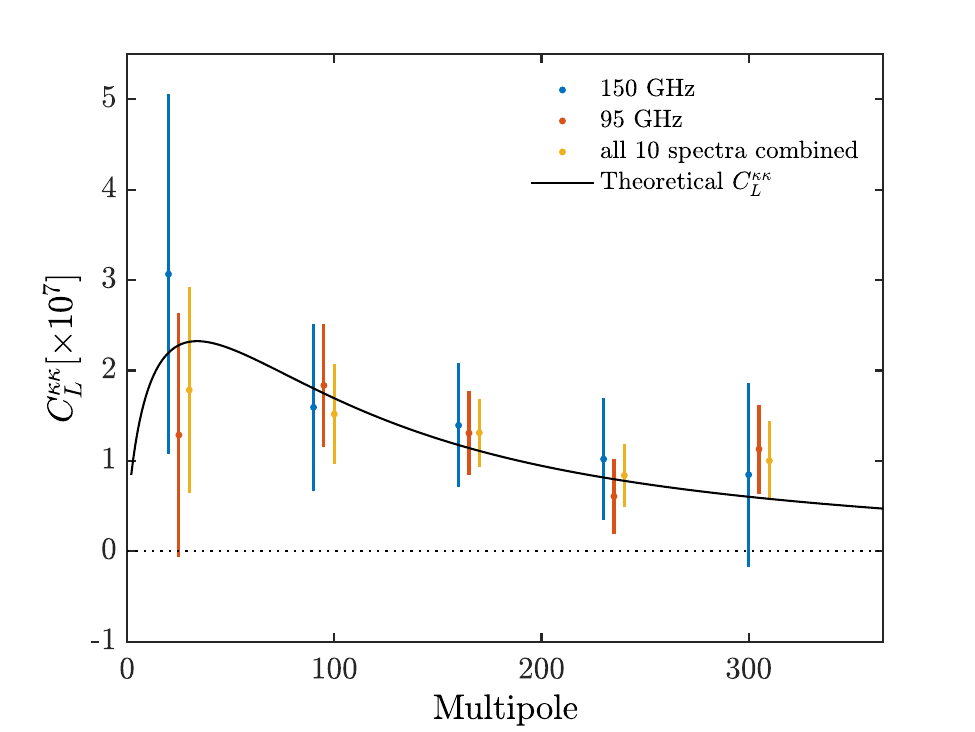}
\caption{The lensing convergence power spectrum $C_L^{\kappa \kappa}$ for 95\,GHz auto-spectrum, 150\,GHz auto-spectrum, and for all 10 auto- and cross-spectra combined. }
\label{Fig: mbs: lensing kappa spectra 95/150/all 10 spectra}
\end{figure}

In Fig.~\ref{Fig: mbs: lensing kappa spectra 95/150/all 10 spectra} we show the reconstructed $\hat{C}_L^{\kappa\kappa}$ of 95\,GHz only, 150\,GHz only, and with all 10 spectra combined. 
With the bandpowers in Fig.~\ref{Fig: mbs: lensing kappa spectra 95/150/all 10 spectra}, we fit for the amplitude of the lensing potential power spectrum by taking a weighted mean of the real bandpowers over the fiducial simulation bandpowers (BK-VIII). With a linear model $\hat{C}_b = A_L^{\phi \phi} C_b^f$ of the noise-debiased power spectrum, where $C_b^f$ is the fiducial model corresponding to the \planck\ \lcdm\ prediction from their 2013 release\footnote{The lensing {\bmode} power from the {\planck} 2013 parameters is around 5\% higher than the \planck\ 2018 results \cite{Planck2018Cosmology}. This difference is small compared to the uncertainties in the present work.} \citep{Planck2013Cosmology}, the least squares fit for $A_L^{\phi \phi}$ is:
\begin{equation}
    A_L^{\phi \phi} = \frac{\sum_{bb'} \hat{C}_b \mathbf{Cov}_{bb'}^{-1} C_{b'}^f}{\sum_{bb'} C_b^f \mathbf{Cov}_{bb'}^{-1} C_{b'}^f} \,,
\label{Eq. mbs: AL^phi phi amplitude}
\end{equation}
where $\hat{C}_b = \hat{C}_L^{\kappa \kappa}$ is the observed lensing convergence bandpowers, $C_b^f$ is the mean bandpowers from the lensed-\lcdm\ simulations, and $\mathbf{Cov}_{bb'}$ is the bandpower covariance matrix from the same lensed-\lcdm\ simulations. \\

\begin{figure}[thb]
\centering
\includegraphics[width=1.0\linewidth]{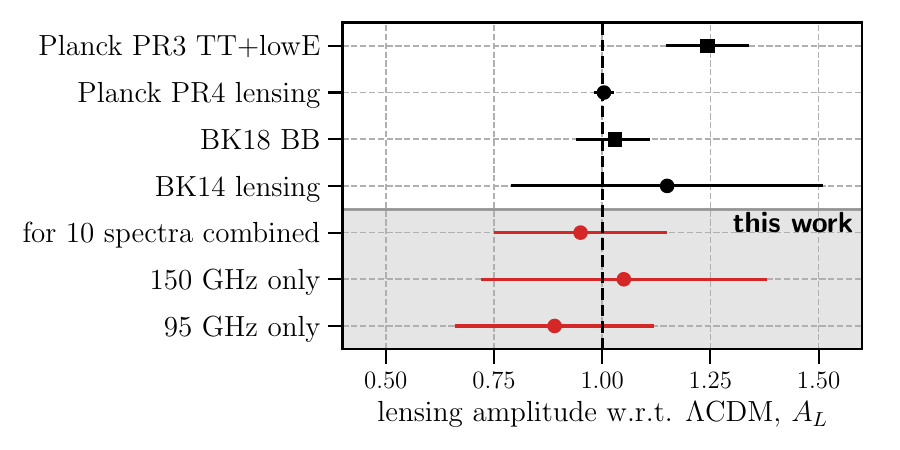}
\caption{Comparison of the constraints of this paper with the latest constraints of the lensing amplitude $A_L$ from the Planck PR3 temperature power spectrum, TT+lowE \citep{Planck2020}, the Planck PR4 lensing reconstruction \citep{Carron2022}, the BK18 B-mode power spectrum measurement (BK-XIII) using the same data set and the previous measurement from the lensing potential reconstruction using BK14 data (BK-VIII). Bullet points denote constraints from the lensing potential auto-power spectrum, squares are used for measurements using CMB two-point functions.}
\label{Fig: lensing constraints}
\end{figure}

The best fit $A_L^{\phi \phi}$ are:
\begin{align}
    & A_L^{\phi \phi} = 0.89\pm 0.23, ~~~~\text{for 95\,GHz only} , \\
    & A_L^{\phi \phi} = 1.05\pm 0.33, ~~~~\text{for 150\,GHz only} ,\\
    & A_L^{\phi \phi} = 0.95\pm 0.20, ~~~~\text{for 10 spectra combined} .
\end{align} 
Since the lensing reconstruction is close to sample variance limited in the central parts of the map, the larger map coverage from {\bicepthree} 95\,GHz produces a tighter $\sigma(A_L^{\phi \phi})=0.23$ compared to the 150\,GHz $\sigma(A_L^{\phi \phi})=0.33$. When all the cross spectra between the two frequencies are combined, we achieve $\sigma(A_L^{\phi \phi})=0.20$, an $\approx$15\% reduction compared to 95\,GHz only. This is around a factor of 2 improvement from the previous BK-VIII result of $A_L^{\phi \phi}=1.15\pm0.36$. However, note that the lensing amplitude is
better constrained by the {\bmode} power spectrum with $A_L^{BB}=1.03^{+0.08}_{-0.09}$ in
BK-XIII. We compile these constraints on the lensing amplitude in Fig.~\ref{Fig: lensing constraints}. \\
\\

\subsection{Patchy Reionization}

\begin{figure}[htb]
\centering
\includegraphics[width=1.0\linewidth]{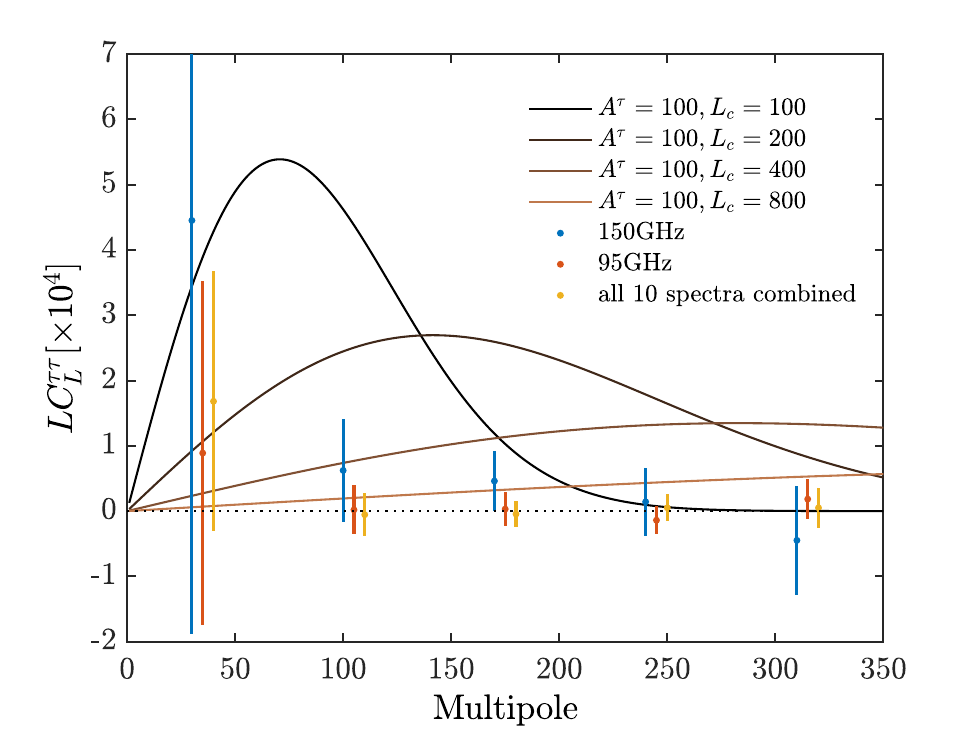}
\caption{Data and model power spectra of patchy reionization.
The solid lines show fiducial model spectra from Eq.~\ref{Eq. mbs: patchy reionization fiducial model spectra} for $L_c=100$, 200, 400 and 800.
The data points show $\hat{C}_L^{\tau \tau}$ for 150\,GHz only, 95\,GHz only, and all 10 spectra combined.}
\label{Fig: mbs: patchy reionization simple Fiducial model spectra}
\end{figure}

Following \cite{Gluscevic2013} and \cite{Namikawa2018}, we use the $\tau(\hn)$ reconstruction from the {\it EB} estimator to constrain a simple crinkly-surface model. The model describes a scenario in which the universe goes suddenly from neutral to ionized but with a reionization surface that is crinkled on a comoving scale of $R_c \approx 200 \text{Mpc} \; (L_c/150)^{-1}$. The predicted power spectrum in Eq.~\ref{Eq. mbs: patchy reionization fiducial model spectra} consists of white noise smoothed on an angular scale of $\theta_C = \pi/L_c$. Fiducial model spectra are shown in Fig.~\ref{Fig: mbs: patchy reionization simple Fiducial model spectra} for $L_c=100$, 200, 400 and 800. The use of this parameter space is only valid in the assumption of this simplified model as it assumes an instantaneous reionization. As such the parameter $L_c$ has no physical meaning and limitations can be evaded with a more realistic model of the reionization history. \\

The amplitude $A^\tau$ is constrained with a log likelihood based on  \cite{Hamimeche2008}:
\begin{equation}
-2\ln L (A^\tau) = \sum_{bb'}g(\hat{R}_{b})C^f_{b} \textbf{Cov}^{-1}_{b b'}C^f_{b'}g(\hat{R}_{b'})
\label{Eq. mbs: Atau HL likelihood} \,,
\end{equation}
where $C^f_b$ are the mean bandpowers from the simulations of the fiducial model, $g(x) = \text{sign}(x-1)\sqrt{2(x-\ln x-1)}$, and $\hat{R}_{b}$ is the per bin ratio of the observed bandpowers over the fiducial bandpowers including the $N^0$, $N^1$, and lensing bias $N^{Lens}$:
\begin{equation}
    \hat{R}_{b}=\frac{\hat{C}^{\tau \tau}_{b}+N_{b}^0+N_{b}^{\text{Lens} }}{A^\tau(C^{f}_{b} + N^1_{b}) + N_{b}^0+N_{b}^{\text{Lens}}} \,.
\label{Eq. bin ratio of observed power for HL likelihood}
\end{equation}
\\

Using the method outlined in Section~\ref{subsection: mbs: combine 95/150 trispectra}, Fig.~\ref{Fig: mbs: patchy reionization simple Fiducial model spectra} shows the reconstructed $\hat{C}_L^{\tau \tau}$ for 150\,GHz auto-spectra, 95\,GHz auto-spectra, and for all 10 auto- and cross-spectra combined. 
We see that the BK data are consistent with zero, offering no evidence for a patchy
reionization signal---consistent with earlier limits derived from WMAP and Planck temperature maps \citep{Gluscevic2013, Namikawa2018}. \\

We proceed to set upper limits on $A^\tau$ in Eq.~\ref{Eq. mbs: patchy reionization fiducial model spectra} using the log likelihood of Eq.~\ref{Eq. mbs: Atau HL likelihood}. Fiducial simulations of $L_c$ at 100, 200, 400, and 800 are used to derive the constraints.  In Table~\ref{table: Atau 2 sigma upper limits for 159/95/all 10 trispectra.}, the $2\sigma$ (95\% C.L.) upper limits on $A^\tau$ for 95\,GHz only, 150\,GHz only, and with all 10 spectra combined are listed. The sensitivity primarily comes from the 95\,GHz map. \\ 

\begin{table}[htb]
\begin{tabular}{c| c c c } 
 \hline \hline 
  \rule{0pt}{3ex}
   & \multicolumn{3}{c}{2$\sigma$ upper limit on $A^\tau$} \\
 \rule{0pt}{2ex}
$L_c$ & 150\,GHz & 95\,GHz & all 10 spectra \\ [0.5ex] 
    \hline
100  & 76  & 24  & 19  \\ [0.5ex]
200  & 51  & 17  & 16  \\ [0.5ex]
400  & 72  & 27  & 26  \\ [0.5ex]
800  & 190  & 77  & 71  \\ [0.5ex]
    \hline \hline
\end{tabular}
\caption{The $2\sigma$ upper limits for $A^\tau (L_c)$ derived from the 95\,GHz auto-spectrum, 150\,GHz auto-spectrum, and the all 10 spectra combined. The corresponding $\hat{C}_L^{\tau \tau}$ is shown in Fig.~\ref{Fig: mbs: patchy reionization simple Fiducial model spectra}, and the fiducial model spectrum is Eq.~\ref{Eq. mbs: patchy reionization fiducial model spectra}. 
}
\label{table: Atau 2 sigma upper limits for 159/95/all 10 trispectra.}
\end{table}

\begin{figure}[thb]
\centering
\includegraphics[width=1.0\linewidth]{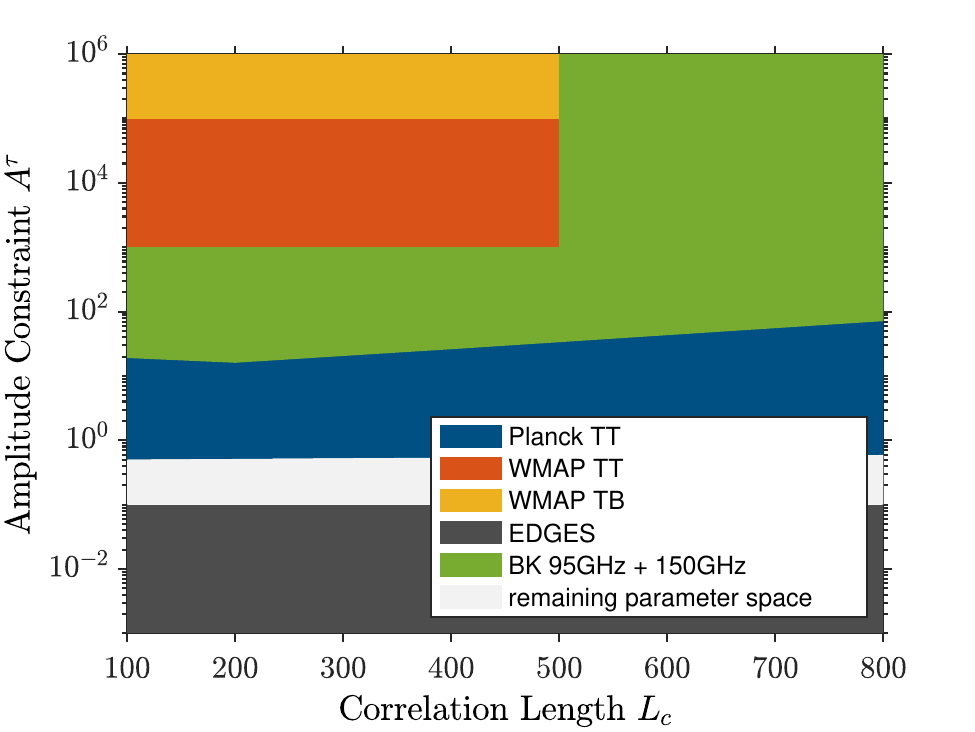}
\caption{Constraints on the amplitude $A^\tau(L_c)$ of Eq.~\ref{Eq. mbs: patchy reionization fiducial model spectra} obtained from this work (green) and in the previous work \citep{Gluscevic2013,Namikawa2018}. The colored regions are excluded.}
\label{Fig: mbs: Atau-Lc parameter space constraint}
\end{figure}

Following \cite{Gluscevic2013} and \cite{Namikawa2018}, we plot the constraints of Table~\ref{table: Atau 2 sigma upper limits for 159/95/all 10 trispectra.} in the $A^\tau$ vs. $L_c$ parameter space. In Fig.~\ref{Fig: mbs: Atau-Lc parameter space constraint}, the constraint derived from our data is seen to be between the constraints from WMAP {\it TT} and Planck {\it TT}. The noise level of the $\hat{\tau}$ reconstruction in the {\bicep} patch is roughly the same as the reconstruction from the {\planck} {\it TT} estimator. However, {\planck}'s wider sky coverage significantly reduces the overall sample variance. According to \cite{Gluscevic2013}, the lower limit on the duration of reionization $\Delta z \gtrsim 0.4$ obtained by EDGES \citep{EDGES2017} can be translated to $A^\tau \gtrsim 0.1$, so only a narrow allowed band remains. \\

\subsection{Cosmic Birefringence}

The two physical processes that can lead to anisotropic cosmic birefringence, parity-violating physics and primordial magnetic fields, produce the predicted power spectra given in Eq.~\ref{Eq. mbs: ACB to axion like particles interaction g_ag} and Eq.~\ref{Eq. mbs: ACB constraints to PMF nG}.
These are both of the form of $L(L+1)C_L^{\alpha \alpha}=\text{constant}$. Following previous conventions (BK-IX, \cite{Namikawa2020,Bianchini2020}), we  parametrize the power spectra with $A_{CB}$,
\begin{equation}
    \frac{L(L+1)C_L^{\alpha \alpha}}{2\pi} = A_{CB}\times 10^{-4} \;\; [\text{rad}^2] \,.
\label{Eq. mbs: ACB fiducial spectra}
\end{equation}
In standard \bicep/\keck\ analysis the overall polarization angle is adjusted to minimize the observed {\it TB} and {\it EB} power spectra. After this self-calibration, the polarization maps lose sensitivity to a uniform polarization rotation but are still sensitive to anisotropic rotations. \\

\subsubsection{Constraints on parity violating physics}

\begin{figure}[thb]
\centering
\includegraphics[width=1.0\linewidth]{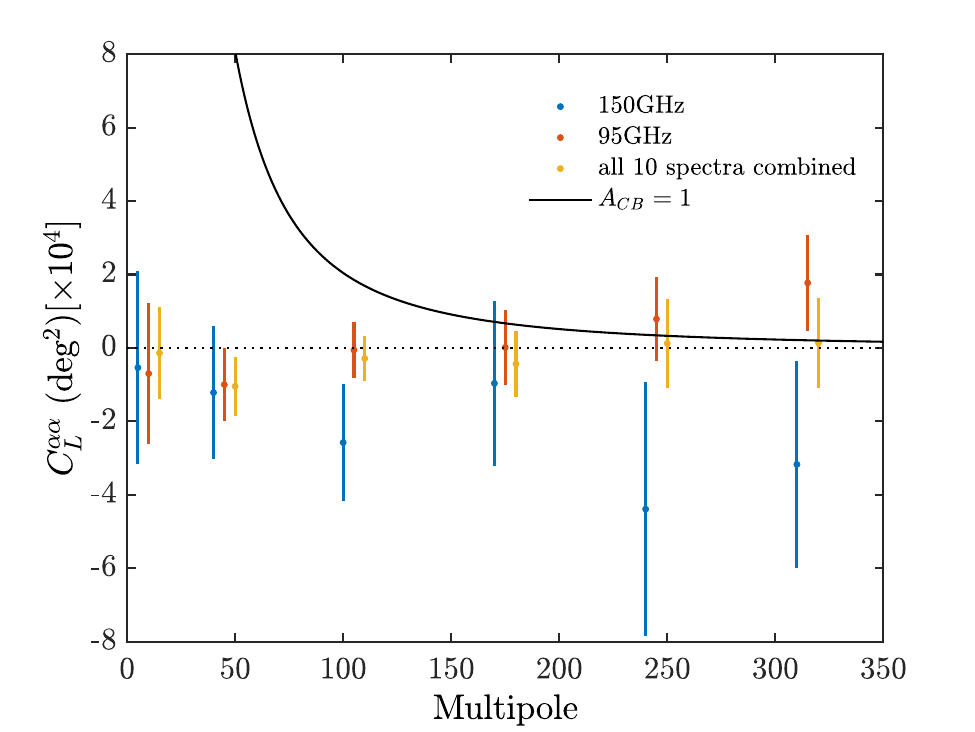}
\caption{Cosmic birefringence power spectra $\hat{C}_L^{\alpha \alpha}$ for 95\,GHz only, 150\,GHz only, and all 10 spectra combined. The black line is the fiducial spectra from Eq.~\ref{Eq. mbs: ACB fiducial spectra} with $A_{CB}=1$. Compared to Fig.~\ref{Fig: mbs: patchy reionization simple Fiducial model spectra} and Fig.~\ref{Fig: mbs: lensing kappa spectra 95/150/all 10 spectra}, one additional bin of $L\in[1,20)$ is separated out from the $L\in[1,70)$ bin since the lowest multipoles are important for constraining $A_{CB}$.}
\label{Fig: mbs: 95+150 CL^aa ACB spectra reconstruction.}
\end{figure}

The best constraints from our data set on the coupling constant $g_{a\gamma}$ between axion-like particles and photons  (Eq.~\ref{Eq. mbs: ACB to axion like particles interaction g_ag}) are derived using the combined minimum variance $\hat{C}_L^{\alpha \alpha}$ of the two frequency maps. Following the method outlined in Section~\ref{subsection: mbs: combine 95/150 trispectra}, the reconstructed $\hat{C}_L^{\alpha \alpha}$ for 95\,GHz, 150\,GHz, and all 10 spectra combined are shown in Fig.~\ref{Fig: mbs: 95+150 CL^aa ACB spectra reconstruction.}. \\

The $\hat{C}_L^{\alpha \alpha}$ in Fig.~\ref{Fig: mbs: 95+150 CL^aa ACB spectra reconstruction.} are consistent with the un-rotated lensed-\lcdm+dust+noise simulations. In a similar approach to \cite{Namikawa2020} and \cite{Bianchini2020}, we use a log likelihood based on \cite{Hamimeche2008} to evaluate the 95\% $2\sigma$ upper limit for $A_{CB}$. This likelihood is the same as Eq.~\ref{Eq. mbs: Atau HL likelihood} and \ref{Eq. bin ratio of observed power for HL likelihood} but with $A_{CB}$ in place of $A^{\tau}$.
We obtain a 95\% confidence upper limit of $A_{CB} \leq 0.044$.
Using Eq.~\ref{Eq. mbs: ACB to axion like particles interaction g_ag}, this corresponds to an upper limit on the coupling constant $g_{a\gamma}$,
\begin{equation}
    g_{a\gamma} \leq \frac{2.6\times10^{-2}}{H_I} \,.
\end{equation}
This is a factor of 3 improvement over our previous results using the BK14 maps in BK-IX: $g_{a\gamma}\leq 7.2\times10^{-2}/H_I$. It is also somewhat better than the constraints from ACT: $g_{a\gamma}\leq 4\times10^{-2}/H_I$ \citep{Namikawa2020} and SPT: $g_{a\gamma}\leq 4\times10^{-2}/H_I$ \citep{Bianchini2020}.\\

\subsubsection{Constraints on Primordial Magnetic Fields}

To derive constraints on PMFs, we study the two frequency maps separately since the polarization angle rotation from Faraday rotation scales  with frequency as $\nu^{-2}$ (Eq.~\ref{Eq. mbs: ACB constraints to PMF nG}). With the 95\,GHz only and 150\,GHz only spectra shown in Fig.~\ref{Fig: mbs: 95+150 CL^aa ACB spectra reconstruction.}, we again use the log likelihood Eq.~\ref{Eq. mbs: Atau HL likelihood} to derive 95\% upper limits on $A_{CB}$, where we obtain $A_{CB} \leq 0.097$ for 95\,GHz and $A_{CB} \leq 0.17$ for 150\,GHz. With Eq.~\ref{Eq. mbs: ACB constraints to PMF nG}, we convert these constraints on $A_{CB}$ to the following constraints on PMFs $B_{1\text{Mpc}}$,
\begin{align}
        B_{1\text{Mpc}} &\leq 6.6 \text{ nG}, \;\;\;\;\; \text{for 95\,GHz}\,, \\
        B_{1\text{Mpc}} &\leq 22 \text{ nG}, \;\;\;\;\; \text{for 150\,GHz}\,.
\end{align}

\begin{figure}[thb]
\centering
\includegraphics[width=1.0\linewidth]{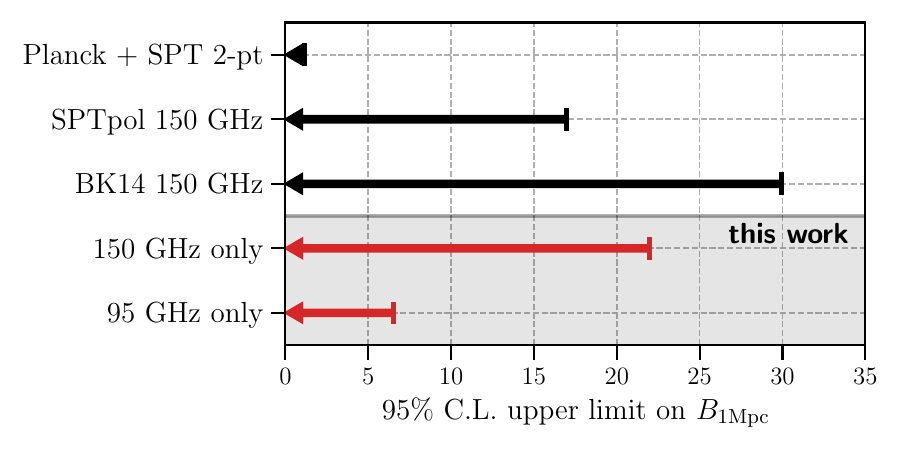}
\caption{Comparing the constraint on the strength of primordial magnetic fields smoothed over $1\ \textrm{Mpc}$, $B_{1\text{Mpc}}$, of this work with constraints from SPTpol anisotropic birefringence reconstruction \citep{Bianchini2020} and Planck and SPTpol CMB power spectra \citep{Zucca_2017}.}
\label{Fig: PMF constraints}
\end{figure}

We show the previously published constraints from CMB four-point function measurements in Fig.~\ref{Fig: PMF constraints}, which are derived from 150\,GHz maps, are SPT: $B_{1\text{Mpc}} \leq 17~\text{nG}$ \citep{Bianchini2020} and BK-IX: $B_{1\text{Mpc}}  \leq 30~\text{nG}$. The leading constraint on this parameter is $B_{1\text{Mpc}}  \leq 1.2~\text{nG}$ for a nearly
scale-invariant PMF, derived from a combination of Planck and SPT two-point power spectra \citep{Zucca_2017}. Through the effect of PMFs on the post-recombination ionization history, \cite{Paoletti_2022} are able to constrain the amplitude of the magnetic fields to $\sqrt{B^2} < 0.69~\text{nG}$. \\

The $A_{CB}$ constraint derived from BICEP3 95\,GHz alone ($A_{CB} \leq 0.097$) is comparable to the upper limits from SPT and ACT, but the resulting constraint on $B_{1\text{Mpc}}$ is considerably better because of the advantage of the lower frequency leverage with the $\nu^{-2}$ scaling. \\

\subsection{Consistency checks and null tests}
In this subsection, we discuss consistency checks and jackknife null tests for the three distortion fields $\tau(\hn), \alpha(\hn), \kappa(\hn)$ that have been used to derive science constraints. We want to emphasize that the BK18 data set has passed a comprehensive set of data validations (BK-I, BK-III, BK-XIII), most importantly the jackknife null tests on the {\it EE/BB} power spectra. In the next section (Section~\ref{section: mbs: Distortion Fields as systematics check}), we study all the distortion effects in Eq.~\ref{Eq.mbs: Characterization of 11 distortion fields} as systematics checks, providing further evidence that the BK18 data set has systematic effects controlled below the level of statistical uncertainty.
In this subsection we focus on demonstrating the robustness of the reconstructed distortion field spectra with different analysis choices, and present some additional null tests for the three distortion fields being used to derive science results. \\

\subsubsection{Consistency checks} \label{subsection: consistency checks}
As consistency checks of the $\kappa(\hn)$, $\tau(\hn)$ and $\alpha(\hn)$ reconstructions, the distortion bandpowers are constructed while altering some choices of the analysis. We summarize the conclusions here and provide detailed PTE values in Appendix \ref{Appendix Alternative choices of analysis}.
\begin{itemize}
    \item Input $E$/{\bmode} multipoles: \\
    Similarly to BK-VIII and BK-IX, we lower the maximum multipole $\ell_{\text{max}}$ from 600 to 400, raise the minimum multipole $\ell_{\text{min}}$ to 200, or lower the {\bmode} maximum multipole $\ell_{\text{max}}^B$ from 600 to 350. The results from these three alternate choices are all consistent with the lensed-{\lcdm}+dust+noise simulations. Additionally, the shift of the observed bandpowers from the three alternate choices vs. the baseline are also consistent with the shift in the simulation bandpowers for both 95 and 150\,GHz and all three fields $\alpha$, $\tau$ and $\kappa$. 
    
    \item Differential beam ellipticity:\\
    In BK analysis the $T$ to $P$ leakage from differential gain and differential pointing are filtered out with the technique we call {\it deprojection} (BK-III). However, the $T$ to $P$ leakage from differential beam ellipticity cannot be treated with a direct filtering operation because the CMB {\it TE} correlation would cause a bias. Instead, we subtract a leakage template derived from the measured differential beam map ellipticity. We repeat the analysis {\it without} the subtraction and find very small changes in the reconstructed spectra, similar to what has been seen by \cite{Mirmelstein2021}.
    
    \item Alternate foreground models:\\
    In Appendix \ref{appendix: alternate dust foreground}, the different foreground models explored in the main line BK18 analysis (BK-XIII) are used instead of the Gaussian dust simulations. We find that with the realization-dependent method and the baseline choice of $\ell_{\text min}^B=100/150$ for 95\,GHz and 150\,GHz, the shifts in the bandpowers when switching to the alternate foreground models, or to no foreground, are negligible.
\end{itemize}

\subsubsection{Effects of absolute calibration error}
Although the distortion fields $\kappa(\hn)$, $\tau(\hn)$ and $\alpha(\hn)$ are dimension-less quantities, the {\it EB} quadratic estimator construction will lead to a distortion spectra $C_L^{DD}$ that depends on the overall amplitude of the polarization map. An absolute calibration uncertainty of $\delta$ on the polarization map will translate to a systematic uncertainty of $4\delta$ on either the $A_{CB}/A^\tau$ upper limits or the amplitude of the lensing potential $A_L^{\phi \phi}$ (BK-VIII). The absolute calibration procedure which correlates the observed $T$ with the {\planck} $T$ map (BK-I) is estimated to have an uncertainty of 0.3\%. The polarization efficiency is high ($\approx 99$\%) with an uncertainty of $\lesssim 0.5$\% (BK-I, BK-II). Therefore, we estimate that the systematic uncertainty on $A_L^{\phi \phi}$ from the absolute calibration of the polarized map is around $4\delta \lesssim 3$\%. \\

\subsubsection{Jackknife null tests} \label{subsection: science field jackknife null tests}
We perform distortion field reconstruction on the 14 flavors of differenced ({\it jackknife}) maps that are designed to target different systematics in the main line analysis (BK-I, BK-III, BK-X, BK-XIII). The distortion field reconstruction is done with the {\it full} {\emode}s and the jackknife {\bmode}s, written $\hat{D}^{E_{\text{full}}, B_{\text{jack}}}$. We are interested in probing systematic effects that can potentially bias the distortion field reconstructions. 
For our B-mode search in the main analysis we are most worried about $E$-to-$B$ leakage terms. Further, {\bmode} systematics that can be interpreted as a distortion field coupled with the full {\emode}s would be the most concerning contamination in terms of biasing the distortion field science results. Hence we focus on the particular combination of $\hat{D}^{E_{\text{full}}, B_{\text{jack}}}$ as opposed to $\hat{D}^{E_{\text{jack}}, B_{\text{full}}}$. While the latter would be more sensitive if Eq.~\ref{Eq.mbs: Characterization of 11 distortion fields} would be a perfect model of our instrumental systematic contamination and systematic effects act symmetrically on $E$ and $B$-modes, we decide to perform a more focused search of $E$-to-$B$ leakage terms with the $\hat{D}^{E_{\text{full}}, B_{\text{jack}}}$ estimator. \\

With the method outlined in Section~\ref{section: mbs: estimating the distortion field power spectra}, we reconstruct the observed $\hat{C}_L^{DD}$ from $\hat{D}^{E_{\text{full}}, B_{\text{jack}}}$ and compare it to $C_L^{DD}$ from the 499 lensed-\lcdm+dust+noise simulations by evaluating $\chi$ and $\chi^2$ values,
\begin{align}
    \chi^2 &= \sum_{bb'} (\hat{C}_b - \braket{C_b}) \textbf{Cov}_{bb'}^{-1} (\hat{C}_{b'} - \braket{C_{b'}}) \,, \label{Eq. chi2 null test definition.}\\
    \chi &= \sum_{b} (\hat{C}_b - \braket{C_b})/\sigma(C_b) \,, 
\label{Eq. chi null test definition.}
\end{align}
where $\textbf{Cov}_{bb'}$ is the bandpower covariance matrix from 499 simulations, $\hat{C}_b$ is the observed distortion field bandpowers, and $\braket{C_b}$ is the mean bandpowers from the simulations. We also compute the $\chi$ and $\chi^2$ values for each of the 499 simulation realizations and evaluate the probability-to-exceed (PTE) or $p$-value by counting the percentage of simulations that have larger $\chi$ or $\chi^2$. \\

\begin{figure}[thb]
\centering
\includegraphics[width=1.0\linewidth]{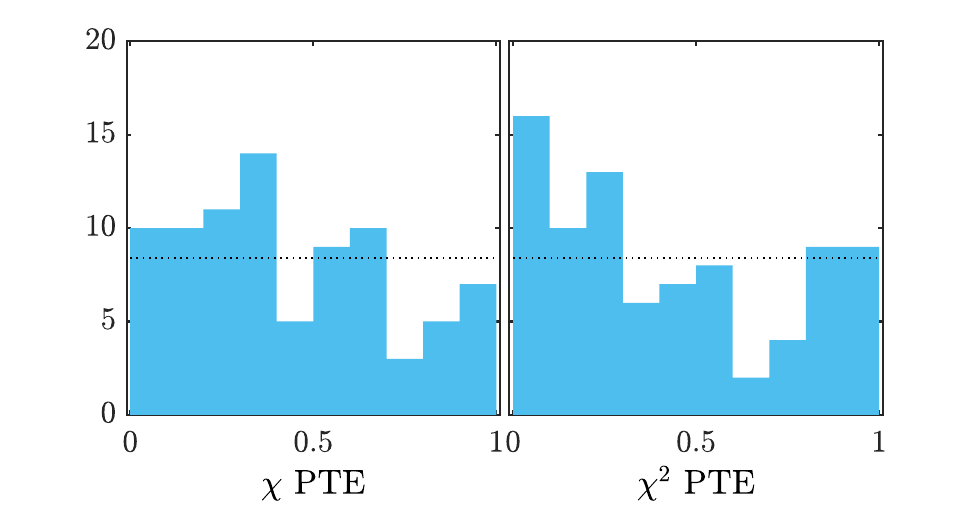}
\caption{The distributions of the 84 $\chi/\chi^2$ PTE values from 14 jackknives, 3 distortion fields of 2 real data frequency maps.} 
\label{Fig: mbs: science jackknife pte histogram}
\end{figure}

There are 14 (jackknives) $\times$ 3 (fields) $\times$ 2 (frequency maps) = 84 PTE values for both $\chi$ and $\chi^2$ statistics.
These values are histogramed in Fig.~\ref{Fig: mbs: science jackknife pte histogram}. The value for $\chi$ PTE closest to zero or unity is 0.006 and the lowest $\chi^2$ PTE is 0.008.
Taking into account the look-elsewhere effect, we can construct a {\it global} statistical test that compares these real data values to the simulations.
The specific procedure is as follows:
\begin{align}
    &\chi^2 \text{ extreme PTE: }p_{\chi^2} = \min_{p}(p)  \,, \label{Eq. chi2 extreme}\\
    &\chi \text{ extreme PTE: }
    p_\chi = \min_{p}\left(\min(p,1-p)\right) \,, \label{Eq. chi extreme}\\
    &\text{overall extreme PTE: }p_{\text{all}} = \min(p_{\chi},p_{\chi^2}) \label{Eq. overall extreme}\,,
\end{align}
where $p$ in Eq.~\ref{Eq. chi2 extreme} are the 84 $\chi^2$ PTE values and $p$ in Eq.~\ref{Eq. chi extreme} are the 84 $\chi$ PTE values for the real data, or a given simulation realization.
The quantities $p_{\chi^2}$ and $p_\chi$ are the most extreme $\chi^2$ and $\chi$ PTEs, and the overall most extreme PTE $p_\text{all}$ is the smaller one of $p_\chi$ and $p_{\chi^2}$.

We find that the most extreme value for the real bandpowers is $p_\text{all}^\text{obs} = 0.006$.
Comparing $p_\text{all}$ between the observation and simulations, the probability to get a value
smaller than the observed value is 0.59.
Therefore, we conclude that there is no evidence of spurious {\bmode}s in the $\kappa(\hn)$, $\tau(\hn)$, and $\alpha(\hn)$ reconstructions from the jackknife maps. \\

\section{Results: Distortion Fields as systematics tests}
\label{section: mbs: Distortion Fields as systematics check}

In this section, we comprehensively investigate the distortion fields potentially caused by systematics and consider different types of instrumental effects that could produce these distortions.
In the main line analysis for tensor-to-scalar ratio $r$ (BK-I, BK-VI, BK-X, BK-XIII), the most fundamental guard against systematics are the map jackknife (or null) tests.
Maps are made splitting the data into (approximate) halves according to criteria which would
be expected to result in nearly equal signal, but potentially different systematic contamination.
The split maps are then differenced and the $EE$, $BB$ and $EB$ spectra of the result compared 
to simulations of signal plus noise.
Well chosen jackknife splits can amplify systematics which cancel in the full coadd map (BK-III).
It must be emphasized again that the published BK measurements of the tensor-to-scalar ratio, including the latest BK18 release, have passed all these null tests. \\

However, systematics detection and mitigation based on the distortion fields may complement and enhance the standard map jackknife tests in several ways.  
When using line-of-sight distortion fields for systematics we are checking for spurious {\bmode}s in the full {\it Q/U} maps.
Therefore, any spurious {\bmode}s detected will indeed be present in the data set used to derive the science results. 
There are systematics that naturally cancel out with many detectors, or with the instrument boresight rotation of the observation strategy (BK-III Section 2.3 and Section 4).
In this case failing a jackknife test might not necessarily mean that there was significant contamination in the full coadd maps.
Conversely, there could hypothetically be systematic contamination in the full coadd map that somehow cancels in all considered jackknife splits. \\

In addition, going beyond the two-point statistics offers more information about the observed maps.
Each of the distortion field corresponds to a certain type of systematics. Therefore, failing the systematics check for a certain distortion field can offer hints as to where to investigate. Furthermore, we will show that the quadratic estimators for distortion fields are usually more sensitive compared to the {\it BB} spectrum at detecting the corresponding distortions. Any spurious {\bmode}s from distortion fields would be detected by its quadratic estimator before it significantly effects the {\it BB} spectrum. \\ 

In the map-making process, the $Q$ and $U$ modes that are potentially contaminated through beam systematics, in particular differential gain and differential pointing, are filtered out by the {\it deprojection} procedure (BK-I). The choice of deprojection time scale of around 10 hours is a compromise---a shorter deprojection time scale guards against systematics that vary over short periods, but at the same time removes more modes and reduces the overall statistical power (BK-I, BK-III). The distortion fields $\gamma_1, \gamma_2$ are sensitive to the modes corresponding to the differential gain, while $d_1$, $d_2$ are sensitive to the modes corresponding to differential pointing. Therefore, the distortion field estimators as systematics checks can guard against beam systematics that vary faster than the 10 hour deprojection time scale, eluding deprojection. \\

However, there are many classes of systematic contamination that do not correspond to any of the distortion fields. Therefore, the distortion field systematics tests should be treated as a useful complementary check for the standard jackknife tests rather than a replacement. \\

For different experiments with different ways of measuring CMB polarization (e.g.\ pair differencing vs.\ rotating half wave plate), the mapping between detector systematics to the final line-of-sight distortion field can vary. We will discuss the case for experiments similar to BK which take the pair difference signal from pairs of detectors with orthogonal polarization directions, and then use boresight angle rotation to get a distribution of polarization angles to be able to solve for $Q/U$ (BK-II). \cite{Hu2003} offers a more general discussion of the connection between instrumental systematics and distortion fields. 

\subsection{Instrumental Systematics and Distortion Fields}
\label{subsection: mbs: instrumental effects that correspond to distortion fields}

Since the telescope is constantly scanning the sky, a time-varying spurious systematic effect will translate into a position-dependent error in the map, which can be associated with different distortions \citep{Hu2003,Yadav2010}. On the other hand, a spurious systematic effect that stays constant in time but varies from detector to detector will also cause a position-dependent error in the map, since different detectors cover different regions of the map. \\

Miscalibration of detector gains (pair sum timestream signal) would be captured in the amplitude modulation field $\tau(\hn)$. The miscalibration can be time-varying, due to the uncertainties in the elevation-nod gain calibration (BK-II) between each hour of observation. There are also gain variations that stay constant in time but vary among detector pairs. When making the full season map, the observed $T$ map is correlated with the {\planck} $T$ map to derive one overall normalization factor to calibrate the amplitude of the map, referred to as the {\it absolute calibration} (BK-II). \\

Variations of the actual absolute calibration values between detectors can translate into spatial amplitude modulation of the coadded maps. This amplitude variation could also be introduced by bandpass mismatches (BK-II) between pairs of detectors. Since the detector gain is calibrated with the atmospheric response, and the atmospheric emission and CMB have different spectra, a mismatch in detector bandpasses will lead to a gain mismatch in the observed CMB signal. The gain mismatches discussed above can also happen between intra pair detectors. In this case, instead of an amplitude modulation distortion $\tau(\hn)$, we will get monopole $T$ to $P$ leakage (or differential gain leakage) which corresponds to the $\gamma_1(\hn)$ and $\gamma_2(\hn)$ fields. \\

Miscalibration of the orientation of the detectors will translate to the rotation of the plane of polarization field $\alpha(\hn)$. The overall rotation of the map is calibrated out by minimizing the {\it EB} and {\it TB} spectra (BK-I). However, variations of the orientation from detector to detector can translate to an anisotropic rotation distortion field whose amplitude can be limited by the quadratic reconstructions. \\

The $f_1(\hn)$ and $f_2(\hn)$ fields can be generated from a coupling between a gain miscalibration and the boresight angle rotation (for a telescope with such capability). For example, if at the boresight angles that contribute more to the $Q$ map, the detectors consistently exhibit a higher gain than the boresight angles that contribute more to the $U$ map, we will effectively get a higher amplitude $Q$ map compared to $U$, which corresponds to an $f_1(\hn)$ distortion. With the observation strategy of BK, different pairs of detectors cover different RA and Dec. ranges on the sky. Therefore, gain variations among detector pairs can stochastically lead to $f_{1/2}(\hn)$ distortion fields. Although there is no physical mechanism known to us that can produce this type of systematic coupling between the detector gains and the boresight angles in the BK experiments, we constrain $f_1/f_2$ for completeness. \\

The $\bm{p}(\hn)$ ($\kappa(\hn)$/$\omega(\hn)$) fields capture changes in the CMB photon directions. The corresponding instrumental systematic is miscalibration of the beam center locations. In the main line analysis, the beam centers are derived from cross correlating the observed $T$ maps with the {\planck} $T$ map (Section 11.9 of BK-II). Any miscalibration or uncertainty that varies from detector pair to detector pair will produce the $\bm{p}(\hn)$ distortion fields. \\

The second line in Eq.~\ref{Eq.mbs: Characterization of 11 distortion fields} involves $T$ to $P$ leakage. These distortions arise from a mismatch of the beams of pairs of orthogonal detectors A and B. Consider a Gaussian beam:
\begin{eqnarray}
    &\mathcal{B}(\hn, \bm{b}, e) = \frac{1}{2\pi \sigma^2 (1-e^2)}
    \times \\ \nonumber
    &\exp[-\frac{1}{2\sigma^2} \left( \frac{(n_1-b_1)^2}{(1+e)^2} 
    + \frac{(n_2-b_2)^2}{(1-e)^2}
    \right)] \,,
\end{eqnarray}
where $\bm{b}$ is the beam offset, $\sigma$ is the mean beam width, and $e$ is the ellipticity in the direction of detector polarization (plus ellipticity). In the BK beam map measurements, the differential cross ellipticites are subdominant compared to the differential plus ellipticites \citep{biceptwoXI}. Therefore, the cross ellipticity and its corresponding distortion field are ignored in this paper. A mismatch of the beam parameters between A and B will translate to the distortion fields as follows:
\begin{align}
        \sigma \bm{p} &= (\bm{b}_A+\bm{b}_B)/2 \,, \\
        \sigma \bm{d} &= (\bm{b}_A - \bm{b}_B)/2 \,, \\
        q &= (e_A-e_B)/2 \,.
\end{align}

In Table~\ref{table: instrumental systematics corresponding to distortion fields. } we summarize the correspondence of instrumental systematics to the distortion fields. To demonstrate the connection between the systematic effects and the distortion fields, we generate simulations that contain some of the systematic effects and reconstruct their distortion field spectra. The systematic effects are added to the {\it pair maps} (BK-I) before the map coaddition step to reduce the cost of computation. The complete analysis is presented in Appendix \ref{appendix: simulations of systematic effects}. Here we present two representative cases, one where the systematics are constant in time but vary over detectors (a 10$\deg$ random Gaussian detector polarization angle rotation), and another where the systematics vary over time (10\% random Gaussian differential gain fluctuation varying from hour to hour). \\

\begin{table*}[htb]
\centering
\begin{tabular}{c c  } 
 \hline \hline 
 \rule{0pt}{4ex}
 fields & instrumental systematics \\ [0.5ex] 
    \hline
    \rule{0pt}{3ex}
    $\tau$ 
    & detector gain miscalibration $(g_A + g_B )/2$  \\ [0.5ex] 
    $\alpha$ 
    & detector polarization orientation miscalibration  \\ [0.5ex] 
    $f_1,\; f_2$ 
    & detector gain miscalibration coupled with boresight angle  \\ [0.5ex]     
    $\bm{p}$ 
    & beam center miscalibration  \\ [0.5ex] 
    $\gamma_1,\; \gamma_2$ 
    & A/B detector differential gain  $(g_A - g_B)/2$  \\ [0.5ex] 
    $d_1,\; d_2$ 
    & A/B detector differential pointing $(\bm{b}_A - \bm{b}_B)/2$  \\ [0.5ex] 
    $q$ 
    & A/B detector differential beam ellipticity  \\ [0.5ex] 
    \hline \hline
\end{tabular}
\caption{A summary of the instrumental systematics that correspond to each distortion field in Eq.~\ref{Eq.mbs: Characterization of 11 distortion fields}.}
\label{table: instrumental systematics corresponding to distortion fields. }
\end{table*}

\begin{figure}[thb]
\centering
\includegraphics[width=1.05\linewidth]{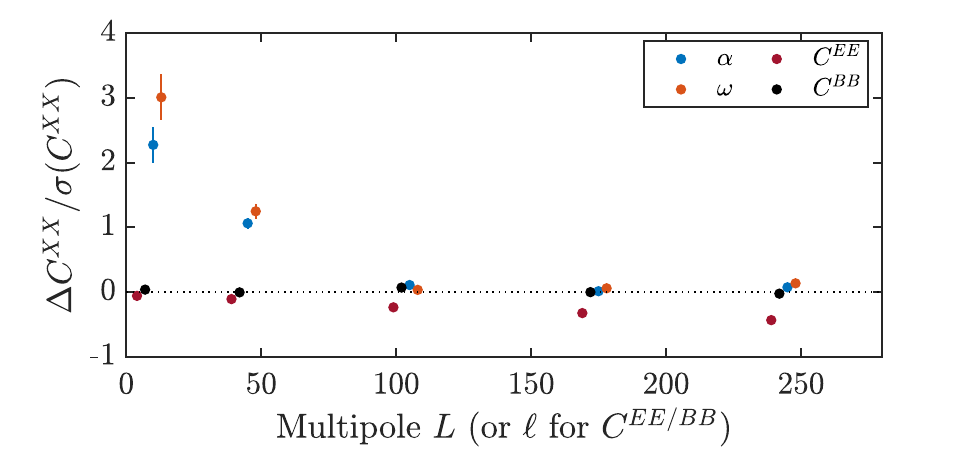}
\caption{Mean shift in $\alpha$, $\omega$, {\it EE}, and {\it BB} spectra in simulations with a 10$\deg$ random detector polarization angle scatter divided by the standard deviation of the non-rotated simulations.
$\alpha$ and $\omega$ pick up a strong signal at low $L$, while $C_\ell^{EE}$ is suppressed at higher $\ell$ because the random per-detector polarization rotations average out near the center of the map and reduces the overall amplitude.}
\label{Fig: mbs: random rotation systematics spectra}
\end{figure}

The shift in the $\alpha$, $\omega$, $EE$, and $BB$ spectra caused by the randomized detector rotation angles are shown in Fig.~\ref{Fig: mbs: random rotation systematics spectra}.
For the extreme 10$\deg$ case simulated $C_L^{\alpha \alpha}$ shows a very strong signal at low $L$ while the $BB$ spectrum remains unaffected. We stress that this is not anymore true if we didn't calibrate the overall rotation of the maps and there would be a non-zero mean angle calibration error, causing a scale dependent signal in the distortion field power spectrum \cite{Mirmelstein2021}. The curl component of the lensing field $\omega$ also detects the rotation field due to the correlation between the $\alpha$ and $\omega$ estimators. \\

\begin{figure}[thb]
\centering
\includegraphics[width=1.05\linewidth]{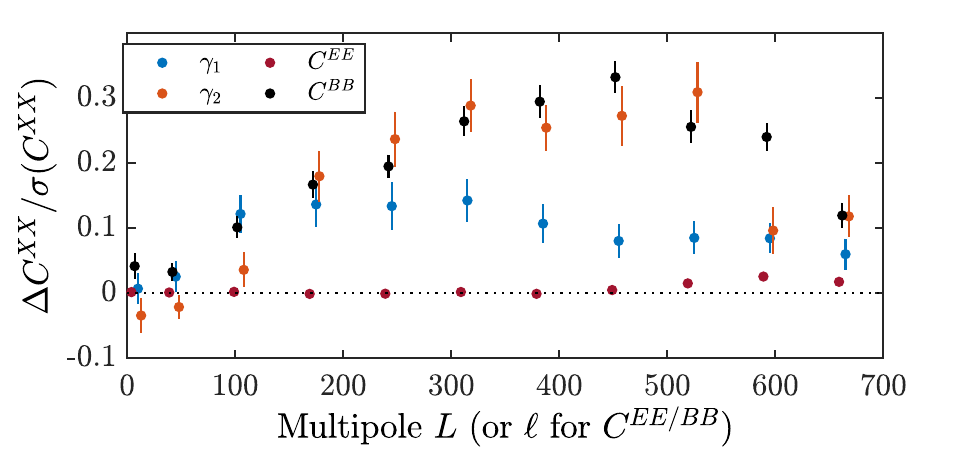}
\caption{Mean shift in $\gamma_1$, $\gamma_2$, $EE$ and $BB$ spectra in simulations with a 10\% Gaussian fluctuation of differential gain that varies from detector pair to detector pair and from hour to hour divided by the standard deviation of the non-fluctuated simulations.}
\label{Fig: mbs: diffgain systematics}
\end{figure}

In Fig.~\ref{Fig: mbs: diffgain systematics}, we show the spectra for $\gamma_1$, $\gamma_2$, {\it EE}, and {\it BB} for the 10\% random differential gain fluctuations. Compared to the detector angle rotation where the instrumental effect is constant in time but varies over detector pairs, the time-varying gain mismatch simulations generate distortion power that are distributed over a wider range of multipoles. This systematic $T$ to $P$ leakage shows up strongly in both $\gamma$ and $BB$. \\

In Appendix \ref{appendix: simulations of systematic effects}, we observe that other kinds of systematics that are constant in time but vary over detectors also generate distortions at large scales (low $L$). This is because each detector pair covers a significant portion of the map, and variations among detector pairs therefore primarily create large scale distortions. On the other hand, time-varying instrumental systematics can generate distortion power over a much wider range of multipoles. Depending on the type of systematic effects being studied, we can design systematics tests that focus on different multipole ranges of the distortion field spectra.

\subsection{Quadratic Estimators vs. {\it BB} power spectra for detecting distortion fields}
\label{subsection: QE vs. BB at detecting distortion fields}

In the main line analysis (BK-I; BK-V; BK-VI; BK-X; BK-XIII) $EE$, $BB$ and $EB$ spectra of map difference splits are used to test for instrumental systematics.
In this section, comparisons are made between $BB$ power spectra and quadratic estimators in their ability to detect various systematics. We highlight cases in which the latter are more sensitive at detecting the spurious $B$-mode produced by the distortion fields. To that end, we generate simulations with Gaussian realizations of distortion fields within a narrow range of multipoles ($\Delta L=50$). Any Gaussian distortion field with a smooth spectrum can be considered as a combination of multiple $\Delta L$ distortions,

\begin{equation}
    C_L^{DD} = \begin{cases} A_D^2 & (L_{\rm min}\leq L\leq L_{\rm max}) \\ 0 & (\text{otherwise}) \end{cases} \,.
\label{Eq. mbs: input deltaL=50 distortion spectra}
\end{equation}

To make sensitivity comparisons between quadratic $EB$/$TB$ estimators vs.\ $BB$ power spectra, we use distortion field simulations with different $L$ range inputs as the fiducial model and see how well the amplitude of that fiducial distortion spectra can be constrained by the standard simulations (un-distorted lensed-\lcdm+dust+noise).
We define the {\it sensitivity ratio} as:
\begin{align}
    &\hat{A}_D^{XX'} = \frac{\sum_{bb'} \hat{C}^{XX'}_b \mathbf{Cov}_{bb'}^{-1} C_{b'}^{XX',f}}{\sum_{bb'} C_b^{XX',f} \mathbf{Cov}_{bb'}^{-1} C_{b'}^{XX',f}} \,,\\
    &\text{sensitivity ratio} = \frac{\sigma(A_D^{BB})}{\sigma(A_D^{EB/TB})} \,,
\label{Eq. mbs: sensitivity comparison between EB/TB vs. BB}
\end{align}
where {\it XX'} represents {\it EB}, {\it TB}, or {\it BB}. $C_b^{EB/TB}$ are the distortion field reconstruction bandpowers (4-point) from quadratic {\it EB/TB} estimators, and $C_b^{BB}$ are the 2-point {\it BB} bandpowers. $C_b^f$ stands for the mean bandpower from the distortion simulations characterized by Eq.~\ref{Eq. mbs: input deltaL=50 distortion spectra}, which we take as the ``signal" of the particular systematic effect that we want to measure or constrain. $\sigma(A_D)$ is the standard deviation of the best fit $A_D$ amplitude for the level of systematics from the 499 un-distorted lensed-\lcdm+dust+noise simulations. The estimator that is more sensitive in detecting the systematics will have a larger signal-to-noise in measuring the amplitude $\hat{A}_D$, and therefore a smaller $\sigma(A_D)$ value. When the sensitivity ratio defined in Eq.~\ref{Eq. mbs: sensitivity comparison between EB/TB vs. BB} is greater than 1, the quadratic estimator is more sensitive than the {\it BB} power spectrum at detecting the particular distortion at that angular scale. \\

\begin{figure}[htb]
\centering
\includegraphics[width=1.05\linewidth]{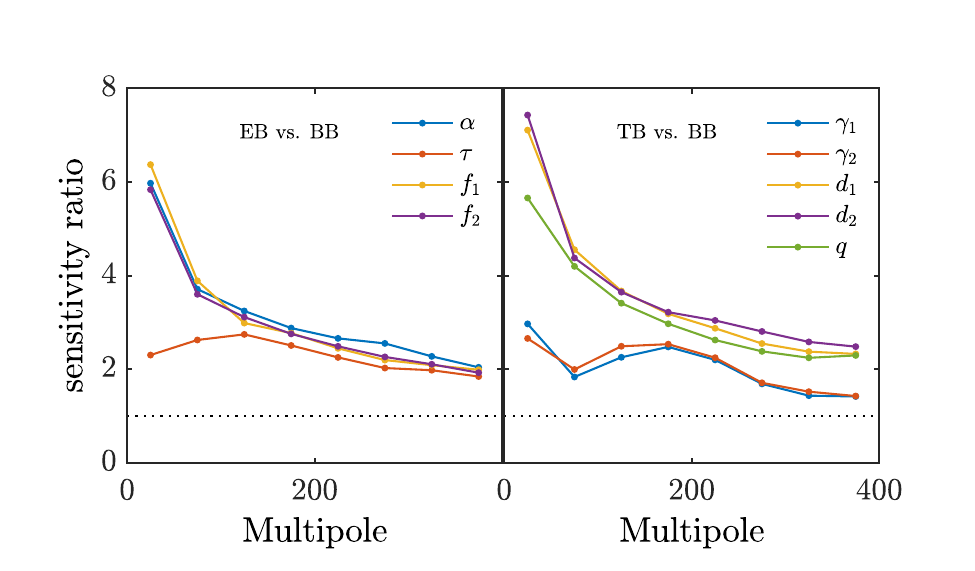}
\caption{sensitivity ratio as defined in Eq.~\ref{Eq. mbs: sensitivity comparison between EB/TB vs. BB}. A sensitivity ratio ${\sigma(A_D^{EB})}/{\sigma(A_D^{BB})}$ above 1 means that the quadratic estimator is more sensitive than the {\it BB} spectrum at detecting the distortion field at that angular scale. The left plot shows the sensitivity ratio for the fields involving only the polarization, while the right plot shows the fields involving $T$.}
\label{Fig: mbs: sensitivity comparison EB/TB vs. BB}
\end{figure}

In Fig.~\ref{Fig: mbs: sensitivity comparison EB/TB vs. BB}, we demonstrate that the quadratic estimators are more sensitive than the {\it BB} power spectrum at detecting the distortion fields between $L=1-400$. For all distortion fields, the quadratic estimators perform better when the distortion power is at larger scale (lower $L$). Among the polarization-only distortions, $\alpha$, $f_1$, and $f_2$ in particular are detected by the {\it EB} quadratic estimators with high sensitivity relative to the {\it BB} spectra. The distortions involving CMB temperature, $d_1$, $d_2$, and $q$ are also very sensitively measured by the {\it TB} quadratic estimators.
The above is as we would like it to be---we can detect systematics using the distortion fields before they significantly bias the $BB$ spectrum. \\

Since we are reconstructing all the distortion field spectra $C_L^{DD}$ simultaneously and the different distortion fields are not necessarily orthogonal to each other, it is important to study whether any spurious signal detected by a particular $D_1$ estimator can be reliably pointed to as an actual distortion signal from that field. To this end, we use the same set of simulations described by Eq.~\ref{Eq. mbs: input deltaL=50 distortion spectra} to test for the cross sensitivity or correlation between the different distortion fields. We define the sensitivity ratio the same way as in Eq.~\ref{Eq. mbs: sensitivity comparison between EB/TB vs. BB}, but in this analysis a different quadratic estimator $D_2$ is applied to try to detect the $D_1$ distortion input. \\

\begin{figure}[htb]
\centering
\includegraphics[width=1.0\linewidth]{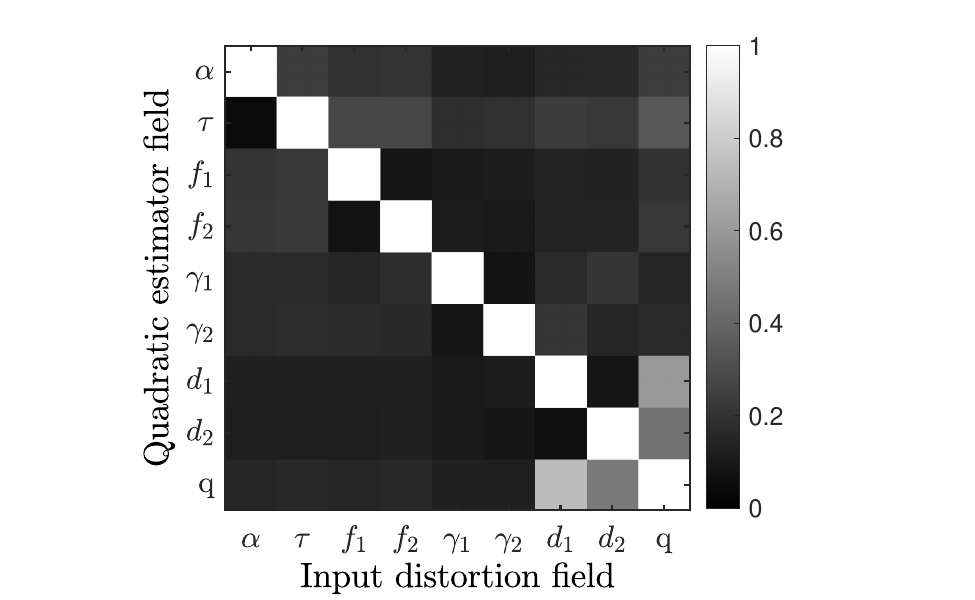}
\caption{The correlation matrix of the sensitivity ratio as defined in Eq.~\ref{Eq. mbs: sensitivity comparison between EB/TB vs. BB} for all combinations of input distortion and quadratic estimators. The results plotted use distortion input from $L=1-50$. The horizontal axis shows the input distortion field injected in the simulations, and the vertical axis shows the quadratic estimators used to detect the signal. 
We see that $q$ and $d_1$/$d_2$ have the strongest correlations, but in general all the distortion signals are best measured with their own estimators. }
\label{Fig: mbs: cross sensitivity between different distortion fields.}
\end{figure}

In Fig.~\ref{Fig: mbs: cross sensitivity between different distortion fields.}, we show the correlation between different distortion fields using the $L=1-50$ distortion simulations with the diagonal normalized to 1. The fact that the diagonal terms are much larger than the off-diagonal terms means the distortions would be much more strongly detected with the corresponding quadratic estimator before they are detected by another estimator. One exception is the correlation of $q$ with $d_1$/$d_2$ at low $L$. The existence of a large $T$ to $P$ dipole leakage can swamp the $q$ estimator as we will see in Section~\ref{subsection: TB null tests without deprojection}. \\

\subsection{Effects of Point Source Contamination}
\label{subsection: Effects of point source contamination}
The brightest point sources at the frequencies relevant for CMB observations are flat-spectrum radio sources \citep{Battye2011}. Since these point sources are brighter at lower frequencies relative to the CMB spectrum, 
the 95\,GHz data set is much more affected by point sources in the distortion field analysis. \\

We estimate the effect of point source contamination in the distortion field analysis by injecting simulated point sources from a preliminary catalog obtained by private communication with the SPT-3G collaboration. The fluxes are taken from preliminary SPT-3G 95\,GHz data, and are on average 2.5\% polarized with an approximately exponential distribution. The $T$, $Q$, and $U$ fluxes are converted to equivalent CMB temperature and added to the pixels closest to the location of the sources in the input map.
The \bicepthree beam and observation matrix $R$ is then applied on to generate a point source simulation map (Eq.~\ref{Eq. observation matrix}).
We find that the mean shift of the $EB$ distortion spectra caused by these point sources is negligible. However, the 95\,GHz {\it TB} estimators strongly detect the point sources, with the brightest few accounting for most of the contribution. \\

From this simulation we determine that the point source contribution becomes negligible after masking the 20 sources with the highest polarized fluxes. These are then added to the apodization mask by injecting Gaussian divots with 0.5$\deg$ width at the 20 locations. In Table~\ref{table: point source mask PTE}, the $\chi/\chi^2$ PTEs for the real data, with and without the point source mask, for the {\it TB} estimators are listed. We find that the point source mask is necessary for the \bicepthree\ 95\,GHz data to pass the distortion field systematics tests, but does not affect the \biceptwo/\keck\ 150\,GHz data much. \\

\begin{table*}[htb]
\centering
\begin{tabular}{c | c c | c c} 
 \hline \hline 
 \rule{0pt}{2ex}
  & \multicolumn{2}{c|}{95\,GHz ($\chi$/$\chi^2$ PTE)} & \multicolumn{2}{c}{150\,GHz ($\chi$/$\chi^2$ PTE)}  \\
 Field & w/o psm & with psm & w/o psm & with psm \\ [0.5ex] 
    \hline
    \rule{0pt}{3ex}
$d_1$ & 0.02 / 0.01 & 0.45 / 0.08 & 0.69 / 0.88 & 0.56 / 0.88\\ [0.5ex]
$d_2$ & 0.03 / 0.40 & 0.54 / 0.63 & 0.84 / 0.99 & 0.97 / 0.83\\ [0.5ex]
$\gamma_1$ & \textbf{7.3e-04}  / 0.13 & 0.09 / 0.41 & 0.07 / 0.12 & 0.18 / 0.06\\ [0.5ex]
$\gamma_2$ & 0.52 / 1.00 & 0.78 / 0.98 & 0.03 / 0.06 & 0.05 / 0.08\\ [0.5ex]
q & 0.22 / 0.04 & 0.48 / 0.63 & 0.22 / 0.53 & 0.63 / 0.80\\ [0.5ex]
    \hline \hline
\end{tabular}
\caption{The $\chi$/$\chi^2$ PTEs with and without point source masks (psm) for the {\it TB}-reconstructed distortion fields. The bolded value is derived from the theoretical $\chi$ distribution since the observed value is outside of the 499 simulation distribution. The point source mask removes the brightest 20 sources in polarization according to a preliminary SPT-3G catalog. Without the point source mask, 95\,GHz would fail the $\gamma_1$ systematics test and in general have lower PTEs. For 150\,GHz, there is no significant change to the PTEs. }
\label{table: point source mask PTE}
\end{table*}

In BK-XIII Appendix F it is estimated that the polarized flux from point sources may produce a bias on $r$ at a level of $\approx 1-3 \times10^{-3}$. While small compared to our present uncertainties, point source contamination and its mitigation will become more important in future analysis, and the {\it TB} quadratic estimators can be a powerful diagnostic tool. In Appendix \ref{appendix: Detection of point sources}, we discuss the reasons why {\it TB} quadratic estimators are sensitive to polarized point sources, derive estimators that are even more powerful for detecting point sources, and compare the performance of the different estimators at point source detection. \\

\subsection{Distortion Field Systematics Tests on BK Real Data} \label{subsection: Distortion field null tests}

\begin{figure*}[htb]
\centering
\includegraphics[width=.8\linewidth]{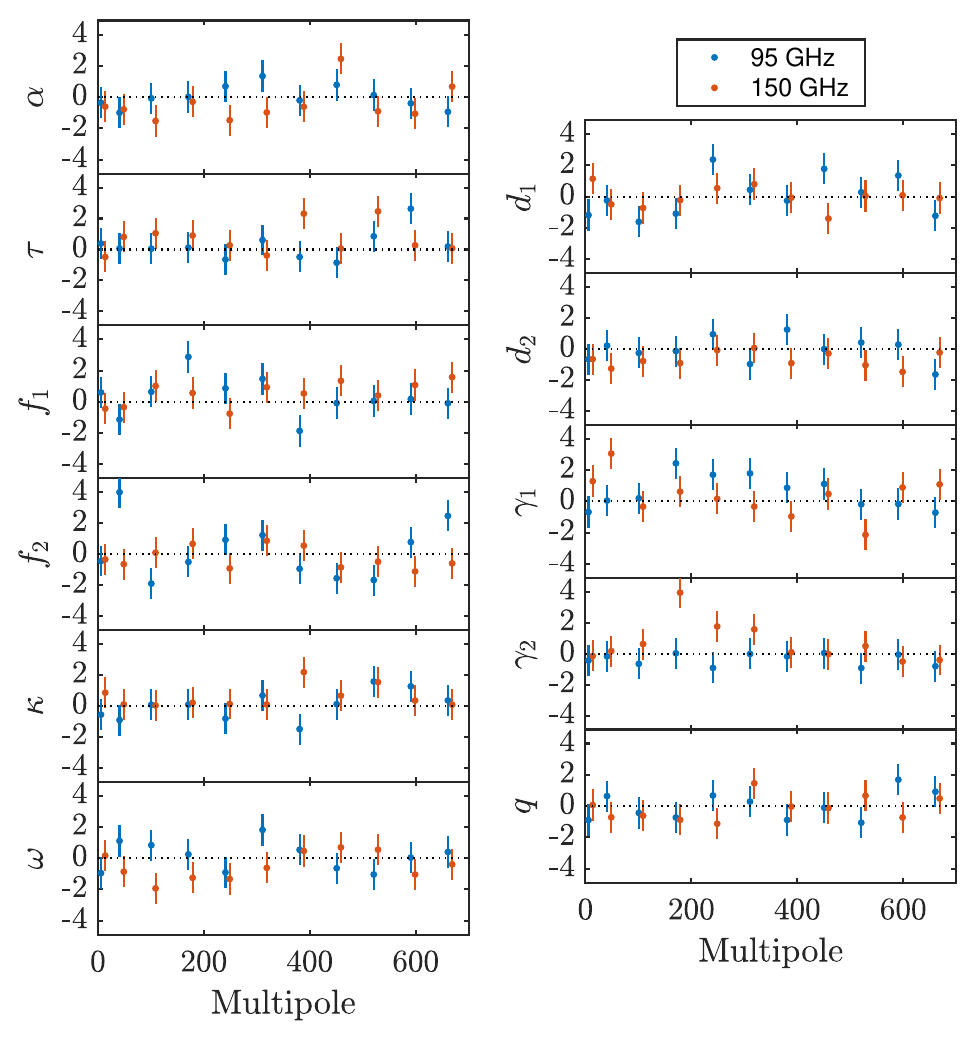}
\caption{The fractional deviations of the 11 real data distortion field bandpowers  $(\hat{C}_L^{DD}-\bar{C}_L^{DD})/\sigma(\hat{C}_L^{DD})$. Plots on the left are reconstructed with the {\it EB} quadratic estimators, while the ones on the right are reconstructed with the {\it TB} estimators, and have the point source mask applied. The corresponding $\chi$ and $\chi^2$ PTEs are listed in Table~\ref{table: chi and chi2 PTE values from jack0 reconstruction spectra.}. }
\label{Fig: mbs: jack0 CLdd over errorbar}
\end{figure*}

With the connections between the various systematics and distortion fields established, in this section we present the results of the distortion field systematics tests for the two real data maps at 95 and 150\,GHz.
The distortion field systematics tests are performed with the same method as in Section~\ref{subsection: science field jackknife null tests} and Eq.~\ref{Eq. chi2 null test definition.}--\ref{Eq. chi null test definition.}. The only difference is that the bandpowers $\hat{C_b}$ here correspond to the reconstructions from the full $E$ and $B$ map instead of the jackknife $B$ map. In Fig.~\ref{Fig: mbs: jack0 CLdd over errorbar} we plot the difference of the real data reconstructed distortion field bandpowers and the mean of simulations, divided by the standard deviation of the simulations to show the significance of detection. \\

The $C_L^{DD}$ spectra use the realization-dependent bias estimation outlined in Section~\ref{section: mbs: estimating the distortion field power spectra}. For the fields reconstructed with {\it TB} estimators (right side of Fig.~\ref{Fig: mbs: jack0 CLdd over errorbar}), we substitute in the same observed $T$ map for all the simulations. This is because the standard lensed-\lcdm simulations are constrained to the real CMB $T$ map (BK-I), and $T$ is measured with such high signal-to-noise that the noise contribution is negligible. Since we are much more interested in testing for systematics in our $B$ map rather than in the $T$ map, we elect to fix to the same extremely well measured observed BK $T$ map for both observation and simulations. \\

\begin{figure}[thb]
\centering
\includegraphics[width=1.0\linewidth]{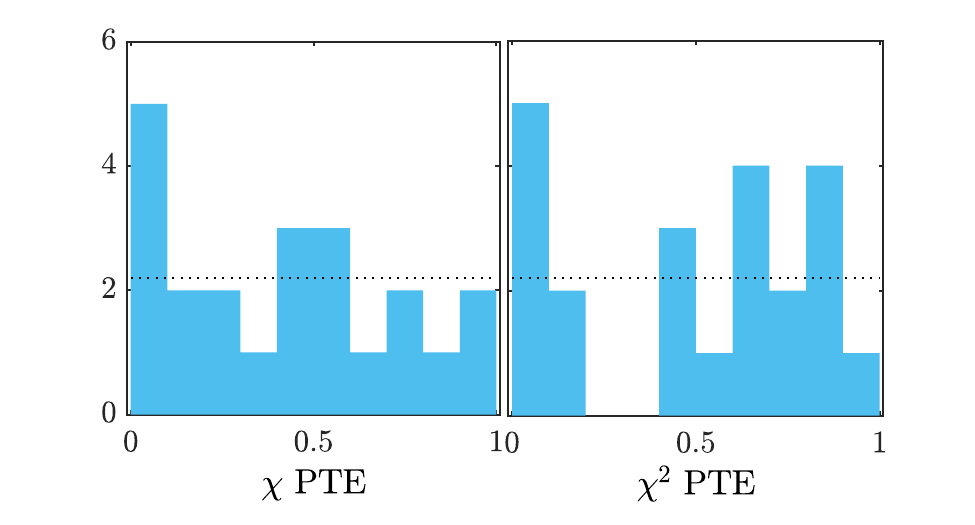}
\caption{Distributions of the distortion systematics test $\chi$ and $\chi^2$ PTEs in Table~\ref{table: chi and chi2 PTE values from jack0 reconstruction spectra.}. 
}
\label{Fig: mbs: chi and chi2 PTE distributions.}
\end{figure}

With the $\hat{C}_L^{DD}$ in Fig.~\ref{Fig: mbs: jack0 CLdd over errorbar}, we evaluate the $\chi$ and $\chi^2$ PTEs.
We list these values in Table~\ref{table: chi and chi2 PTE values from jack0 reconstruction spectra.},
and show histograms in Fig.~\ref{Fig: mbs: chi and chi2 PTE distributions.}.
All the $\chi$ and $\chi^2$ values lie within the 499 simulation distributions. There is one low PTE at 0.002 for the $\chi^2$ of $f_2$ from 95\,GHz. Examining the spectra in Fig.~\ref{Fig: mbs: jack0 CLdd over errorbar}, the low $\chi^2$ PTE can be traced to the second band power that fluctuates high. With the same method as Eq.~\ref{Eq. chi2 extreme}--\ref{Eq. overall extreme}, we take into account the look-elsewhere effect and evaluate the global PTE statistic that compares the most extreme value among the 44 numbers in Table~\ref{table: chi and chi2 PTE values from jack0 reconstruction spectra.} to the simulations.
We find that the probability to get a global value less than 0.002 is 0.08, offering no
evidence for contamination in the data. \\ \vspace{1cm}

\begin{table}[htb]
\centering
\hspace*{-20ex}
\begin{tabular}{c | c c | c c} 
 \hline \hline 
 \rule{0pt}{4ex}
  & \multicolumn{2}{c|}{95\,GHz PTE} & \multicolumn{2}{c}{150\,GHz PTE}  \\
Field & $\chi$ & $\chi^2$ & $\chi$ & $\chi^2$ \\ [0.5ex] 
    \hline
    \rule{0pt}{3ex}
$\alpha$& 0.50& 0.90& 0.90& 0.18\\ [0.5ex]
$\tau$& 0.22& 0.43& 0.04& 0.19\\ [0.5ex]
$f_1$& 0.20& 0.09& 0.08& 0.65\\ [0.5ex]
$f_2$& 0.26& 0.002& 0.75& 0.84\\ [0.5ex]
$\kappa$& 0.48& 0.53& 0.08& 0.73\\ [0.5ex]
$\omega$& 0.36& 0.48& 0.92& 0.62\\ [0.5ex]
$d_1$& 0.45& 0.08& 0.56& 0.88\\ [0.5ex]
$d_2$& 0.54& 0.63& 0.97& 0.83\\ [0.5ex]
$\gamma_1$& 0.09& 0.41& 0.18& 0.06\\ [0.5ex]
$\gamma_2$& 0.78& 0.98& 0.05& 0.08\\ [0.5ex]
$q$& 0.48& 0.63& 0.63& 0.80\\ [0.5ex]
    \hline \hline
\end{tabular}
\caption{$\chi$ and $\chi^2$ PTEs derived from the distortion field spectra shown in Fig.~\ref{Fig: mbs: jack0 CLdd over errorbar}. The results of {\it TB} estimators are derived with the point source mask applied, therefore the PTEs from $d_1$ to $q$ are identical to the ``with psm" case of Table~\ref{table: point source mask PTE}. The global PTE for the most extreme $\chi$/$\chi^2$ PTE (0.002 here) is 0.08. }
\label{table: chi and chi2 PTE values from jack0 reconstruction spectra.}
\end{table}

\subsection{Effectiveness of Deprojection} \label{subsection: TB null tests without deprojection}

As discussed in Section~\ref{subsection: mbs: instrumental effects that correspond to distortion fields}, many of the distortion fields correspond to specific forms of beam mismatch. Beam systematics have been very important and well studied in the BK experiments (BK-III). Starting from {\biceptwo}, the deprojection method has been developed to filter out potentially spurious signals that correspond to $T$ to $P$ leakage modes (BK-I, BK-VII). We have also carried out extensive far field beam measurement campaigns every year as well as published a beam systematics paper to model and quantify how the beam systematics can affect the measurement of the tensor-to-scalar ratio $r$ \citep{biceptwoXI}. \\

In Section~\ref{subsection: Distortion field null tests}, with the standard differential gain and differential pointing deprojection, the distortion field {\it TB} systematics tests pass with no evidence of any residual $T$ to $P$ leakage. We will now investigate whether the distortion field quadratic estimators can detect any spurious signal if we do not perform the differential gain and differential pointing deprojections. With the data products available on disk, it is simple to add the components that are filtered out by the deprojections back in, and construct maps without deprojection. We then perform the same $\chi$/$\chi^2$ distortion field systematics tests by comparing the observed $\hat{C}_L^{DD}$ with the simulations. \\

\begin{figure*}[b]
\centering
\includegraphics[width=.8\linewidth]{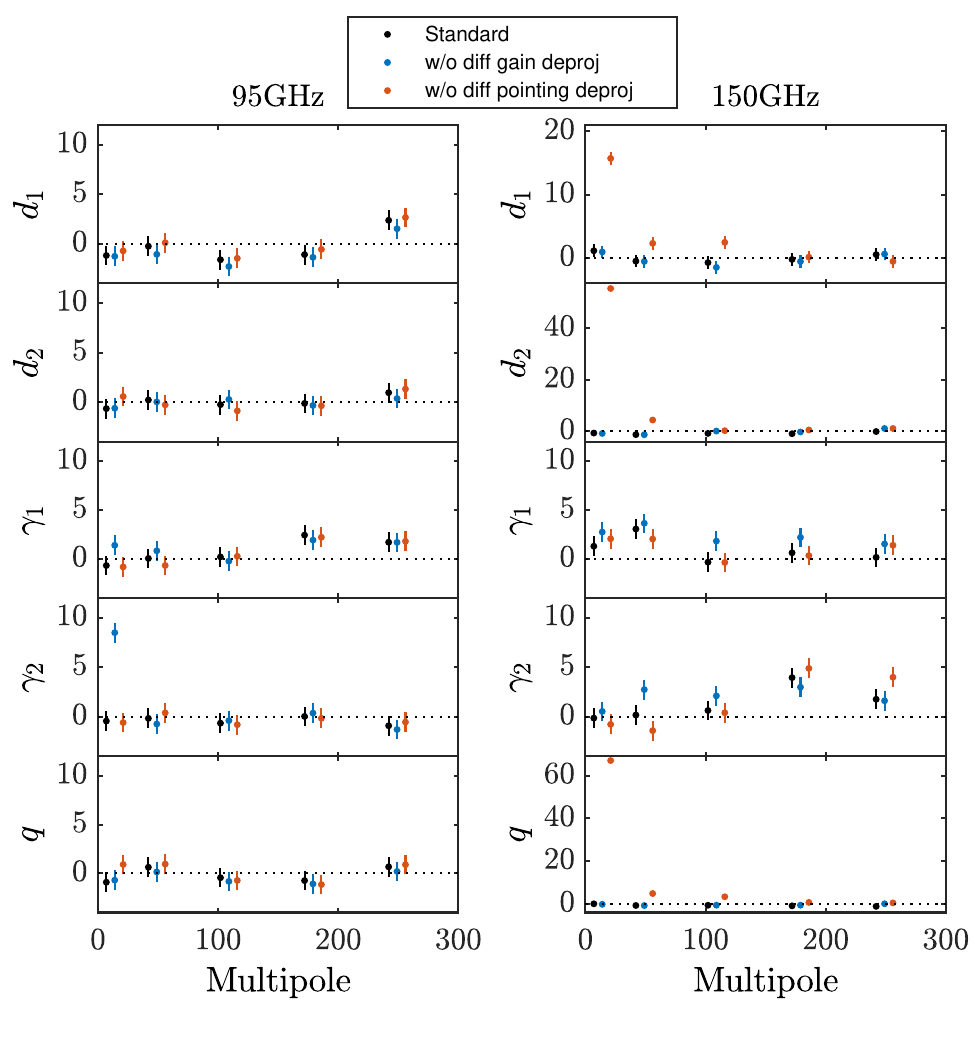}
\caption{The fractional deviations of the 11 real data distortion field bandpowers $(\hat{C}_L^{DD}-\bar{C}_L^{DD})/\sigma(\hat{C}_L^{DD})$ with and without the differential gain and differential pointing deprojections. 95\,GHz fails the $\gamma_2$ systematics test without differential gain deprojection. 
150\,GHz fails the $d_1$, $d_2$, and $q$ systematics tests without differential pointing deprojection.}
\label{Fig: mbs: TB null spectra with different deprojection options}
\end{figure*}

In Fig.~\ref{Fig: mbs: TB null spectra with different deprojection options}, the \biceptwo/\keck\ 150\,GHz maps fail the $d_1$, $d_2$, and $q$ systematics tests spectacularly without differential pointing deprojection. On the other hand, the \bicepthree\ 95\,GHz maps have much lower differential pointing and do not see much of a change in the reconstructed spectra when differential pointing deprojection is turned off. Without differential gain however, we would detect a strong large scale $\gamma_2$ distortion in 95\,GHz.
We note that the differential pointing systematic in 150\,GHz is also strongly detected by the $q$ {\it TB} estimator. This is consistent with the results in Fig.~\ref{Fig: mbs: cross sensitivity between different distortion fields.}, where we showed that $q$ has significant correlation with $d_1$ and $d_2$ at large scale. \\

In Table~\ref{table: chi and chi2 PTE values for different deprojection options}, we show the $\chi$ and $\chi^2$ PTEs for different deprojection options. In general, the $\chi$ and $\chi^2$ PTEs decrease when either differential gain or differential pointing deprojection is disabled. This offers strong evidence that the deprojections are indeed successful in filtering out the monopole and dipole $T$ to $P$ leakage in the real data when compared to simulations with no such systematics. We note that the suite of jackknife tests for the power spectrum analysis includes targeted tests to detect these types of systematic contamination, which would lead to the same conclusion. \\

 \begin{table}[htb]
\centering
\begin{tabular}{c | c c c } 
 \hline \hline 
 \rule{0pt}{4ex}
  & \multicolumn{3}{c}{95\,GHz PTE ($\chi/\chi^2$)} \\
Field & Standard & No diff gain & No diff point \\ [0.5ex] 
    \hline
    \rule{0pt}{3ex}
$d_1$& 0.45  $/$ 0.08 & 0.69  $/$ 0.06 & 0.24  $/$ 0.14 \\ [0.5ex]
$d_2$& 0.54  $/$ 0.63 & 0.56  $/$ 0.63 & 0.45  $/$ 0.24 \\ [0.5ex]
$\gamma_1$& 0.09  $/$ 0.41 & 0.08  $/$ 0.49 & 0.13  $/$ 0.43 \\ [0.5ex]
$\gamma_2$& 0.78  $/$ 0.98 & 0.22  $/$ \textbf{2e-13}& 0.76  $/$ 0.99 \\ [0.5ex]
$q$& 0.48  $/$ 0.63 & 0.51  $/$ 0.46 & 0.34  $/$ 0.36 \\ [0.5ex]
    \hline \hline
 \rule{0pt}{4ex}
 &  \multicolumn{3}{c}{150\,GHz PTE ($\chi/\chi^2$)}  \\
 field & Standard & No diff gain & No diff point \\ [0.5ex] 
 \hline
$d_1$& 0.56  $/$ 0.88 & 0.11  $/$ 0.42 & \textbf{7e-05}  $/$ \textbf{1e-49}\\ [0.5ex]
$d_2$& 0.97  $/$ 0.83 & 0.38  $/$ 0.88 & \textbf{2e-39}  $/$ $<$ \textbf{1e-99}\\ [0.5ex]
$\gamma_1$& 0.18  $/$ 0.06 & 0.002  $/$ 0.008 & 0.03  $/$ 0.04 \\ [0.5ex]
$\gamma_2$& 0.05  $/$ 0.08 & 0.01  $/$ 0.07 & 0.006  $/$ 0.002 \\ [0.5ex]
$q$& 0.63  $/$ 0.80 & 0.06  $/$ 0.27 & \textbf{7e-70}  $/$ $<$ \textbf{1e-99}\\ [0.5ex]
    \hline \hline
\end{tabular}
\caption{$\chi$ and $\chi^2$ PTEs derived from the distortion field spectra shown in Fig.~\ref{Fig: mbs: TB null spectra with different deprojection options}.
The bolded numbers are derived from theoretical $\chi$ and $\chi^2$ distributions when the real values are outside of the 499 sim distributions.}
\label{table: chi and chi2 PTE values for different deprojection options}
\end{table}

\section{Conclusions}
The line-of-sight distortion effects of the CMB can be characterized to first order with 11 fields. Three of these correspond to known or conjectured cosmological signals: gravitational lensing and $\kappa(\hn)$, patchy reionization and $\tau(\hn)$, and cosmic birefringence and $\alpha(\hn)$. Combining the sensitivity from our two deepest maps: 150\,GHz from {\biceptwo}/{\keck}, and 95\,GHz from {\bicepthree}, we constrained physical models that can generate these distortion fields.  For gravitational lensing, we measured the lensing amplitude to be $A_L^{\phi \phi}=0.97\pm 0.19$, which is a factor of two improvement from our previous results in BK-VIII. For cosmic birefringence, we constrained the amplitude of cosmic birefringence and the related cosmological parameters to $A_{CB}\leq 0.044$, $g_{a\gamma} \leq 2.6\times10^{-2}/H_I$, and $B_{1\text{Mpc}} \leq 6.6 \text{nG}$. This is a factor of three improvement on $g_{a\gamma}$ and a factor of four improvement for $B_{1\text{Mpc}}$ compared to our previous BK-IX (BK14) analysis, resulting in the tightest constraint to date from the CMB four-point function.
For patchy reionization, while not competitive compared to the {\planck} {\it TT} results \citep{Namikawa2018}, we achieve the best constraint with CMB polarization on the $\tau(\hn)$ amplitude. \\

Treating the distortion fields as systematics tests, we demonstrated with simulations the connections between the distortion fields and the various instrumental effects in experiments with similar designs to BK. We further show that the {\it EB/TB} distortion field estimators are more sensitive than the {\it BB} spectrum at detecting random Gaussian realizations of distortions especially at larger scales. Additionally, we find that the {\it TB} estimators are very sensitive to contamination from polarized point sources. \\

We perform instrumental systematics tests on the 95\,GHz and 150\,GHz maps by comparing the distortion field bandpowers of the real data to lensed-\lcdm+dust+noise simulations, and confirm that the 11 observed distortion spectra are consistent with the simulations. We also verify that the differential gain and differential pointing deprojections in our standard map-making pipeline are effective at filtering out the $T$ to $P$ leakage. Without the differential gain deprojection, we would detect an excess $\gamma_2$ power in the 95\,GHz map, while without differential pointing deprojection, we would detect a very strong excess in the $q$, $d_1$ and $d_2$ fields of the 150\,GHz map. 
This also confirms that the quadratic estimators are powerful tools to guard against $T$ to $P$ leakage systematics in the absence of differential gain and pointing deprojections. \\

With this first demonstration of quadratic estimators as instrumental systematics diagnostics on real data, we pave the way towards their future application as tools to self-calibrate upcoming data sets \citep{Yadav2010,Williams2021}. \\

\vspace{11mm}

The \bicep/\keck\ projects have been made possible through
a series of grants from the National Science Foundation
including 0742818, 0742592, 1044978, 1110087, 1145172, 1145143, 1145248,
1639040, 1638957, 1638978, \& 1638970, and by the Keck Foundation.
The development of antenna-coupled detector technology was supported
by the JPL Research and Technology Development Fund, and by NASA Grants
06-ARPA206-0040, 10-SAT10-0017, 12-SAT12-0031, 14-SAT14-0009
\& 16-SAT-16-0002.
The development and testing of focal planes were supported
by the Gordon and Betty Moore Foundation at Caltech.
Readout electronics were supported by a Canada Foundation
for Innovation grant to UBC.
Support for quasi-optical filtering was provided by UK STFC grant ST/N000706/1.
The computations in this paper were run on the Odyssey/Cannon cluster
supported by the FAS Science Division Research Computing Group at
Harvard University.
The analysis effort at Stanford and SLAC is partially supported by
the U.S. DOE Office of Science.
We thank the staff of the U.S. Antarctic Program and in particular
the South Pole Station without whose help this research would not
have been possible.
Most special thanks go to our heroic winter-overs Robert Schwarz,
Steffen Richter, Sam Harrison, Grantland Hall and Hans Boenish.
We thank all those who have contributed past efforts to the \bicep/\keck\
series of experiments, including the \bicepone\ team.
We also thank the \planck\ and \wmap\ teams for the use of their
data, and are grateful to the \planck\ team for helpful discussions.\\

\vspace{5cm}

\appendix

\section{Minimal Quadratic Estimators of Distortion Fields}
\label{section: Appendix: Minimal QE}
We follow \cite{Yadav2010} and construct a minimum variance quadratic estimator of the distortion field from {\it EB} and {\it TB} correlations. See Eq.~\ref{Eq.mbs: Characterization of 11 distortion fields} for the definitions of the distortions. A scalar field such as CMB temperature $T$ can be expanded in the Fourier basis as:
\begin{equation}
    T_\bl = \int d\hn T(\hn) e^{-i \bm{l}\cdot \hn} \,.
\end{equation}

A complex field $(S_1 \pm iS_2)(\hn)$ of spin $\pm s$ can be expanded in the Fourier harmonics basis as:
\begin{equation}
    [S_a \pm iS_b]_\bl = (\pm 1)^s \int d\hn [S_1(\hn)\pm iS_2(\hn)] e^{\mp si\phi_\bl} e^{-i\bl\cdot \hn} \,,
\end{equation}
where $\phi_\bl = \cos^{-1}(\hn \cdot \hat{\bl})$. We can also directly Fourier transform the $S_1 \pm iS_2(\hn)$ fields as:
\begin{equation}
    [S_1 \pm iS_2]_\bl = \int d\hn [S_1(\hn)\pm iS_2(\hn)]  e^{-i\bl \cdot \hn} \,.
\label{Eq. mbs: Fourier component of distortion fields.}
\end{equation}

One well known example is the transformation from $Q,U$ in map space to the $Q_\bl,U_\bl,E_\bl,B_\bl$ Fourier modes. In this case $(S_1\pm iS_2)(\hn) = (Q \pm iU)(\hn)$ is a spin $\pm 2$ field. Its Fourier transform is $(Q_\bl \pm iU_\bl)$ and its Fourier harmonics are $(E_\bl \pm iB_\bl)$. The Fourier harmonics are not dependent on the coordinates, whereas the direct Fourier transform of the individual spin fields will transform into each other with a rotation of the coordinates. \\

For $f_1/f_2$, $d_1/d_2$, $\gamma_1/\gamma_2$, we reconstruct Fourier transform quantities which are coordinate dependent. They are used as systematics checks, and it is convenient to be able to connect them directly to the $Q$ and $U$ maps. For the lensing deflection $\bm{p}$, we reconstruct the curl ($\Omega$) and gradient ($\Phi$) components, which are independent of the coordinates. Assuming zero primordial {\bmode}, we write down to leading order the $E_\Bl$ and $B_\Bl$ with a distortion field $D$,
\begin{align}
    B_\Bl &= \int \frac{d^2 \bl_1}{(2\pi)^2} D_{\bl_1} \Tilde{E}_{\bl_2} W^B_{\bl_1, \bl_2} \label{Eq. mbs: EB distortion} \,,\\
    E_\Bl &= \Tilde{E}_\Bl + \int \frac{d^2 \bl_1}{(2\pi)^2} D_{\bl_1} \Tilde{E}_{\bl_2} W^{E}_{\bl_1, \bl_2} \,,
\end{align}
where $\bl_2 = \Bl - \bl_1$ and $W^B$, $W^E$ are weights that can be derived for the individual distortion fields. Similarly, the $E_\Bl$ and $B_\Bl$ generated by the distortions that can be sourced by $T$ to $P$ leakage ($\gamma_{1/2}, d_{1/2}, q$) are written as:
\begin{align}
    B_\Bl &= \int \frac{d^2 \bl_1}{(2\pi)^2} D_{\bl_1} \Tilde{T}_{\bl_2} W^B_{\bl_1, \bl_2} \label{Eq. mbs: TB distortion} \,,\\
    E_\Bl &= \Tilde{E}_\Bl + \int \frac{d^2 \bl_1}{(2\pi)^2} D_{\bl_1} \Tilde{T}_{\bl_2} W^E_{\bl_1, \bl_2} \,,
\end{align}
where the weights $W^B_{\bl_1,\bl_2}$ for generating {\bmode}s are listed in Table~\ref{table: TB/EB quadratic estimator weights appendix}. \\

\begin{table*}[htb]
\centering\scriptsize
\begin{tabular}{c c c c c} 
 \hline \hline 
 \rule{0pt}{4ex}
$D$ & $f^{EB}_{\bl_1,\bl_2}$ & $f^{TB}_{\bl_1,\bl_2}$ & $W^B_{\bl_1,\bl_2}$ & $W^E_{\bl_1,\bl_2}$ \\ [0.5ex] 
    \hline
    \rule{0pt}{3ex}
    $\tau$         & $\CE \sin 2(\phi_{\bl_1}-\phi_{\bl_2})$  
                & $\CTE \sin 2(\phi_{\bl_1}-\phi_{\bl_2})$
                & $\sin 2(\phi_{\bl_2}-\phi_\Bl)$ 
                & $\cos 2(\phi_{\bl_2}-\phi_\Bl)$ \\ [0.5ex] 
    $\alpha$    & $2 \CE \cos 2(\phi_{\bl_1}-\phi_{\bl_2})$ 
                & $2 \CTE \cos 2(\phi_{\bl_1}-\phi_{\bl_2})$ 
                & $2 \cos 2(\phi_{\bl_2}-\phi_\Bl)$
                & $-2 \sin 2(\phi_{\bl_2}-\phi_\Bl)$ \\[0.5ex] 
    $\gamma_a$  & $\CTE \sin 2(\phi_{\Bl}-\phi_{\bl_2})$ 
                & $\CT \sin 2(\phi_{\Bl}-\phi_{\bl_2})$  
                & $\sin 2(\phi_{\bl_1}-\phi_\Bl)$ 
                & $\cos 2(\phi_{\bl_1}-\phi_\Bl)$ \\[0.5ex] 
    $\gamma_b$  & $\CTE \cos 2(\phi_{\Bl}-\phi_{\bl_2})$ 
                & $\CT \cos 2(\phi_{\Bl}-\phi_{\bl_2})$ 
                & $\cos 2(\phi_{\bl_1}-\phi_\Bl)$  
                & $-\sin 2(\phi_{\bl_1}-\phi_\Bl)$ \\ [0.5ex] 
    $f_a$       & $\CE \sin 2(2\phi_{\Bl}-\phi_{\bl_1}-\phi_{\bl_2})$ 
                & $\CTE \sin 2(2\phi_{\Bl}-\phi_{\bl_1}-\phi_{\bl_2})$  
                & $ \sin 2(2\phi_{\bl_1}-\phi_{\bl_2}-\phi_\Bl)$  
                & $ \cos 2(2\phi_{\bl_1}-\phi_{\bl_2}-\phi_\Bl)$ \\[0.5ex] 
    $f_b$       & $\CE \cos 2(2\phi_{\Bl}-\phi_{\bl_1}-\phi_{\bl_2})$ 
                & $\CTE \cos 2(2\phi_{\Bl}-\phi_{\bl_1}-\phi_{\bl_2})$  
                & $ \cos 2(2\phi_{\bl_1}-\phi_{\bl_1}-\phi_\Bl)$  
                & $ -\sin 2(2\phi_{\bl_1}-\phi_{\bl_1}-\phi_\Bl)$ \\[0.5ex] 
    $\Omega$    & $-\CE \sigma (\bl_1 \times \hat{\Bl}) \sin 2(\phi_{\bl_1}-\phi_{\bl_2})$ 
                & $-\CTE \sigma (\bl_1 \times \hat{\Bl}) \sin 2(\phi_{\bl_1}-\phi_{\bl_2})$  
                & $\sigma (\bl_2\times \hat{\bl}_1)\cdot\hat{\textbf{z}} \; \sin 2(\phi_{\bl_2}-\phi_\Bl)$ 
                & $\sigma (\bl_2 \cdot \hat{\bl}_1) \; \sin 2(\phi_{\bl_2}-\phi_\Bl)$  \\[0.5ex] 
    $\Phi$ & $-\CE \sigma (\bl_1 \cdot \hat{\Bl}) \sin 2(\phi_{\bl_1}-\phi_{\bl_2})$ 
                & $-\CTE \sigma (\bl_1 \cdot \hat{\Bl}) \sin 2(\phi_{\bl_1}-\phi_{\bl_2})$ 
                & $\sigma (\bl_2 \cdot \hat{\bl}_1) \; \sin 2(\phi_{\bl_2}-\phi_\Bl)$  
                & $\sigma (\bl_2\times \hat{\bl}_1)\cdot\hat{\textbf{z}} \; \sin 2(\phi_{\bl_2}-\phi_\Bl)$\\[0.5ex] 
    $d_a$       & $\CTE (\bl_1 \sigma) \cos 2(\phi_{\Bl}+\phi_{\bl_1}-2\phi_{\bl_2})$ 
                & $\CT (\bl_1 \sigma) \cos 2(\phi_{\Bl}+\phi_{\bl_1}-2\phi_{\bl_2})$ 
                & $-(l_2\sigma) \cos(\phi_{\bl_1}+\phi_{\bl_2}-2\phi_\Bl)$
                & $-(l_2\sigma) \sin(\phi_{\bl_1}+\phi_{\bl_2}-2\phi_\Bl)$ \\[0.5ex] 
    $d_b$       & $-\CTE (\bl_1 \sigma) \sin 2(\phi_{\Bl}+\phi_{\bl_1}-2\phi_{\bl_2})$ 
                & $-\CT (\bl_1 \sigma) \sin 2(\phi_{\Bl}+\phi_{\bl_1}-2\phi_{\bl_2})$ 
                & $(l_2\sigma) \sin(\phi_{\bl_1}+\phi_{\bl_2}-2\phi_\Bl)$ 
                & $(l_2\sigma) \cos(\phi_{\bl_1}+\phi_{\bl_2}-2\phi_\Bl)$ \\[0.5ex]
    $q$         & $-\CTE (\bl_1\sigma)^2 \sin 2(\phi_{\bl_1}-\phi_{\bl_2})$ 
                & $-\CT (\bl_1\sigma)^2 \sin 2(\phi_{\bl_1}-\phi_{\bl_2})$  
                & $-(l_2\sigma)^2 \sin 2(\phi_{\bl_2}-\phi_\Bl)$   
                & $-(l_2\sigma)^2 \cos 2(\phi_{\bl_2}-\phi_\Bl)$  \\ [1ex] 
    \hline \hline
\end{tabular}
\caption{Weights and filters for the different distortion fields, where $\phi_\bl = \cos^{-1}(\hn \cdot \hat{\bl})$ \citep{Yadav2010}. $f^{D,XB}_{\bl_1, \bl_2}$ are filter functions in Eq.~\ref{Eq. mbs: general quadratic estimator}, and the weight functions $W^B_{\bl_1,\bl_2}$ describe the {\bmode}s generated from the distortions in Eq.~\ref{Eq. mbs: EB distortion} and \ref{Eq. mbs: TB distortion}. See Eq.~\ref{Eq. mbs: Harmonics component of distortion fields.} and Eq.  \ref{Eq. mbs: Fourier component of distortion fields.} for the relation between the Fourier harmonic basis with subscript $a/b$ and the Fourier transform of the distortion fields with subscript $1/2$. }
\label{table: TB/EB quadratic estimator weights appendix}
\end{table*}

For the primordial un-distorted CMB fields, the power spectra are:
\begin{equation}
    \braket{X_{\bl_1}X'_{\bl_2}} = (2\pi)^2\delta(\bl_1+\bl_2)\Tilde{C}_{l_1}^{XX'} \,,
\end{equation}
where $X,X'=T,E,B$, and $\braket{\Tilde{E}_{\bl_1}\Tilde{B}_{\bl_2}}=0$, $\braket{\Tilde{T}_{\bl_1}\Tilde{B}_{\bl_2}}=0$. From Eq.~\ref{Eq. mbs: EB distortion} and Eq.~\ref{Eq. mbs: TB distortion}, we can calculate the ensemble $\braket{X_{\bl_1}B_{\bl_2}}$ correlation averaged over CMB realizations,
\begin{equation}
    \braket{X_{\bl_1}B_{\bl_2}}_{\text{CMB}}=f^{D,XB}_{\bl_1, \bl_2} D_{\Bl} \,,
\label{Eq. mbs: XX' average over CMB}
\end{equation}
where the filter functions $f^{D,XB}$ are listed in Table~\ref{table: TB/EB quadratic estimator weights} for $X=T/E$, and  $\Bl = \bl_1 + \bl_2$.

There is only one universe and one CMB realization available for observation. However, given a $\Bl$, there are many different combinations of $\bl_1$ and $\bl_2$ that satisfies $\bl_1+\bl_2 = \Bl$. Therefore, we can write down a linear combination of $X_{\bl_1}B_{\bl_2}$ with some weight factor $F^{D,XB}_{\bl_1,\bl_2}$ that would minimize the variance of the $\hat{D}_\Bl \propto \int d^2 \bl_1 X_{\bl_1}B_{\bl_2} F^{D,XB}_{\bl_1,\bl_2}$ estimator. The weight $F^{D,XB}_{\bl_1,\bl_2}$ can be derived from:
\begin{equation}
    \frac{\partial}{\partial F^{D,XB}_{\bl_1,\bl_2}}\braket{|\hat{D}^{XB} - D|^2}_{\text{CMB, distortions}}=0 \,,
\end{equation}
where the bracket stands for the average over both CMB and distortion fields realizations. \\

For $X=E$ or $T$, $\Tilde{C}_l^{XB} = 0$. In this case, 
\begin{equation}
    F^{D,XB}_{\bl_1,\bl_2}=\frac{f^{D,XB}_{\bl_1,\bl_2}}{C_{l_1}^{XX} C_{l_2}^{BB}} \,,
\end{equation}
where $f^{D,XB}$ is exactly the factor in Eq.~\ref{Eq. mbs: XX' average over CMB}, and the $C_{l_1}^{XX}$, $C_{l_2}^{BB}$ are the total observed power including contributions from the noise and the distortion fields (usually just lensing). Up to a normalization factor $A^{D,XB}_L$, the quadratic estimator for the distortion field can be written as:
\begin{equation}
    \bar{D}^{XB}_\Bl=A^{D,XB}_L \int \frac{d^2\bl_1}{(2\pi)^2}X_{\bl_1}B_{\bl_2}F^{D,XB}_{\bl_1, \bl_2} \,,
\end{equation}
where $\Bl=\bl_1+\bl_2$, and the analytical form for the normalization factor $A^{D, XB}_L$ is:  
\begin{equation}
    A^{D,XB}_L=\left[ \int \frac{d^2 \bl_1}{(2\pi)^2} f^{D,XB}_{\bl_1,\bl_2} F^{D,XB}_{\bl_1,\bl_2}  \right]^{-1}
    =\left[ \int \frac{d^2 \bl_1}{(2\pi)^2} \frac{(f^{D,XB}_{\bl_1,\bl_2})^2}{C_{l_1}^{XX} C_{l_2}^{BB}}   \right]^{-1} \,.
\end{equation}
The mean-field bias $\braket{\bar{D}^{XB}_\Bl}$ is estimated from the simulations. After applying the correction for the mean-field bias, we have:
\begin{equation}
\hat{D}^{XB}_\Bl = \bar{D}^{XB}_\Bl - \braket{\bar{D}^{XB}_\Bl} \,.
\end{equation}
\vspace{2cm}

\section{Simulations of systematic effects that generate line-of-sight distortions}
\label{appendix: simulations of systematic effects}

\begin{figure}[thb]
\centering
\gridline{\fig{random_rotation_systematics_spectra.pdf}{0.45\textwidth}{(a)10 \% random polarization angle rotation}
\fig{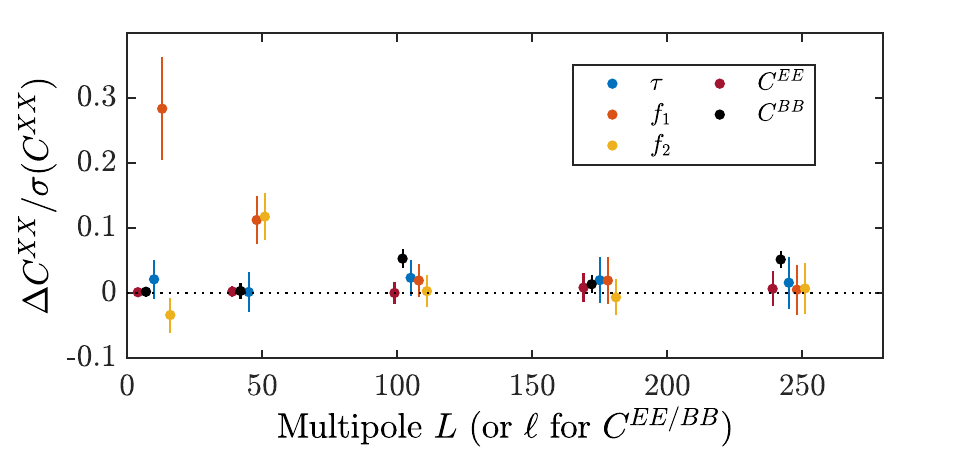}{0.45\textwidth}{(b) 20\% random pair averaged gain fluctuation}
          }
\gridline{\fig{diffgain_systematics_spectra.pdf}{0.45\textwidth}{(c)10 \% hour to hour differential gain fluctuation}
\fig{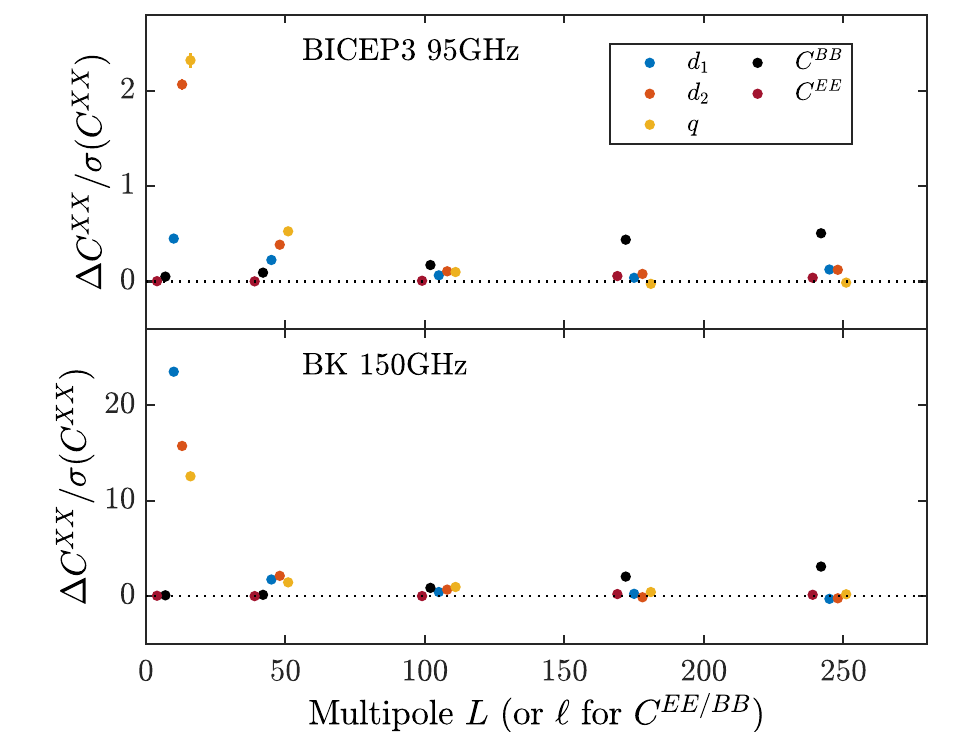}{0.45\textwidth}{(d) Dipole $T$ to $P$ leakage from beam map simulations with differential pointing deprojection turned off.}
          }
\caption{The relevant distortion field, {\it EE} and {\it BB} difference spectra over the errorbar $\Delta C^{XX}/\sigma(C^{XX})$ from the different systematics simulations. Panels (a) and (c) are identical to Figs. \ref{Fig: mbs: random rotation systematics spectra} and
\ref{Fig: mbs: diffgain systematics}
of Sec. \ref{subsection: mbs: instrumental effects that correspond to distortion fields}.
}
\label{Fig: 4 systematics simulations}
\end{figure}

In Section~\ref{subsection: mbs: instrumental effects that correspond to distortion fields}, we presented results for systematics simulations of random polarization and differential gain fluctuations. In this Appendix, we show the results from two more systematics simulations and offer a more in-depth discussion of each of the systematic effects. The four systematics simulations are: 
\begin{enumerate}
    \item 10$\deg$ random detector polarization angle rotation (Fig.~\ref{Fig: 4 systematics simulations}(a)).
    \item 20\% random pair averaged detector gain fluctuations (Fig.~\ref{Fig: 4 systematics simulations}(b)). 
    \item 10\% random differential gain fluctuation varying from hour to hour (Fig.~\ref{Fig: 4 systematics simulations}(c)). 
    \item Dipole component of the $T$ to $P$ leakage from beam map simulations (Fig.~\ref{Fig: 4 systematics simulations}(d)).
\end{enumerate}

The well-designed BK observation strategy leads to a very high degree of cancellation of systematics with an increasing number of detectors and observing time. 
Since the purpose of these simulations are to clearly establish the connections between the different instrument systematics and the distortion fields, we inject large systematic errors in the polarization rotation, gain fluctuation, and differential gain. The simulations are for {\bicepthree} 95\,GHz except for the beam map simulations where we have simulations for both 95\,GHz and 150\,GHz. In each figure, the error bars represent the scatter on the mean spectra over 49 realizations of the systematics simulations. For clarity, we only show the distortion field spectra that are expected to detect the injected systematics. The distortion spectra that are not plotted do not show an elevated signal. \\

For the detector polarization angle errors, we generate misestimated angles by drawing from a Gaussian distribution of mean zero and standard deviation 10$\deg$ for each detector pair for the entire observing season. We rotate the detector polarization angle assumed in the map-making by these angles to produce a set of simulations including polarization angle systematics. The shift in the $\alpha$, $\omega$, {\it EE}, and {\it BB} spectra caused by the random detector rotation are shown in Fig.~\ref{Fig: 4 systematics simulations}(a). Unsurprisingly, the random detector angle rotation creates a {\it washout} effect that reduces $C_\ell^{EE}$. However, the reduction caused by the washout effect and the distortion generated {\it BB} power roughly cancel, leaving the total $C_\ell^{BB}$ largely unchanged. For the distortion reconstruction spectra of the random rotation simulations, $C_L^{\alpha \alpha}$ shows a strong signal at low $L$ as expected. However, $\omega$ also detects the rotation field due to the strong correlation between $\alpha$ and $\omega$ estimators. \\

For the pair-averaged gain systematics, we simulate a 20\% random Gaussian fluctuation on the pair gain. The injected relative gain error is constant over time and only varies from detector pair to detector pair. Due to the observation strategy, most of the detector pairs do not cover the same sky area at multiple boresight rotation angles. This means that a gain fluctuation over detector pairs also sporadically generates $f_1$ and $f_2$ distortions in addition to the amplitude modulation $\tau$ field. We show the $f_1$, $f_2$, $\tau$, $EE$, and $BB$ spectra generated by the gain fluctuation in Fig.~\ref{Fig: 4 systematics simulations}(b). We observe a clear signal in $f_1$ and $f_2$ that corresponds to the injected systematics. However, we do not detect excess power in $\tau$ as one naively expects. One reason is that the $\tau$ {\it EB} estimator is not sensitive enough to detect the distortion effect at the level of 20\% gain fluctuation with only 49 realizations of the systematics simulation. Another reason is the smooth apodization mask combined with the purification matrix which degrades the sensitivity to $\tau$ at the lowest multipole range where most of the $\tau$ distortion power is expected. \\

For the gain mismatches (differential gain) between detector pairs, we simulate a 10\% random Gaussian fluctuation in $(g_A-g_B)/2$ for every detector pair and for every hour of observation. One possible contribution to gain mismatch is the uncertainties in the elevation-nod derived calibration factors (BK-I). With the differential gain deprojection that removes $T$ to $P$ leakage over 10 hour time scales, residual systematics can still arise from a differential gain that varies over shorter periods. Differential gain systematics lead to a monopole $T$ to $P$ leakage that corresponds to $\gamma_1$ and $\gamma_2$. In Fig.~\ref{Fig: 4 systematics simulations}(c), we show the spectra for $\gamma_1$, $\gamma_2$, {\it EE}, and {\it BB}. Compared to the detector angle rotation and gain variation simulations (Fig.~\ref{Fig: 4 systematics simulations}(a) and (b)) where the instrumental effects are constant in time but vary over detector pairs, time-varying gain mismatches generate distortion power that is distributed over a larger range of multipoles. \\

For the instrumental systematics caused by the beam mismatch between orthogonal pairs of detectors, we make use of the $T$ to $P$ leakage template from the beam map simulations described in BK-III and BK-XI. With the high signal-to-noise far-field beam map measurements, the expected $T$ to $P$ leakage signal from the measured beam mismatch is simulated for both 95\,GHz and 150\,GHz. Since the beam map simulations are constant in time, the standard differential pointing deprojection completely removes the dipole leakage in the template and no signal is detected with $d_1$ and $d_2$. As a sanity check, and to demonstrate the power of the quadratic estimators, in Fig.~\ref{Fig: 4 systematics simulations}(d), we show the $d_1$, $d_2$, $q$, {\it BB}, and {\it EE} spectra generated by the dipole component of the leakage signal {\it without} deploying the differential pointing deprojection filter. Without deprojection, $d_1$ and $d_2$ spectra can detect the leakage signal at the lowest multipole with much higher signal-to-noise compared to the {\it EE} and {\it BB} spectra. In addition, $q$ also detects the dipole leakage signal strongly because of its high correlation with $d_1$ and $d_2$ field. \\

In Table~\ref{table: systematics sim summary}, we quantify the impact on the {\it BB} power from the four systematic simulations with an estimator $\rho$ that represents the equivalent tensor-to-scalar ratio $r$ level of the contamination \citep{biceptwoXI}. The estimator is constructed in a similar way to the estimator for the lensing amplitude in Eq.~\ref{Eq. mbs: AL^phi phi amplitude},
\begin{equation}
    \rho = \frac{\sum_{bb'} \hat{C}_b \mathbf{Cov}_{bb'}^{-1} C_{b'}^{r=1}}{\sum_{bb'} C_b^{r=1} \mathbf{Cov}_{bb'}^{-1} C_{b'}^{r=1}} \,,
\label{Eq. equivalent level or r contamination}
\end{equation}
where $\hat{C}_b$ are the {\it BB} bandpowers from the systematics simulations, $C_b^{r=1}$ is the mean {\it BB} bandpowers for an $r=1$ signal, and $\mathbf{Cov}_{bb'}$ is the {\it BB} bandpower covariance matrix of the lensed-\lcdm+dust+noise simulations. Even with the conservatively high levels of systematics in our simulations, the $\rho$ estimates are relatively low at $\lesssim 1\times 10^{-3}$. Note that the large $\rho$ value for $T$ to $P$ dipole leakage are for the case without differential pointing deprojection shown just for demonstration. In the main line analysis with deprojection enabled, the dipole $T$ to $P$ leakage will be entirely filtered out.\\

\begin{table}[htb]
\centering
\begin{tabular}{c c c c } 
 \hline \hline 
 \rule{0pt}{4ex}
Systematics & sensitive distortion fields & equivalent level of $r$ ($\rho$) & sensitivity ratio \\ [0.5ex] 
    \hline
    \rule{0pt}{3ex}
B3 10$^{\deg}$ random polarization angle rotation & $\alpha$, $\omega$ &  $< 5.4\times 10^{-4}$ & $20$ \\ [0.5ex]
B3 20\% pair averaged gain fluctuation & $f_1$, $f_2$ & $< 4.7\times 10^{-4}$ & $3.9$ \\ [0.5ex]
B3 10\% time-varying differential gain & $\gamma_1$, $\gamma_2$ & $< 1.2\times 10^{-3}$ & $1.2$ \\ [0.5ex]
B3 95\,GHz dipole $T$ to $P$ leakage (no deproj.) & $d_1$, $d_2$, $q$ & $6.4\times 10^{-3}$ & $2.6$ \\ [0.5ex]
 150\,GHz dipole $T$ to $P$ leakage (no deproj.) & $d_1$, $d_2$, $q$ & $8.6\times 10^{-2}$ & $4.2$ \\ [0.5ex]
    \hline \hline
\end{tabular}
\caption{A summary of the four systematics simulations. The level of {\it BB} power from the systematics simulations are characterized by $\rho$, the equivalent level of $r$ contamination. The sensitivity ratio shows the detection significance of the distortion field quadratic estimators vs. {\it BB} spectrum at measuring the systematics. A ratio larger than 1 means that the quadratic estimators are more sensitive to the systematic effect. Note that the dipole $T$ to $P$ leakage shown here is the case without the differential pointing deprojection. }
\label{table: systematics sim summary}
\end{table}

From the distortion and {\it BB} bandpowers in Fig.~\ref{Fig: 4 systematics simulations}, it is evident the relevant distortion spectra are able to detect the systematics with higher significance compared to the {\it BB} spectrum. Applying the same formalism described by Eq.~\ref{Eq. mbs: sensitivity comparison between EB/TB vs. BB} and using the mean systematics bandpowers as the fiduical $C_b^f$, we again evaluate the sensitivity ratio to compare the performance of quadratic estimators vs.\ {\it BB} spectra. When the sensitivity ratio is greater than 1, the quadratic estimator is more sensitive than the {\it BB} power spectra in detecting the systematics. In Table~\ref{table: systematics sim summary}, we show the sensitivity ratio of the combined sensitivity of all relevant distortion spectra vs.\ {\it BB} spectra. The sensitivity ratio for the four systematics considered here are all larger than 1, which means that the quadratic estimators for distortion fields are more sensitive than {\it BB} at detecting the spurious {\bmode}s from these systematics. The ratio for random polarization angle rotation is particular striking at 20 due to the fact that the random polarization rotation alters the {\bmode} while keeping the overall {\it BB} power roughly unchanged. \\

\section{PTE values for alternate choices of analysis}
\label{Appendix Alternative choices of analysis}

\begin{table}[htb]
\centering
\begin{tabular}{c | c c c | c c c} 
 \hline \hline 
 \rule{0pt}{2ex}
 & \multicolumn{3}{c|}{95\,GHz} & \multicolumn{3}{c}{150\,GHz}  \\
  \rule{0pt}{2ex}
  & $A_L^{\phi \phi}$ &  vs. lensed-\lcdm &  vs. baseline & $A_L^{\phi \phi}$ &  vs. lensed-\lcdm &  vs. baseline \\ [0.5ex] 
 
  &  & ($\alpha/\tau/\kappa$) & ($\alpha/\tau/\kappa$) & &($\alpha/\tau/\kappa$)  &($\alpha/\tau/\kappa$) \\
    \hline
    \rule{0pt}{3ex}
Baseline & $0.88 \pm 0.23$ & 0.66/0.96/0.71 & N/A & $1.10 \pm 0.33$ & 0.35/0.83/1.00 & N/A \\ [0.5ex]
\hline
$\ell_{\text{min}}=200$ & $0.88 \pm 0.26$ & 0.86/0.58/0.58 & 0.64/0.39/0.36 & $0.78 \pm 0.36$ & 0.57/0.45/0.80 & 0.73/0.31/0.26 \\ [0.5ex]
$\ell_{\text{max}}^B=350$ & $0.89 \pm 0.29$ & 0.57/0.99/0.98 & 0.55/0.99/0.88 & $1.24 \pm 0.42$ & 0.38/0.95/1.00 & 0.44/0.99/0.93 \\ [0.5ex]
$\ell_{\text{max}}=400$ & $0.65 \pm 0.36$ & 0.41/0.95/0.94 & 0.44/0.83/0.93 & $1.68 \pm 0.48$ & 0.15/0.25/0.20 & 0.49/0.12/0.08 \\ [0.5ex]
no diff. ellipticity & $0.88 \pm 0.23$ & 0.64/0.94/0.73 & N/A & $1.07 \pm 0.33$ & 0.42/0.88/1.00 & N/A \\ [0.5ex]
    \hline \hline
\end{tabular}
\caption{The results for alternate analysis choices are shown for $\alpha(\hn), \tau(\hn)$, and $\kappa(\hn)$. $A_L^{\phi \phi}$ is the measured amplitude of the lensing potential, ``vs.\ lensed-{\lcdm}'' shows the $\chi^2$ PTE values of the observed bandpowers compared to the bandpowers from lensed-{\lcdm}+dust+noise simulations, and ``vs. baseline'' shows the $\chi^2$ PTE values when comparing the bandpowers of the alternate analysis with the baseline analysis.}
\label{table: alt analysis choices}
\end{table}

In this appendix, we present the details and the PTE values for the consistency checks listed in Section~\ref{subsection: consistency checks}. The two frequency maps are examined independently for the consistency checks. For $\kappa(\hn)$, we derive an amplitude of the lensing potential for every analysis scenario. For all three fields $\tau(\hn)/ \alpha(\hn)/ \kappa(\hn)$, we evaluate the $\chi^2$ PTEs of whether the different choices of analysis lead to the same conclusion as the baseline result, i.e. consistent with the lensed-{\lcdm} + dust + noise simulations, and also whether the different choices of analysis are consistent with the baseline. \\

The $\chi^2$ PTE for comparing with lensed-\lcdm simulations is exactly the same as Eq.~\ref{Eq. chi2 null test definition.}--\ref{Eq. chi null test definition.}, therefore the PTE values for the baseline case are the same as the numbers in Table~\ref{table: chi and chi2 PTE values from jack0 reconstruction spectra.}. All the PTEs in the ``vs.\ lensed-{\lcdm}" columns in Table~\ref{table: alt analysis choices} are reasonable, which means that the main science result, i.e. that $C_L^{\alpha \alpha}$, $C_L^{\tau \tau}$, and $C_L^{\kappa \kappa}$ are consistent with lensed-{\lcdm}, is not sensitive to the different choices of analyses. For the consistency checks of alternate analysis choices vs. baseline, we evaluate the difference of the bandpowers $\hat{C}_b^{dd}$ from the two analysis,
\begin{equation}
    \Delta \hat{C}_b^{dd} = \hat{C}_b^{dd, \text{alt}} - \hat{C}_b^{dd, \text{baseline}} \,,
\end{equation}
where $\hat{C}_b^{dd, \text{alt}}$ are the reconstructed bandpowers from the alternate analysis and $\hat{C}_b^{dd, \text{baseline}}$ are the bandpowers from the baseline analysis. The $\chi^2$ statistics is constructed as:
\begin{equation}
    \chi_{\text{alt}}^2 = \sum_{bb'} (\Delta \hat{C}_b - \braket{\Delta C_b}) \textbf{Cov}_{bb'}^{-1} (\Delta \hat{C}_{b'} - \braket{\Delta C_{b'}}) \,,
\end{equation}
where $\textbf{Cov}_{bb'}$ is the bandpower covariance matrix from the difference bandpowers from simulations, and $\braket{\Delta C_{b}}$ is the mean difference bandpowers of simulations. The PTE values are then calculated by comparing $\chi_{\text{alt}}^2$ of the data vs.\ simulations.

\section{Exploration of Alternate Foreground Models}
\label{appendix: alternate dust foreground}


\begin{figure*}[htb]
\centering
\includegraphics[width=.49\linewidth]{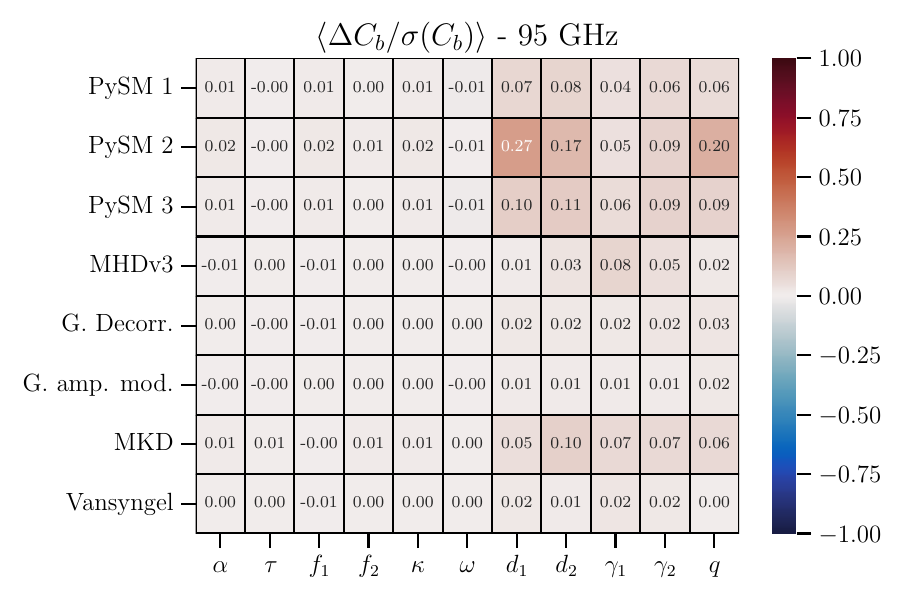}
\includegraphics[width=.49\linewidth]{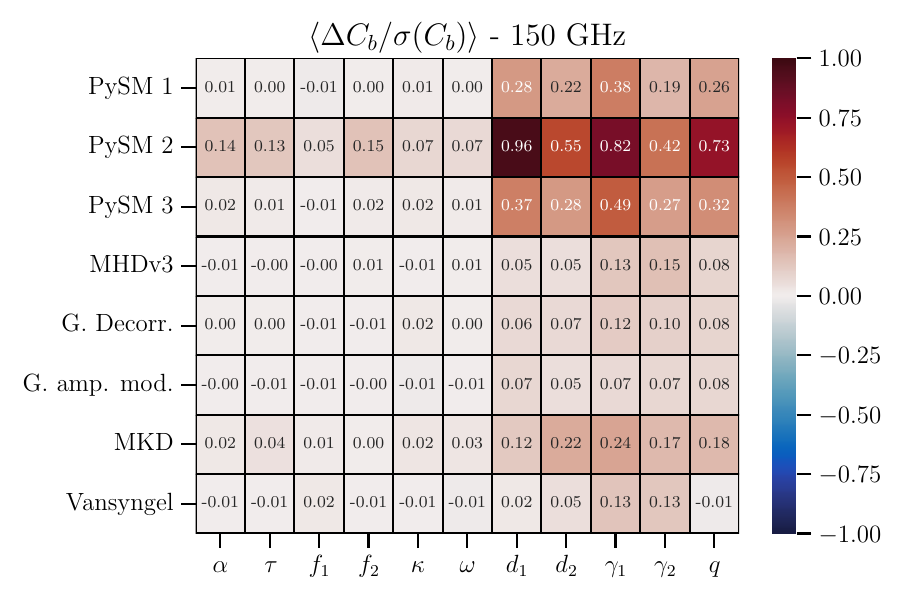}
\caption{The mean band power deviation $\braket{\Delta C_b/\sigma(C_b)}$ averaged over all bins for each of the 11 distortion fields. The choice of bins are the same as Fig.~\ref{Fig: mbs: jack0 CLdd over errorbar} and Fig.~\ref{Fig: mbs: TB null spectra with different deprojection options}.}
\label{figure: alt dust models}
\end{figure*}

In Section~\ref{section: Input E and B modes for reconstruction}, we set a lower bound on the {\bmode} multipole input with $\ell_{\text{min}}^B=100,150$ for 95\,GHz and 150\,GHz respectively to avoid {\bmode}s that have significant contribution from galactic foreground. Here we extend the analysis to simulations using some alternate foreground models to explore the effects of different models of galactic foreground on distortion field reconstructions. The models considered are described in Appendix E.4 of BK-XIII and BK-X. \\

Our basic set of simulations includes Gaussian realizations of dust. One model extends the Gaussian foreground to contain frequency decorrelation (labeled G. Decorr.), which should have no effect in this assessment since we study one frequency at a time. We also extend the Gaussian foreground to have amplitude modulation, where the Gaussian full sky realizations are multiplied by the square root of maps of degree scale {\it BB} power measured from small patches of the {\planck} 353GHz map (labeled G. amp. mod.). \\

A suite of third-party foreground models  with only one realization available are also considered. The PySM models 1, 2, and 3 \citep{thorne17}, the MHD model \citep{kritsuk17,kritsuk18}, the MKD model \citep{mkd}, and the Vansyngel model \citep{vansyngel16}. We find that many of the alternate foreground models would cause significant bias on the reconstructed spectra without the realization-dependent method. However, with the realization-dependent method, the reconstructed spectra to first order are not sensitive to a change in the {\it EE} and {\it BB} power, and the shifts become negligible. This suggests that the dust models considered here do not have a significant impact on the reconstructed distortion fields, but mainly affect the distortion field analysis through altering the overall level of {\it EE} and {\it BB} power. In Fig.~\ref{figure: alt dust models}, we summarize the mean bandpower shift in the reconstructed distortion spectra over bins for each distortion field, 
$\braket{\Delta C_b^{DD}/\sigma(C^{DD}_b)}_{b}$. Note that the PySM models predict considerably higher dust power in the {\bicep}/{\keck} field than is actually observed. The mean shift in bandpowers for distortion fields for which we expect and measure a cosmological signal are negligible.
\\

\section{Detection of polarized point sources}
\label{appendix: Detection of point sources}

In this appendix, we provide an explanation of why the {\it TB} distortion spectra are sensitive to polarized point sources. We further derive $EB/BB/TB$ point source estimators (labelled as $\braket{EBEB}/\braket{BBBB}/\braket{TBTB}$ later in this appendix) of the same quadratic form as Eq.~\ref{Eq. mbs: general quadratic estimator} that are designed specifically to detect point sources. Compared to detecting point sources with excess {\bmode} power, the point source quadratic estimators are more sensitive when the flux from a few individual sources dominates. Additionally, the $TB$ point source estimator is more sensitive than the polarization-only estimators when the polarization fraction is low. Since most of the polarized point source flux comes from the few brightest sources with low polarization fraction ($\approx 2-3$\%) \citep{Tucci2012}, we expect the $TB$ estimators to be most sensitive at detecting polarized point sources in the BK data. \\

There has not been much discussion in the literature about the detection of point sources from temperature and polarization correlations. The main reason is that the random polarization angles of point sources imply that the 2-point functions $C_\ell^{TQ}$ and $C_\ell^{TU}$ from point sources are zero on average \citep{Tucci2012}. However, if we go beyond the 2-point functions, the 4-point functions such as $C_L^{\gamma_1 \gamma_1}$ and $C_L^{\gamma_2 \gamma_2}$ that are constructed from $\braket{TBTB}$ correlations do not cancel out when averaged over random point source polarization angle orientations, and they are evidently sensitive at detecting polarized point sources as shown in Table~\ref{table: point source mask PTE}. \\

A single linearly polarized point source generates $\braket{T(\hn)Q(\hn)}$ or $\braket{T(\hn)U(\hn)}$ correlations at the location of the source. On the other hand, it produces no $\braket{T(\hn)B(\hn)}$ correlations if we assume a radially symmetric and thus even-parity profile. However, with the filter functions in Table~\ref{table: TB/EB quadratic estimator weights} applied, the contribution from a polarized point source to $\gamma_1$ and $\gamma_2$ will be non zero. The power spectra of the quadratic {\it TB} estimators for $\gamma_1$ and $\gamma_2$ are effectively measuring the 4-point $\braket{TQTQ}$ and $\braket{TUTU}$ correlations only using the {\bmode} and not the {\emode} component of the CMB polarization signal. Compared to a direct correlation of the full $\braket{TQTQ}$ and $\braket{TUTU}$, the $\gamma_{1/2}$ {\it TB} estimators will be more sensitive to the point sources because the sample variance is much lower without the contributions from the bright \lcdm {\emode}s. \\

The {\it TB} quadratic estimators with filter functions in Table~\ref{table: TB/EB quadratic estimator weights} are designed to measure the distortion fields and not point sources. It is possible to design better estimators of the same quadratic form as Eq.~\ref{Eq. mbs: general quadratic estimator} that specifically target point sources. The point source estimators from the CMB temperature signal are described in \cite{TTpointsource}, and \cite{NamikawaPolLensing} extend the formalism to include polarization-only point source estimators. Here we extend the point source estimators described in Section 3.1.2 in \cite{NamikawaPolLensing} to include temperature and polarization correlation. We will only consider the 1 source terms (see Section III of \cite{TTpointsource}), ignoring any contribution from clustering of the sources. \\

Let us consider a point source model with sky signal $[T^p(\hn), U^p(\hn), Q^p(\hn)]$ that is uncorrelated between pixels. Assuming the points sources are partially polarized with random orientations, we have:
\begin{align}
\Q (\hn) &= g_1(\hn) \T (\hn) \,, \\
\U (\hn) &= g_2(\hn) \T (\hn) \,,
\end{align}
where 
\begin{align}
\left< g_1(\hn) g_2(\hn')  \right>_{\text{src}} &= 0 \,, \\
\left< g_1(\hn) \T(\hn')  \right>_{\text{src}} &= 0 \,.
\end{align}

With $S(\hn) = \T(\hn)^2$, $\sigma_1(\hn) = g_1(\hn)^2$, and $\sigma_2(\hn) = g_1(\hn)^2$,
\begin{align}
        \braket{ \T(\hn) \T(\hn')}_{\text{src}} &= \braket{S(\hn)}_{\text{src}}\delta(\hn-\hn') \,, \\
\left< g_1(\hn) g_1(\hn')  \right>_{\text{src}} &= \braket{\sigma_1(\hn)}_{\text{src}}\delta(\hn-\hn')\,, \\
\left< g_2(\hn) g_2(\hn')  \right>_{\text{src}} &= \braket{\sigma_2(\hn)}_{\text{src}}\delta(\hn-\hn')\,,
\end{align}
where the bracket stands for mean over point source realizations given our point source model, and $\braket{\sigma_1(\hn)}_\text{src} = \braket{\sigma_2{\hn}}_\text{src}$.

\begin{figure}[htb]
\centering
\includegraphics[width=0.5\linewidth]{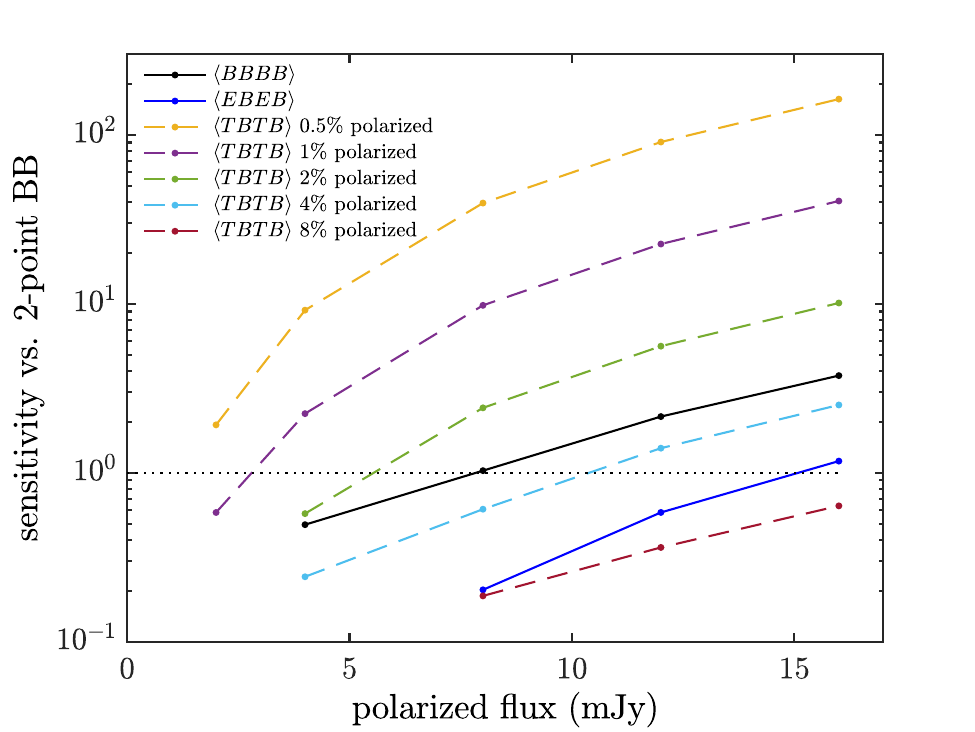}
\caption{The signal-to-noise for detecting point sources of the different 4-point estimators compared to 2-point $C_\ell^{BB}$ power in the BK18 95\,GHz simulations. A sensitivity ratio of larger than 1 on the y-axis means the 4-point estimator is more sensitive than the $BB$ power spectrum at detecting that population (polarized flux and fraction) of point sources. $\braket{EEEE}$ has too low sensitivity to be probed by our simulations and is therefore not shown.}
\label{Fig: point source EB BB TB vs. 2 point BB}
\end{figure}

Recall that a correlation of Eq.~\ref{Eq. mbs: XX' average over CMB} would lead to a minimum variance quadratic estimator of Eq.~\ref{Eq. mbs: general quadratic estimator}. Therefore, we evaluate $\braket{X_{\bl_1}X'_{\bl_2}}_{\text{CMB}}$ with $XX' = EE, EB, BB, TB$. Note that we are taking the ensemble average over the CMB realizations here. With $[E^p_\bl \pm i B^p_\bl] = \int d^2 \hn e^{-i\hn \cdot \bl} [Q^p \pm i U^p](\hn) e^{\mp 2i \phi_\bl}$, we have:
\begin{align}
        \braket{T_{\bl_1} T_{\bl_2}}_{\text{CMB}} &= 
        S_\Bl \,,\\
        \braket{  E_{\bl_1} E_{\bl_2}    }_{\text{CMB}} &= 
        [S\sigma_1]_\Bl \cos(2\phi_{\bl_1})\cos(2\phi_{\bl_2}) + 
        [S\sigma_2]_\Bl \sin(2\phi_{\bl_1})\sin(2\phi_{\bl_2})\,,\\
        \left<  B_{\bl_1} B_{\bl_2}    \right>_{\text{CMB}} &= 
        [S\sigma_1]_\Bl \cos(2\phi_{\bl_1})\cos(2\phi_{\bl_2}) + 
        [S\sigma_2]_\Bl \sin(2\phi_{\bl_1})\sin(2\phi_{\bl_2})\,,\\
        \left<  E_{\bl_1} B_{\bl_2}    \right>_{\text{CMB}} &= 
        - [S\sigma_1]_\Bl \cos(2\phi_{\bl_1})\sin(2\phi_{\bl_2})
        + [S\sigma_2]_\Bl \sin(2\phi_{\bl_1})\cos(2\phi_{\bl_2}) \,,\\
        \braket{  T_{\bl_1} B_{\bl_2}    }_{\text{CMB}} &= 
        [Sg_2]_\Bl \cos(2\phi_{\bl_2}) - [Sg_1]_\Bl \sin(2\phi_{\bl_2}) \,,
\end{align}
where $[S\sigma_i]_\Bl = \int d^2 \hn e^{-i\hn \cdot \Bl} S(\hn) \sigma_i(\hn)$ and similarly for $[Sg_i]_\Bl$. If we consider only the common modes of $S\sigma_1$ and $S\sigma_2$, the filter functions of the $EE, EB, BB$ point source estimators are $f_{\bl_1,\bl_2}^{EE} = f_{\bl_1,\bl_2}^{BB} = \cos(2\phi_{\bl_1}-2\phi_{\bl_2})$ and $f_{\bl_1,\bl_2}^{EB} = \sin(2\phi_{\bl_1}-2\phi_{\bl_2})$. For the {\it TB} estimator, we have $f_{\bl_1,\bl_2}^{TB,1} = -\sin(2\phi_{\bl_2})$ and $f_{\bl_1,\bl_2}^{TB,2} = \cos(2\phi_{\bl_2})$ that probe $\braket{TQTQ}$ and $\braket{TUTU}$ respectively. It is also possible to compute the cross spectrum of the two {\it TB} estimators to probe $\braket{TQTU}$. Note that the two {\it TB} point source estimators for $Sg_1$ and $Sg_2$ have the same geometric terms ($\sin(2\phi_{\bl_2})$ and $\cos(2\phi_{\bl_2})$) as the distortion field estimators for $\gamma_1(\hn)$ and $\gamma_2(\hn)$. The only difference is that there is no longer the $C_\ell^{TT}$ factor in the filter function $f$. This change in the weights result in a factor of $\sim 2$ sensitivity improvement at detecting point sources in the {\bicepthree} 95\,GHz maps. \\

The polarized point sources also produce excess $C_{\ell}^{BB}$ power especially at higher $\ell$. With the same method as Section~\ref{subsection: QE vs. BB at detecting distortion fields}, we compare the signal-to-noise of detecting the polarized point sources with the 4-point point source estimators vs. the 2-point $C_\ell^{BB}$ by running simulations. The parameter space explored are the polarized point source flux for individual sources (from 2 to 16 mJy) and the fraction of polarization (from 0.5\% to 8\%). For each set of parameters, we generated 99 realizations of 40 point sources with the same flux at random locations in the map and with random polarization orientations. In Fig.~\ref{Fig: point source EB BB TB vs. 2 point BB}, we find that all the point source estimators (4-point functions) have sensitivity that scales as $\text{flux}^2$ relative to the 2-point function $C_\ell^{BB}$, since the signal in a 4-point functions is $\propto \text{flux}^4$ while the signal in a 2-point function is $\propto \text{flux}^2$. The $\braket{BBBB}$ estimator is much more sensitive than the $\braket{EBEB}$ and $\braket{EEEE}$ estimators, since $\braket{BBBB}$ does not have the large sample variance contribution from {\lcdm} {\emode}s. Additionally, the $\braket{TBTB}$ estimator is more sensitive than the polarization only point source estimators when the polarization fraction is small. When the point sources are less than $\approx 3\%$ polarized, the $\braket{TBTB}$ estimator is more sensitive compared to the $\braket{BBBB}$ estimator. 

\bibliography{ms}{}
\bibliographystyle{aasjournal}

\end{document}

%% file: defs.tex





\newcommand{\CE}[0]{\Tilde{C}{}^{EE}_{l_1}}
\newcommand{\CX}[0]{\Tilde{C}{}^{XX}_{l_1}}
\newcommand{\CT}[0]{\Tilde{C}{}^{TT}_{l_1}}
\newcommand{\CTE}[0]{\Tilde{C}{}^{TE}_{l_1}}
\newcommand{\CXE}[0]{\Tilde{C}{}^{XE}_{l_1}}
\newcommand{\CTX}[0]{\Tilde{C}{}^{TX}_{l_1}}
\newcommand{\Q}[0]{Q^p}
\newcommand{\U}[0]{U^p}
\newcommand{\T}[0]{T^p}

\newcommand{\vdag}{(v)^\dagger}
\newcommand\aastex{AAS\TeX}
\newcommand\latex{La\TeX}
\newcommand{\Tt}[0]{\tilde{T}}
\newcommand{\Qt}[0]{\tilde{Q}}
\newcommand{\Ut}[0]{\tilde{U}}
\newcommand{\hn}[0]{\bm{\hat{n}}}
\newcommand{\bl}[0]{{\bm l}}
\newcommand{\Bl}[0]{{\bm L}}

\def\act{ACT} 
\def\bicep{BICEP}
\def\bolocam{{\sc Bolocam}}
\def\dasi{DASI}
\def\python{{\sc Python}}
\def\bicepone{{\sc BICEP1}}
\def\biceptwo{{\sc BICEP2}}
\def\bicepthree{{\sc BICEP3}}
\def\biceparray{{\sc BICEP} Array}
\def\spud{{\sc Spud}}
\def\spudone{{\sc Spud1}}
\def\spudsix{{\sc Spud6}}
\def\keck{{\it Keck}}
\def\keckarray{{\it Keck Array}}
\def\bk{\bicep/\keck}
\def\BKfourteen{{BK14}}
\def\BKthirteen{{BK13}}
\def\planck{{\it Planck}} 
\def\quiet{QUIET}
\def\spider{SPIDER}
\def\acbar{{\sc Acbar}}
\def\QUAD{QUAD}
\def\cbi{CBI}
\def\capmap{CAPMAP}
\def\maxipol{MAXIPOL}
\def\archeops{{\it Archeops}}
\def\cmbpol{{\sc CMB}pol}
\def\ebex{EBEX}
\def\boom{BOOMERANG}
\def\wmap{WMAP}
\def\spt{SPT}
\def\sza{{\sc SZA}}
\def\polarbear{POLARBEAR}
\def\abs{ABS}
\def\actpol{ACTPOL}
\def\sptpol{{\sc SPTpol}}
\def\class{CLASS}
\def\piper{PIPER}
\def\spass{S-PASS}

\def\cmbfast{{\tt CMBFAST}}
\def\camb{CAMB}
\def\xfaster{{\tt XFASTER}}
\def\anafast{{\tt anafast}}
\def\synfast{{\sc synfast}}
\def\healpix{{\sc healp}ix}
\def\lenspix{{\sc lenspix}}
\def\master{MASTER}
\def\spice{{\tt Spice}}
\def\gcp{{\tt gcp}}

\newcommand{\ukrts}{ $\mu\mathrm{K}_{\mathrm{\mbox{\tiny\sc cmb}}}\sqrt{\mathrm{s}}$}
\newcommand{\ukrjrts}{ $\mu\mathrm{K}_{\mathrm{\mbox{\tiny\sc rj}}}\sqrt{\mathrm{s}}$ }
\newcommand{\ukcmbrts}{ $\mu\mathrm{K}_{\mathrm{\mbox{\tiny\sc cmb}}}\sqrt{\mathrm{s}}$ }
\def\uk{$\mu{\mathrm K}$}
\def\uksq{$\mu{\mathrm K^2}$}
\def\ukcmb{$\mu{\mathrm K}_{\mathrm{\mbox{\tiny\sc cmb}}}$}
\def\ukrj{${{\mu\mathrm{K}_{\mathrm{\mbox{\tiny\sc rj}}}}}$}
\def\deg{^\circ}
\def\emode{$E$-mode}
\def\bmode{$B$-mode}
\newcommand{\cl}{$C_\ell$ }
\def\clstar{\ell \left( \ell + 1 \right) C_\ell / 2 \pi}
\def\lcdm{$\Lambda$CDM}
\def\grad{{\vec \nabla}}
\def\hii{H{\sc ii}}
\def\tp{$T \rightarrow P$}

\def\etal{{\em et al.}}

\def\bI{BK-I}
\def\bV{BK-V}
\def\piXXX{PIP-XXX}

\def\ieeesc{IEEE Trans.\ Appl.\ Supercon.}
\def\sovast{Sov.\ Astron.}
\def\aap{Astron.\ Astrophys.}
\def\ap{Appl.\  Phys.}
\def\apl{Appl.\ Phys.\ Lett.}
\def\aj{Astron.\ J.}
\def\apj{Astrophys.\ J.}
\def\apjl{Astrophys.\ J.\ Lett.}
\def\apjs{Astrophys.\ J.\ Suppl.\ Ser.}
\def\astropart{Astropart.\ Phys.}
\def\baas{Bull. Am. Astron. Soc.}
\def\jap{J.\ Appl.\  Phys.}
\def\jcap{J.\ Cosmol.\ Astropart.\ Phys.}
\def\jetplett{JETP\ Lett.}
\def\nat{Nature}
\def\mnras{Mon.\ Not.\ R.\ Astron.\ Soc.}
\def\nima{Nucl.\ Instrum.\ Methods\ Phys.\ Res.,\ Sect.\ A}
\def\newastr{New\ Astron.}
\def\newastrev{New\ Astron.\ Rev.}
\def\nucphysb{Nucl.\ Phys.\ B}
\def\nucphysbsupp{Nucl.\ Phys.\ B\ Proc. Supp.}
\def\nuovocim{Il\ Nuovo\ Cim.}
\def\pl{Phys.\ Lett.}
\def\plb{Phys.\ Lett.\ B}
\def\physrep{Phys.\ Rep.}
\def\physscripta{Phys.\ Scr.}
\def\pra{Phys.\ Rev.\ A}
\def\prb{Phys.\ Rev.\ B}
\def\prc{Phys.\ Rev.\ C}
\def\prd{Phys.\ Rev.\ D}
\def\pre{Phys.\ Rev.\ E}
\def\prl{Phys.\ Rev.\ Lett.}
\def\prthphys{Prog.\ Theor.\ Phys.}
\def\pspie{Proc.\ Soc.\ Photo-Opt.\ Instrum.\ Eng.}
\def\rmp{Rev.\ Mod.\ Phys.}
\def\rsi{Rev.\ Sci.\ Instr.}
\def\sci{Science}

%% file: authors.tex
\author{BICEP/$Keck$ Collaboration: P.~A.~R.~Ade}
\affiliation{School of Physics and Astronomy, Cardiff University, Cardiff, CF24 3AA, United Kingdom}

\author{Z.~Ahmed}
\affiliation{Kavli Institute for Particle Astrophysics and Cosmology, SLAC National Accelerator Laboratory, 2575 Sand Hill Rd, Menlo Park, CA 94025, USA}

\author{M.~Amiri}
\affiliation{Department of Physics and Astronomy, University of British Columbia, Vancouver, British Columbia, V6T 1Z1, Canada}

\author{D.~Barkats}
\affiliation{Center for Astrophysics, Harvard \& Smithsonian, Cambridge, MA 02138, USA}

\author{R.~Basu~Thakur}
\affiliation{Department of Physics, California Institute of Technology, Pasadena, CA 91125, USA}

\author{C.~A.~Bischoff}
\affiliation{Department of Physics, University of Cincinnati, Cincinnati, OH 45221, USA}

\author{D.~Beck}
\affiliation{Department of Physics, Stanford University, Stanford, CA 94305, USA}
\affiliation{Kavli Institute for Particle Astrophysics and Cosmology, SLAC National Accelerator Laboratory, 2575 Sand Hill Rd, Menlo Park, CA 94025, USA}

\author{J.~J.~Bock}
\affiliation{Department of Physics, California Institute of Technology, Pasadena, CA 91125, USA}
\affiliation{Jet Propulsion Laboratory, Pasadena, CA 91109, USA}

\author{H.~Boenish}
\affiliation{Center for Astrophysics, Harvard \& Smithsonian, Cambridge, MA 02138, USA}

\author{E.~Bullock}
\affiliation{Minnesota Institute for Astrophysics, University of Minnesota, Minneapolis, MN 55455, USA}

\author{V.~Buza}
\affiliation{Kavli Institute for Cosmological Physics, University of Chicago, Chicago, IL 60637, USA}

\author{J.~R.~Cheshire~IV}
\affiliation{Minnesota Institute for Astrophysics, University of Minnesota, Minneapolis, MN 55455, USA}

\author{J.~Connors}
\affiliation{Center for Astrophysics, Harvard \& Smithsonian, Cambridge, MA 02138, USA}

\author{J.~Cornelison}
\affiliation{Center for Astrophysics, Harvard \& Smithsonian, Cambridge, MA 02138, USA}

\author{M.~Crumrine}
\affiliation{School of Physics and Astronomy, University of Minnesota, Minneapolis, MN 55455, USA}

\author{A.~Cukierman}
\affiliation{Department of Physics, Stanford University, Stanford, CA 94305, USA}
\affiliation{Kavli Institute for Particle Astrophysics and Cosmology, SLAC National Accelerator Laboratory, 2575 Sand Hill Rd, Menlo Park, CA 94025, USA}
\affiliation{Department of Physics, California Institute of Technology, Pasadena, CA 91125, USA}

\author{E.~V.~Denison}
\affiliation{National Institute of Standards and Technology, Boulder, CO 80305, USA}

\author{M.~Dierickx}
\affiliation{Center for Astrophysics, Harvard \& Smithsonian, Cambridge, MA 02138, USA}

\author{L.~Duband}
\affiliation{Service des Basses Temp\'{e}ratures, Commissariat \`{a} l'Energie Atomique, 38054 Grenoble, France}

\author{M.~Eiben}
\affiliation{Center for Astrophysics, Harvard \& Smithsonian, Cambridge, MA 02138, USA}

\author{S.~Fatigoni}
\affiliation{Department of Physics and Astronomy, University of British Columbia, Vancouver, British Columbia, V6T 1Z1, Canada}

\author{J.~P.~Filippini}
\affiliation{Department of Physics, University of Illinois at Urbana-Champaign, Urbana, IL 61801, USA}
\affiliation{Department of Astronomy, University of Illinois at Urbana-Champaign, Urbana, IL 61801, USA}

\author{S.~Fliescher}
\affiliation{School of Physics and Astronomy, University of Minnesota, Minneapolis, MN 55455, USA}

\author{C.~Giannakopoulos}
\affiliation{Department of Physics, University of Cincinnati, Cincinnati, OH 45221, USA}

\author{N.~Goeckner-Wald}
\affiliation{Department of Physics, Stanford University, Stanford, CA 94305, USA}

\author{D.~C.~Goldfinger}
\affiliation{Center for Astrophysics, Harvard \& Smithsonian, Cambridge, MA 02138, USA}

\author{J.~Grayson}
\affiliation{Department of Physics, Stanford University, Stanford, CA 94305, USA}

\author{P.~Grimes}
\affiliation{Center for Astrophysics, Harvard \& Smithsonian, Cambridge, MA 02138, USA}

\author{G.~Hall}
\affiliation{School of Physics and Astronomy, University of Minnesota, Minneapolis, MN 55455, USA}

\author{G.~Halal}
\affiliation{Department of Physics, Stanford University, Stanford, CA 94305, USA}

\author{M.~Halpern}
\affiliation{Department of Physics and Astronomy, University of British Columbia, Vancouver, British Columbia, V6T 1Z1, Canada}

\author{E.~Hand}
\affiliation{Department of Physics, University of Cincinnati, Cincinnati, OH 45221, USA}

\author{S.~Harrison}
\affiliation{Center for Astrophysics, Harvard \& Smithsonian, Cambridge, MA 02138, USA}

\author{S.~Henderson}
\affiliation{Kavli Institute for Particle Astrophysics and Cosmology, SLAC National Accelerator Laboratory, 2575 Sand Hill Rd, Menlo Park, CA 94025, USA}

\author{S.~R.~Hildebrandt}
\affiliation{Department of Physics, California Institute of Technology, Pasadena, CA 91125, USA}
\affiliation{Jet Propulsion Laboratory, Pasadena, CA 91109, USA}


\author{J.~Hubmayr}
\affiliation{National Institute of Standards and Technology, Boulder, CO 80305, USA}

\author{H.~Hui}
\affiliation{Department of Physics, California Institute of Technology, Pasadena, CA 91125, USA}

\author{K.~D.~Irwin}
\affiliation{Department of Physics, Stanford University, Stanford, CA 94305, USA}
\affiliation{Kavli Institute for Particle Astrophysics and Cosmology, SLAC National Accelerator Laboratory, 2575 Sand Hill Rd, Menlo Park, CA 94025, USA}
\affiliation{National Institute of Standards and Technology, Boulder, CO 80305, USA}

\author{J.~Kang}
\affiliation{Department of Physics, Stanford University, Stanford, CA 94305, USA}
\affiliation{Department of Physics, California Institute of Technology, Pasadena, CA 91125, USA}

\author{K.~S.~Karkare}
\affiliation{Center for Astrophysics, Harvard \& Smithsonian, Cambridge, MA 02138, USA}
\affiliation{Kavli Institute for Cosmological Physics, University of Chicago, Chicago, IL 60637, USA}

\author{E.~Karpel}
\affiliation{Department of Physics, Stanford University, Stanford, CA 94305, USA}

\author{S.~Kefeli}
\affiliation{Department of Physics, California Institute of Technology, Pasadena, CA 91125, USA}

\author{S.~A.~Kernasovskiy}
\affiliation{Department of Physics, Stanford University, Stanford, CA 94305, USA}

\author{J.~M.~Kovac}
\affiliation{Center for Astrophysics, Harvard \& Smithsonian, Cambridge, MA 02138, USA}
\affiliation{Department of Physics, Harvard University, Cambridge, MA 02138, USA}

\author{C.~L.~Kuo}
\affiliation{Department of Physics, Stanford University, Stanford, CA 94305, USA}
\affiliation{Kavli Institute for Particle Astrophysics and Cosmology, SLAC National Accelerator Laboratory, 2575 Sand Hill Rd, Menlo Park, CA 94025, USA}

\author{K.~Lau}
\affiliation{School of Physics and Astronomy, University of Minnesota, Minneapolis, MN 55455, USA}

\author{E.~M.~Leitch}
\affiliation{Kavli Institute for Cosmological Physics, University of Chicago, Chicago, IL 60637, USA}

\author{A.~Lennox}
\affiliation{Department of Physics, University of Illinois at Urbana-Champaign, Urbana, IL 61801, USA}

\author{K.~G.~Megerian}
\affiliation{Jet Propulsion Laboratory, Pasadena, CA 91109, USA}

\author{L.~Minutolo}
\affiliation{Department of Physics, California Institute of Technology, Pasadena, CA 91125, USA}

\author{L.~Moncelsi}
\affiliation{Department of Physics, California Institute of Technology, Pasadena, CA 91125, USA}

\author{Y.~Nakato}
\affiliation{Department of Physics, Stanford University, Stanford, CA 94305, USA}

\author{T.~Namikawa}
\affiliation{Kavli Institute for the Physics and Mathematics of the Universe (WPI), UTIAS, The~University~of~Tokyo, Kashiwa, Chiba 277-8583, Japan}

\author{H.~T.~Nguyen}
\affiliation{Jet Propulsion Laboratory, Pasadena, CA 91109, USA}

\author{R.~O'Brient}
\affiliation{Department of Physics, California Institute of Technology, Pasadena, CA 91125, USA}
\affiliation{Jet Propulsion Laboratory, Pasadena, CA 91109, USA}

\author{R.~W.~Ogburn~IV}
\affiliation{Department of Physics, Stanford University, Stanford, CA 94305, USA}
\affiliation{Kavli Institute for Particle Astrophysics and Cosmology, SLAC National Accelerator Laboratory, 2575 Sand Hill Rd, Menlo Park, CA 94025, USA}

\author{S.~Palladino}
\affiliation{Department of Physics, University of Cincinnati, Cincinnati, OH 45221, USA}

\author{M.~Petroff}
\affiliation{Center for Astrophysics, Harvard \& Smithsonian, Cambridge, MA 02138, USA}

\author{T.~Prouve}
\affiliation{Service des Basses Temp\'{e}ratures, Commissariat \`{a} l'Energie Atomique, 38054 Grenoble, France}

\author{C.~Pryke}
\affiliation{School of Physics and Astronomy, University of Minnesota, Minneapolis, MN 55455, USA}
\affiliation{Minnesota Institute for Astrophysics, University of Minnesota, Minneapolis, MN 55455, USA}

\author{B.~Racine}
\affiliation{Center for Astrophysics, Harvard \& Smithsonian, Cambridge, MA 02138, USA}
\affiliation{Aix-Marseille  Universit\'{e},  CNRS/IN2P3,  CPPM,  13288 Marseille,  France}

\author{C.~D.~Reintsema}
\affiliation{National Institute of Standards and Technology, Boulder, CO 80305, USA}

\author{S.~Richter}
\affiliation{Center for Astrophysics, Harvard \& Smithsonian, Cambridge, MA 02138, USA}

\author{A.~Schillaci}
\affiliation{Department of Physics, California Institute of Technology, Pasadena, CA 91125, USA}

\author{R.~Schwarz}
\affiliation{School of Physics and Astronomy, University of Minnesota, Minneapolis, MN 55455, USA}

\author{B.~L.~Schmitt}
\affiliation{Center for Astrophysics, Harvard \& Smithsonian, Cambridge, MA 02138, USA}

\author{C.~D.~Sheehy}
\affiliation{Physics Department, Brookhaven National Laboratory, Upton, NY 11973, USA}

\author{B.~Singari}
\affiliation{Minnesota Institute for Astrophysics, University of Minnesota, Minneapolis, MN 55455, USA}

\author{A.~Soliman}
\affiliation{Department of Physics, California Institute of Technology, Pasadena, CA 91125, USA}

\author{T.~St.~Germaine}
\affiliation{Center for Astrophysics, Harvard \& Smithsonian, Cambridge, MA 02138, USA}
\affiliation{Department of Physics, Harvard University, Cambridge, MA 02138, USA}

\author{B.~Steinbach}
\affiliation{Department of Physics, California Institute of Technology, Pasadena, CA 91125, USA}

\author{R.~V.~Sudiwala}
\affiliation{School of Physics and Astronomy, Cardiff University, Cardiff, CF24 3AA, United Kingdom}

\author{G.~P.~Teply}
\affiliation{Department of Physics, California Institute of Technology, Pasadena, CA 91125, USA}

\author{K.~L.~Thompson}
\affiliation{Department of Physics, Stanford University, Stanford, CA 94305, USA}
\affiliation{Kavli Institute for Particle Astrophysics and Cosmology, SLAC National Accelerator Laboratory, 2575 Sand Hill Rd, Menlo Park, CA 94025, USA}

\author{J.~E.~Tolan}
\affiliation{Department of Physics, Stanford University, Stanford, CA 94305, USA}

\author{C.~Tucker}
\affiliation{School of Physics and Astronomy, Cardiff University, Cardiff, CF24 3AA, United Kingdom}

\author{A.~D.~Turner}
\affiliation{Jet Propulsion Laboratory, Pasadena, CA 91109, USA}

\author{C.~Umilt\`{a}}
\affiliation{Department of Physics, University of Cincinnati, Cincinnati, OH 45221, USA}
\affiliation{Department of Physics, University of Illinois at Urbana-Champaign, Urbana, IL 61801, USA}

\author{C.~Verg\`{e}s}
\affiliation{Center for Astrophysics, Harvard \& Smithsonian, Cambridge, MA 02138, USA}

\author{A.~G.~Vieregg}
\affiliation{Department of Physics, Enrico Fermi Institute, University of Chicago, Chicago, IL 60637, USA}
\affiliation{Kavli Institute for Cosmological Physics, University of Chicago, Chicago, IL 60637, USA}

\author{A.~Wandui}
\affiliation{Department of Physics, California Institute of Technology, Pasadena, CA 91125, USA}

\author{A.~C.~Weber}
\affiliation{Jet Propulsion Laboratory, Pasadena, CA 91109, USA}

\author{D.~V.~Wiebe}
\affiliation{Department of Physics and Astronomy, University of British Columbia, Vancouver, British Columbia, V6T 1Z1, Canada}

\author{J.~Willmert}
\affiliation{School of Physics and Astronomy, University of Minnesota, Minneapolis, MN 55455, USA}

\author{C.~L.~Wong}
\affiliation{Center for Astrophysics, Harvard \& Smithsonian, Cambridge, MA 02138, USA}
\affiliation{Department of Physics, Harvard University, Cambridge, MA 02138, USA}

\author{W.~L.~K.~Wu}
\affiliation{Kavli Institute for Particle Astrophysics and Cosmology, SLAC National Accelerator Laboratory, 2575 Sand Hill Rd, Menlo Park, CA 94025, USA}

\author{H.~Yang}
\affiliation{Department of Physics, Stanford University, Stanford, CA 94305, USA}

\author{K.~W.~Yoon}
\affiliation{Department of Physics, Stanford University, Stanford, CA 94305, USA}
\affiliation{Kavli Institute for Particle Astrophysics and Cosmology, SLAC National Accelerator Laboratory, 2575 Sand Hill Rd, Menlo Park, CA 94025, USA}

\author{E.~Young}
\affiliation{Department of Physics, Stanford University, Stanford, CA 94305, USA}
\affiliation{Kavli Institute for Particle Astrophysics and Cosmology, SLAC National Accelerator Laboratory, 2575 Sand Hill Rd, Menlo Park, CA 94025, USA}

\author{C.~Yu}
\affiliation{Department of Physics, Stanford University, Stanford, CA 94305, USA}

\author{L.~Zeng}
\affiliation{Center for Astrophysics, Harvard \& Smithsonian, Cambridge, MA 02138, USA}

\author{C.~Zhang}
\affiliation{Department of Physics, California Institute of Technology, Pasadena, CA 91125, USA}

\author{S.~Zhang}
\affiliation{Department of Physics, California Institute of Technology, Pasadena, CA 91125, USA}

\correspondingauthor{H.~Yang}
\email{iameric@stanford.edu}